\pdfoutput=1

\newcommand*{\ATLASLATEXPATH}{}
\newif\ifIsCONF
\IsCONFfalse

\documentclass[USenglish,texlive=2016,cernpreprint]{\ATLASLATEXPATH atlasdoc}
\usepackage[subfigure]{\ATLASLATEXPATH atlaspackage}
\usepackage{multirow,pdflscape}
\usepackage{\ATLASLATEXPATH atlasbiblatex}

\usepackage[xtab]{\ATLASLATEXPATH atlascontribute}
\usepackage{changepage}
\usepackage{comment}
\usepackage{fixme}

\usepackage{\ATLASLATEXPATH atlasphysics}

\graphicspath{}

\usepackage{ttHMLRun2Paper-defs}
\makeatletter
\def\blx@maxline{77}
\makeatother

\newif\ifIsCOM 
\IsCOMtrue
\newif\ifIsML
\IsMLtrue

\AtlasTitle{Evidence for the associated production of the Higgs boson and a top quark pair with the ATLAS detector}

\AtlasRefCode{HIGG-2017-02}

\ifIsCONF
\AtlasNote{ATLAS-CONF-2017-77}
\else
\AtlasNote{HIGG-2017-02}
\fi

\PreprintIdNumber{CERN-EP-2017-281}

\AtlasJournal{Phys.\ Rev.\ D.}
 \AtlasJournalRef{Phys.\ Rev.\ D. 97 (2018) 072003}
 \AtlasDOI{10.1103/PhysRevD.97.072003}

\AtlasAbstract{%
A search for the associated production of the Higgs boson with a top quark pair (\ttH) is reported.
The search is performed in multilepton final states using a dataset corresponding to an integrated luminosity
of 36.1~\ifb\ of proton--proton collision data recorded by the ATLAS experiment at a center-of-mass energy \ThirTeV\ at the Large Hadron Collider. Higgs boson decays to ${WW^*}$, \ensuremath{\tau\tau}, and $\ensuremath{ZZ^*}$ are targeted. Seven final states, categorized by the number and flavor of charged-lepton candidates, are examined for the presence of the Standard Model Higgs boson with a mass of 125 GeV and a pair of top quarks. An excess of events over the expected background from Standard Model processes is found with an observed significance of 4.1 standard deviations, compared to an expectation of 2.8 standard deviations.
The best fit for the \ttH production cross section is $\sigma(\ttH) = 790^{+230}_{-210}$~fb, in agreement with the Standard Model prediction of $507^{+35}_{-50}$~fb.
The combination of this result with other \ttH searches from the ATLAS experiment using the Higgs boson decay modes to $b\bbar$, $\gamma\gamma$ and $ZZ^* \to 4\ell$, has an observed significance of 4.2 standard deviations, compared to an expectation of 3.8 standard deviations. This provides evidence for the \ttH production mode.
}

\author{The ATLAS Collaboration}

\hypersetup{pdftitle={Evidence for the associated production of the Higgs boson and a top quark pair with the ATLAS detector},pdfauthor={ATLAS Collaboration}}

\begin{document}

\maketitle

\ifIsCONF
\else
\tableofcontents
\fi

\section{Introduction}
\label{sec:intro}

The study of the origin of electroweak symmetry breaking is one of the key goals of the Large Hadron Collider (LHC)~\cite{Evans:2008zzb}.
In the Standard Model (SM)~\cite{Glashow:1961tr, Weinberg:1967tq, Salam:1968rm, tHooft:1972tcz}, the symmetry is broken through the introduction of a complex scalar field doublet,
leading to the prediction of the existence of one physical neutral scalar particle, commonly known as the
Higgs boson~\cite{Englert:1964et,Higgs:1964pj,Higgs:1964pj,Guralnik:1964eu,Higgs:1966ev,Kibble:1967sv}.
The discovery of a Higgs boson with a mass of approximately 125 GeV by the ATLAS~\cite{HIGG-2012-27} and CMS~\cite{CMS-HIG-12-028}
collaborations was a crucial milestone. Measurements of its properties performed so far~\cite{HIGG-2014-06,CMS-HIG-14-009,CMS-HIG-12-041,HIGG-2013-01,CMS-HIG-14-018,HIGG-2015-07}
are consistent with the predictions for the SM Higgs boson.

These measurements rely primarily on studies of the bosonic decay modes, \Hgg, $H \to ZZ^*$, and $H \to WW^*$; therefore it is crucial to also measure the Yukawa interactions, which are predicted to account for the fermion masses~\cite{Weinberg:1967tq,Nambu:1961tp}. Thus far, only the Yukawa coupling of the Higgs boson to $\tau$ leptons has been observed~\cite{HIGG-2015-07,HIGG-2013-32,CMS-HIG-13-004,Sirunyan:2017khh} and evidence for the Yukawa coupling of the Higgs boson to $b$-quarks has been found through direct searches~\cite{vhbb,Sirunyan:2017elk,TevatronBB}.
The Yukawa coupling of the Higgs boson to the top quark, the heaviest particle in the SM, is expected to be of the order of unity, and
could be particularly sensitive to effects beyond the SM (BSM).
A measurement of the ratio of this coupling to the SM prediction of $0.87\pm0.15$ has been obtained from the combined fit of the ATLAS and CMS Higgs boson measurements~\cite{HIGG-2015-07}.
This depends largely on the indirect measurement using the top quark contribution to gluon--gluon fusion production and diphoton decay loops for which no BSM contribution is assumed.
Therefore, a direct measurement of the coupling of the Higgs boson to top quarks is highly desirable to disentangle any deviation in the top quark's Yukawa coupling due to couplings to new particles and to significantly
reduce the model dependence in the extraction of the top quark's Yukawa coupling.

A direct measurement can be achieved by measuring the rate of the process in which the Higgs boson is produced in association with a pair of top quarks,
$gg/q\bar q \to t\bar t H$, which is a tree-level process at lowest order in perturbation theory.
Although the \ttH production cross section at the LHC is two orders of magnitude smaller than the total Higgs boson production cross section, the distinctive signature from the top quarks
in the final state gives access to many Higgs boson decay modes.
The ATLAS and CMS collaborations have searched for \ttH production using proton--proton ($pp$) collision data collected during LHC Run~1 at center-of-mass energies of \SevenTeV~and \EightTeV, with analyses
mainly sensitive to $H \to WW^*$, $H \to \tautau$, \Hbb and \Hgg~\cite{HIGG-2013-26,HIGG-2013-27,HIGG-2013-25,CMS-HIG-12-035,CMS-HIG-13-029}.
The combination of these results yields a best fit of the ratio of observed and SM cross sections, $\mu = \sigma/\sigma_{\mathrm{SM}}$ of $2.3^{+0.7}_{-0.6}$~\cite{HIGG-2015-07}.

The ongoing data-taking at the LHC at an increased center-of-mass energy of \ThirTeV\ allows the collection of a larger dataset because of an increased \ttH production cross section
relative to Run~1~\cite{deFlorian:2016spz,Beenakker:2002nc,Dawson:2003zu,Yu:2014cka,Frixione:2014qaa}.
This article reports the results of a search for \ttH production using a dataset corresponding to an integrated luminosity
of 36.1~\ifb\ collected with the ATLAS detector at $\sqrt{s} = 13$ TeV during 2015 and 2016. Examples of tree-level Feynman diagrams are given in Figure~\ref{fig:feynman} where the Higgs boson is shown decaying to $WW^*/ZZ^*$ or $\tau\tau$.
The search uses seven final states distinguished by the number and flavor of charged-lepton (electron, muon and hadronically decaying $\tau$ lepton) candidates, denoted $l$. In the following, the term ``light lepton'', denoted $\ell$, refers to either electrons or muons and is understood to mean both particle and antiparticle as appropriate.
These signatures are primarily sensitive to the decays $H \to WW^*$ (with subsequent decay to $l \nu l \nu$ or $l \nu q q$), $H \to \tautau$ and $H \to ZZ^*$ (with subsequent decay to $ll\nu\nu$ or $ll q q$), and their selection is designed to avoid any overlap with the ATLAS searches for \ttH production with $H \to b\bbar$~\cite{ttHbb}, $H \to \gamma\gamma$~\cite{yy} and $H \to ZZ^* \to 4\ell$~\cite{4l} decays. Backgrounds to the signal arise from associated production of a top quark pair and a $W$ or $Z$ (henceforth $V$) boson. Additional backgrounds arise from $t\bar{t}$ production with leptons from heavy-flavor hadron decays and additional jets (non-prompt leptons), and other processes where the electron charge is incorrectly assigned (labeled as ``q mis-id") or where jets are incorrectly identified as $\tau$ candidates. Backgrounds are estimated with a combination of simulation and data-driven techniques (labeled as ``Pre-Fit''), and then a global fit
to the data, in all final states, is used to extract the best estimate for the \ttH production rate and adjust the background predictions (labeled as ``Post-Fit'').

\begin{figure}[h!]
  \centering
    \subfigure[]{\includegraphics[width=.3\textwidth]{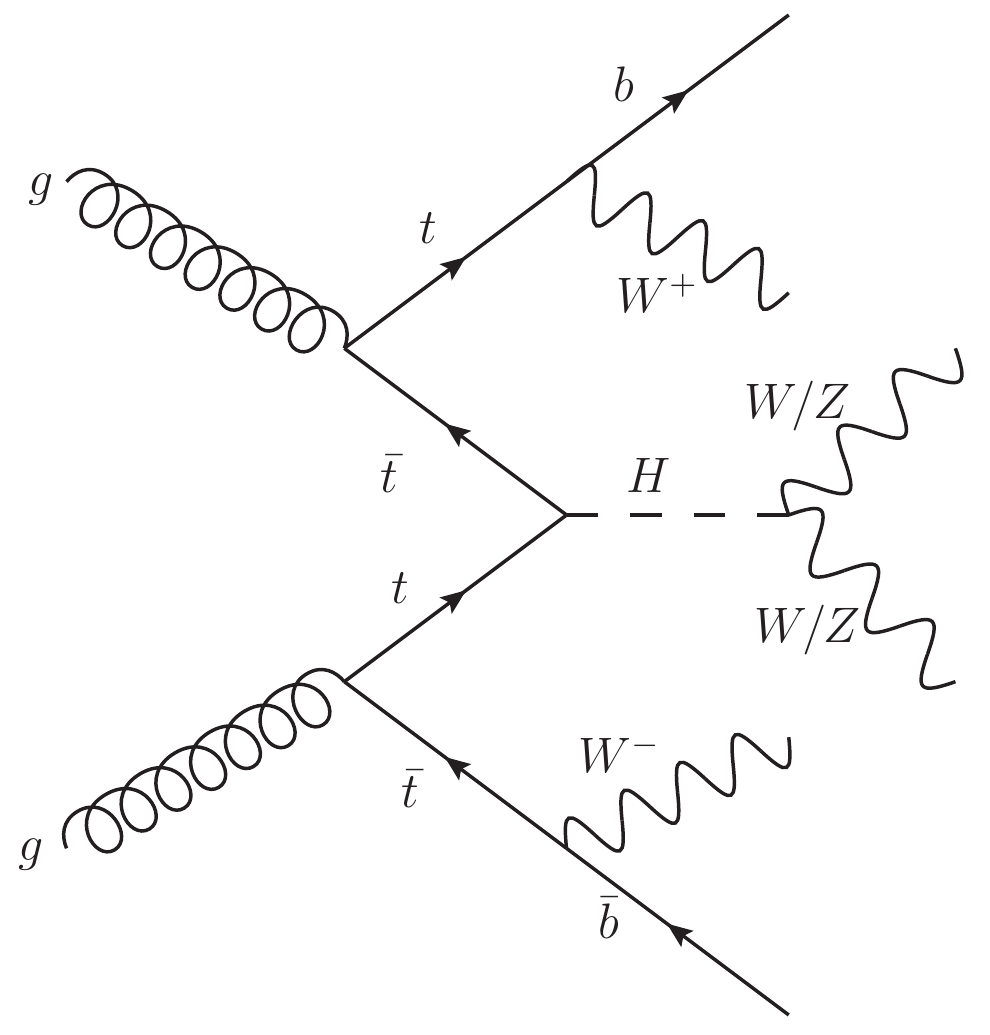}}
    \hspace*{1.5cm}
    \subfigure[]{\includegraphics[width=.3\textwidth]{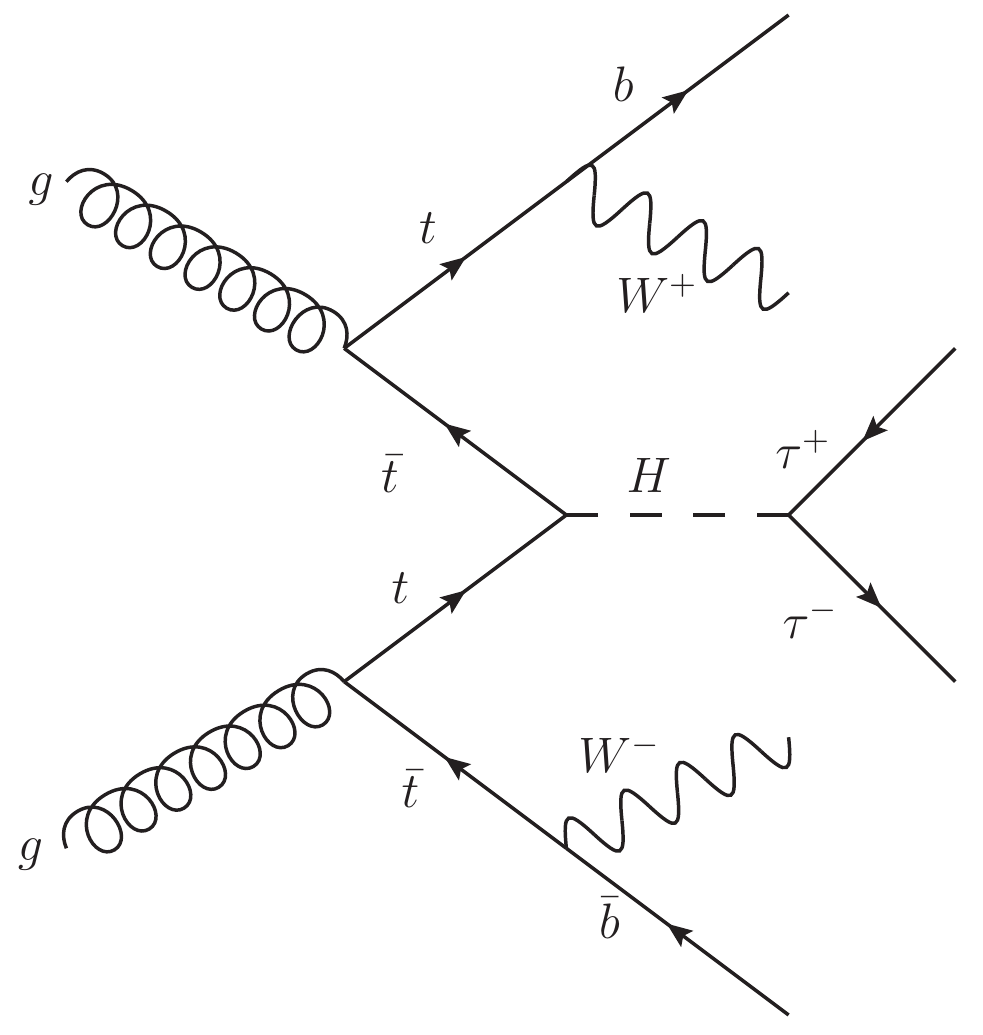}}
  \caption{\label{fig:feynman} Examples of tree-level Feynman diagrams for the production of the Higgs boson in association with a pair of top quarks. Higgs boson decays to (left) $WW/ZZ$ or (right) $\tau\tau$ are shown.}
\end{figure}

The article is organized as follows. Section~\ref{sec:detector} introduces the ATLAS detector; Section~\ref{sec:mc} describes the Monte Carlo (MC) simulation samples as well as
the recorded data used for this analysis. The reconstruction and identification of the physics objects are discussed in Section~\ref{sec:object}.
The event selection and classification are explained in Section~\ref{sec:event}. Section~\ref{sec:bkg} describes the methods used to estimate the backgrounds.
The theoretical and experimental uncertainties are discussed in Section~\ref{sec:syst}.
The results are presented in Section~\ref{sec:result}, and the combination with the three other ATLAS searches for \ttH production mentioned above is reported in Section~\ref{sec:combo}.

\section{ATLAS detector}
\label{sec:detector}

\newcommand{\AtlasCoordFootnote}{%
ATLAS uses a right-handed coordinate system with its origin at the nominal interaction point (IP)
in the center of the detector and the $z$-axis along the beam pipe.
The $x$-axis points from the IP to the center of the LHC ring,
and the $y$-axis points upwards.
Cylindrical coordinates $(r,\phi)$ are used in the transverse plane,
$\phi$ being the azimuthal angle around the $z$-axis.
The pseudorapidity is defined in terms of the polar angle $\theta$ as $\eta = -\ln \tan(\theta/2)$.
Angular distance is measured in units of $\Delta R \equiv \sqrt{(\Delta\eta)^{2} + (\Delta\phi)^{2}}$.}

The ATLAS experiment~\cite{PERF-2007-01} at the LHC is a multipurpose particle detector
with a forward-backward symmetric cylindrical geometry and a near $4\pi$ coverage in
solid angle.\footnote{\AtlasCoordFootnote}
It consists of an inner tracking detector surrounded by a superconducting solenoid
providing a \SI{2}{\tesla} axial magnetic field, electromagnetic and hadron calorimeters, and a muon spectrometer.
The inner tracking detector, covering the pseudorapidity range $|\eta| < 2.5$,
consists of silicon pixel and silicon microstrip tracking detectors inside a transition-radiation tracker that covers $|\eta| = 2.0$.
It includes, for the $\sqrt{s} = 13$~TeV running period, a newly installed innermost pixel layer, the insertable B-layer~\cite{CERN-LHCC-2010-013}.
Lead/liquid-argon (LAr) sampling calorimeters provide electromagnetic (EM) energy measurements for $|\eta| < 2.5$
with high granularity and longitudinal segmentation.
A hadron calorimeter consisting of steel and scintillator tiles covers the central pseudorapidity range ($|\eta| < 1.7$).
The endcap and forward regions are instrumented with LAr calorimeters for EM and hadronic energy measurements up to $|\eta| = 4.9$.
The muon spectrometer surrounds the calorimeters and is based on three large air-core toroid superconducting magnets with eight coils each.
It includes a system of precision tracking chambers ($|\eta| < 2.7$) and fast detectors for triggering ($|\eta| < 2.4$).
A two-level trigger system is used to select events~\cite{Trigger2015}.
The first-level trigger is implemented in hardware and uses a subset of the detector information
to reduce the accepted rate to a design maximum of 100~kHz.
This is followed by a software-based trigger with a sustained average accepted event rate of about 1~kHz.

\section{Data and Monte Carlo samples}
\label{sec:mc}

The data were collected by the ATLAS detector during 2015 and 2016 with a peak instantaneous luminosity of $L~=~1.4\times~10^{34}$~cm$^{-2}$s$^{-1}$. The mean number of $pp$ interactions per bunch crossing in the dataset is 24 and the bunch spacing is 25~ns. After the application of beam and data-quality requirements, the integrated luminosity considered corresponds to 36.1~fb$^{-1}$.

Monte Carlo simulation samples were produced for signal and background processes using the full ATLAS detector simulation~\cite{SOFT-2010-01} based on \textsc{Geant4}~\cite{geant4} or, for selected smaller backgrounds, a fast simulation using a parameterization of the calorimeter response and \textsc{Geant4} for tracking systems~\cite{ATL-PHYS-PUB-2010-013}.
To simulate the effects of additional $pp$ collisions in the same and nearby bunch crossings (pileup), additional interactions were generated using the low-momentum strong-interaction processes of \PYTHIA 8.186~\cite{Sjostrand:2006za,Sjostrand:2007gs} with a set of tuned parameters referred to as the A2 tune~\cite{ATL-PHYS-PUB-2012-003} and the MSTW2008LO set of parton distribution functions (PDF)~\cite{Martin:2009iq}, and overlaid onto the simulated hard-scatter event. The simulated events are reweighted to match the pileup conditions observed in the data and are reconstructed using the same procedure as for the data. The event generators used for each signal and background sample, together with the program and the set of tuned parameters used for the modeling of the parton shower, hadronization and underlying event are listed in Table~\ref{tbl:evgen}. The simulation samples for \ttH, \ttV, $VV$ and \ttbar are described in Refs.~\cite{ATL-PHYS-PUB-2016-005,ATL-PHYS-PUB-2016-002,ATL-PHYS-PUB-2016-004}.
The samples used to estimate the systematic uncertainties are indicated in between parentheses in Table~\ref{tbl:evgen}.

A Higgs boson mass of 125~GeV, from the combined ATLAS and CMS Run~1 measurements~\cite{HIGG-2014-14}, and a top quark mass of 172.5~GeV are assumed. The overall \ttH cross section is 507~fb, which is computed at next-to-leading order (NLO) in quantum chromodynamics (QCD) with NLO electroweak corrections~\cite{deFlorian:2016spz,Beenakker:2002nc,Dawson:2003zu,Yu:2014cka,Frixione:2014qaa}. Uncertainties include $^{+5.8\%}_{-9.2\%}$ due to the QCD factorization and renormalization scales and $\pm3.6\%$ due to the PDFs and the strong coupling \alphas.
The cross sections for \ttV production, including the process $pp \to t\bar t l^+ l^- + X$ over the
full $Z/\gamma^*$ mass spectrum, are computed at NLO in QCD and electroweak couplings following Refs.~\cite{Alwall:2014hca,Frixione:2015zaa}. The cross section for $t\bar t l^+ l^-$, with $m(l^+ l^-) > 5$ GeV, is 124 fb, and 601 fb for $t\bar t W^\pm$~\cite{deFlorian:2016spz}. The QCD scale uncertainties are $\pm12\%$ and uncertainties from PDF and \alphas variations are $\pm4\%$.

Events in the \ttbar sample with radiated photons of high transverse momentum (\pT) are vetoed to avoid overlap with those from the $\ttbar\gamma$ sample.
Dedicated samples are included to account for backgrounds from $\ttbar (Z/\gamma^*)$ where the $Z/\gamma^*$ has low invariant mass but the leptons enter the analysis phase space via asymmetric internal conversions, or rare $t \rightarrow Wb\ell\ell$ radiative decays (referred to as ``rare top decay'' in the following).

\begin{table}
\begin{center}
\caption{\label{tbl:evgen} The configurations used for event generation of signal and background processes.  The samples used to estimate the systematic uncertainties are indicated in between parentheses. ``$V$'' refers to production of an electroweak boson ($W$ or $Z/\gamma^*$).  ``Tune'' refers to the underlying-event tuned parameters of the parton shower program.  The parton distribution function (PDF) shown in the table is the one used for the matrix element (ME). The PDF used for the parton shower is either NNPDF 2.3 LO~\cite{Ball:2012cx}  for samples using the A14~\cite{ATLASUETune4}  tune or CTEQ6L1~\cite{cteq6l1,cteq6} for samples using either the UE-EE-5~\cite{Seymour:2013qka} or the Perugia2012~\cite{perugia} tune. ``\textsc{MG5\_aMC}'' refers to \textsc{MadGraph5\_aMC@NLO} with several versions from 2.1.0 to 2.3.3~\cite{Alwall:2014hca}; ``\textsc{Pythia} 6'' refers to version 6.427~\cite{Pythia6}; ``\textsc{Pythia} 8'' refers to version 8.210 or 8.212~\cite{Sjostrand:2007gs}; ``\textsc{Herwig++}'' refers to version 2.7~\cite{Bahr:2008pv}; ``MEPS'' refers to the method used in \textsc{Sherpa}~\cite{sherpa,Cascioli:2011va,Gleisberg:2008fv,Schumann:2007mg,Hoeche:2012yf} to match the matrix element to the parton shower.  Samples using \textsc{Pythia}~6 or \textsc{Pythia}~8  have heavy-flavor hadron decays modeled by \textsc{EvtGen} 1.2.0~\cite{Lange:2001uf}.  All samples include leading-logarithm photon emission, either modeled by the parton shower program or by \textsc{PHOTOS}~\cite{Golonka:2005pn}.}
 \resizebox{\textwidth}{!}{
\begin{tabular}{lllllll}
\hline\hline
Process & Event generator & ME order & Parton Shower & PDF & Tune  \\
\hline
\ttH & \textsc{MG5\_aMC} & NLO & \textsc{Pythia} 8\ & NNPDF 3.0 NLO \cite{Ball:2014uwa} & A14 \\
     & (\textsc{MG5\_aMC}) & (NLO) & (\textsc{Herwig++}) & (CT10 \cite{ct10})            & (UE-EE-5) & \\
$tHqb$ & \textsc{MG5\_aMC} & LO & \textsc{Pythia} 8 & CT10 & A14  \\
$tHW$ & \textsc{MG5\_aMC} & NLO & \textsc{Herwig++}  & CT10 & UE-EE-5  \\
$\ttbar W$ & \textsc{MG5\_aMC} & NLO & \textsc{Pythia} 8 & NNPDF 3.0 NLO & A14 \\
& (\textsc{Sherpa} 2.1.1) & (LO multileg) & (\textsc{Sherpa}) & (NNPDF 3.0 NLO) & (\textsc{Sherpa} default) & \\
$\ttbar (Z/\gamma^* \to ll)$ & \textsc{MG5\_aMC} & NLO & \textsc{Pythia} 8 & NNPDF 3.0 NLO & A14  \\
& (\textsc{Sherpa} 2.1.1) & (LO multileg) & (\textsc{Sherpa}) & (NNPDF 3.0 NLO) & (\textsc{Sherpa} default) \\
$t Z$ & \textsc{MG5\_aMC} & LO & \textsc{Pythia} 6  & CTEQ6L1 & Perugia2012 \\
$t W Z$ & \textsc{MG5\_aMC} & NLO & \textsc{Pythia} 8 & NNPDF 2.3 LO  & A14 \\
$t\bar t t$, $t\bar t t\bar t$ & \textsc{MG5\_aMC} & LO & \textsc{Pythia} 8 & NNPDF 2.3 LO & A14 \\
$t\bar t W^+ W^-$ & \textsc{MG5\_aMC} & LO & \textsc{Pythia} 8 & NNPDF 2.3 LO & A14 \\
$\ttbar$ & \textsc{Powheg-BOX v2} \cite{powhegtt} & NLO & \textsc{Pythia} 8 & NNPDF 3.0 NLO & A14  \\
$\ttbar\gamma$ & \textsc{MG5\_aMC} & LO & \textsc{Pythia} 8 & NNPDF 2.3 LO & A14 \\
$s$-, $t$-channel, & \textsc{Powheg-BOX v1} \cite{powhegstp,powhegstp2,powhegstp3}& NLO & \textsc{Pythia} 6 & CT10 & Perugia2012 \\
 $Wt$ single top & & & & & \\
$VV (\to ll XX)$, & \textsc{Sherpa} 2.1.1 & MEPS NLO & \textsc{Sherpa} & CT10 & \textsc{Sherpa} default\\
$qqVV$, $VVV$ & & &  \\
$Z \to l^+l^-$ & \textsc{Sherpa} 2.2.1 & MEPS NLO  & \textsc{Sherpa} & NNPDF 3.0 NLO & \textsc{Sherpa} default\\
\hline\hline
\end{tabular}
}
\end{center}
\end{table}

\section{Object reconstruction and identification}
\label{sec:object}

All analysis channels share a common trigger, jet, lepton and overall event preselection. The selections are detailed here and the lepton selection is summarized in Table~\ref{tbl:tightleps}. Unless otherwise specified, light leptons are required to pass the loose lepton selection.
Further channel-specific requirements are discussed in Section~\ref{sec:event}.

The selection of events is based on the presence of light leptons, with either single-lepton or dilepton triggers.
For data recorded in 2015, the single-electron (single-muon) trigger required a candidate with transverse momentum $\pt >$ 24 (20)~\GeV{}~\cite{Trigger2015};
in 2016 the lepton $\pt$ threshold was raised to 26~GeV.
The trigger $\pt$ thresholds for the 2015 (2016) data-taking were 12+12 (17+17) \GeV{} for dielectron and 18+8 (22+8) \GeV{} for dimuon triggers. For the electron+muon triggers, they were 17+14 \GeV{} for both datasets.
The trigger requirement has an efficiency of 82\% to 99\%, depending on the final state and the dataset, for signal events passing the final signal-region selections. The reconstructed light leptons are required to be matched to the trigger signatures. The primary vertex of an event is chosen as the vertex with the highest sum of squared transverse momenta of the associated tracks with $\pt >$ 400~\MeV{}~\cite{ATL-PHYS-PUB-2015-026}.

Muon candidates are reconstructed by combining inner detector tracks with track segments or full tracks in the muon spectrometer~\cite{Aad:2016jkr}.
In the region $|\eta| < 0.1$, where muon spectrometer coverage is reduced, muon candidates are also reconstructed from inner detector tracks matched to isolated energy deposits
in the calorimeters consistent with the passage of a minimum-ionizing particle.
Candidates are required to satisfy $\pt > 10$~GeV and $|\eta| < 2.5$ and to pass loose identification requirements~\cite{Aad:2016jkr}. To reduce the non-prompt muon contribution, the track is required to originate from the primary vertex
by imposing a requirement on its transverse impact parameter significance $|d_0|/\sigma_{d_0} < 3$ and on its longitudinal impact parameter multiplied by the sine of the polar angle $|z_0 \sin \theta| < 0.5$~mm. Additionally, muons are required to be separated by $\Delta R > \min(0.4, 0.04+(\mathrm{10}~\gev)/p_{\mathrm{T},\mu})$ from any selected jets (see below for details
of jet reconstruction and selection). The requirement is chosen to maximize the acceptance for prompt muons at a fixed rejection factor for non-prompt and fake muon candidates.

Electron candidates are reconstructed from energy clusters in the electromagnetic calorimeter that are associated with charged-particle tracks reconstructed in the inner detector~\cite{Aaboud:2016vfy,ATLAS-CONF-2016-024}.
They are required to have a transverse momentum $\pt > 10$ GeV and $|\eta_\textrm{cluster}| < 2.47$, and
the transition region between the barrel and endcap electromagnetic calorimeters, $1.37 < |\eta_\textrm{cluster}| < 1.52$, is excluded.
A multivariate likelihood discriminant combining shower shape and track information is used to distinguish real prompt electrons from electron candidates from hadronic jets, photon conversions and heavy-flavor (HF) hadron decays (fake and non-prompt electrons).
Loose and tight electron discriminant working points are used~\cite{ATLAS-CONF-2016-024}, both including the number of hits in the innermost pixel layer to discriminate between electrons and converted photons.
The same longitudinal impact parameter selection as for muons is applied, while the transverse impact parameter significance is required to be $|d_0|/\sigma_{d_0} < 5$.
If two electrons closer than $\Delta R=0.1$ are preselected, only the one with the higher \pt\ is considered.
An electron is rejected if, after passing all the above selections, it lies within $\Delta R=0.1$ of a selected muon.

Hadronically decaying $\tau$-lepton candidates ($\tauh$) are reconstructed from clusters in the calorimeters and associated inner detector tracks~\cite{ATL-PHYS-PUB-2015-045}.
Candidates are required to have either one or three associated tracks, with a total charge of $\pm 1$.
Candidates are required to have a transverse momentum $\pt > 25$~GeV and $|\eta| < 2.5$, excluding the electromagnetic calorimeter's transition region.
A boosted decision tree (BDT) discriminant using calorimeter- and tracking-based variables is used to identify \tauh candidates and reject jet backgrounds.
Three types of \tauh candidates are used in the analysis, referred to as loose, medium and tight: the latter two are defined by working points with a combined reconstruction and identification efficiency
of 55\% and 45\% (40\% and 30\%) for one (three)-prong \tauh decays, respectively~\cite{ATLAS-CONF-2017-029}, while the first one has a more relaxed selection and is only used for background estimates. The corresponding expected rejection factors against light-quark/gluon jets vary from 30 for loose candidates to 300 for tight candidates~\cite{ATL-PHYS-PUB-2015-045}.
Electrons that are reconstructed as one-prong \tauh candidates are removed via a BDT trained to reject electrons.
Additionally, \tauh candidates are required to be separated by $\Delta R > 0.2$ from any selected electrons and muons. The contribution of fake \tauh from $b$-jets is removed by vetoing the candidates that are also $b$-tagged, which rejects a large fraction of the \ttbar background. The contribution of fake \tauh from muons is removed by vetoing the candidates that overlap with low-\pT reconstructed muons.
Finally, the vertex matched to the tracks of the \tauh candidate is required to be the primary vertex of the event, in order to reject fake candidates arising from pileup collisions.

Jets are reconstructed from three-dimensional topological clusters built from energy deposits in the calorimeters~\cite{PERF-2014-07-2,ATL-PHYS-PUB-2015-036},
using the anti-$k_t$ algorithm with a radius parameter $R=0.4$~\cite{Cacciari:2008gp,Cacciari:2011ma}.
Their calibration is based on simulation with additional corrections obtained using in situ techniques~\cite{PERF-2016-04} to account for differences between simulation and data.
Jets are required to satisfy $\pt > 25$~GeV and $|\eta| < 2.5$.
In order to reject jets arising from pileup collisions, a significant fraction of the total summed scalar $\pt$ of the tracks in jets with $\pt < 60$~GeV and $|\eta| < 2.4$ must originate from tracks that are associated with the primary vertex~\cite{PERF-2014-03-2}.
The average efficiency of this requirement is 92\% per jet from the hard scatter.
The calorimeter energy deposits from electrons are typically also reconstructed as jets; in order to eliminate double counting, any jets within $\Delta R =$ 0.3
of a selected electron are not considered. This is also the case for any jets within $\Delta R = 0.3$ of a \tauh candidate.

Jets containing $b$-hadrons are identified ($b$-tagged) via a multivariate discriminant combining information from algorithms using track impact parameters
and secondary vertices reconstructed within the jet~\cite{PERF-2012-04_corr,ATL-PHYS-PUB-2015-022}. These $b$-tagged jets will henceforth be referred to as $b$-jets. The working point used for this search corresponds to an average efficiency of 70\% for jets containing $b$-hadrons with $\pt > 20$~GeV and $|\eta| < 2.5$ in $t\bar t$ events.
The expected rejection factors against light-quark/gluon jets, $c$-quark jets and hadronically decaying $\tau$ leptons are 380, 12 and 55, respectively~\cite{ATL-PHYS-PUB-2015-022,ATL-PHYS-PUB-2016-012}.
To compensate for differences between data and simulation in the $b$-tagging efficiencies and mis-tagging rates, correction factors are applied to the simulated samples~\cite{ATL-PHYS-PUB-2015-022}.

The lepton requirements are summarized in Table~\ref{tbl:tightleps}. Isolation requirements are applied to all lepton types except the loose definition. Two isolation variables, based on calorimetric and tracking variables, are computed.
Calorimetric isolation uses the scalar sum of transverse energies of clusters within a cone of size $\Delta R = 0.3$ around the light-lepton candidate.
This excludes the electron candidate's cluster itself and clusters within $\Delta R = 0.1$ of the muon candidate's track, respectively,
and is corrected for leakage from the electron's shower and for the ambient energy in the event~\cite{STDM-2010-08,Cacciari:2008gn}.
Track isolation uses the sum of transverse momenta of tracks with $\pt > 1$~\gev\ consistent with originating at the primary vertex,
excluding the light-lepton candidate's track, within a cone of $\Delta R = \min(0.3, 10\textrm{ GeV}/\pt(\ell))$.
Calorimeter- and track-based isolation criteria are applied to electrons and muons to obtain a 99\% efficiency in $Z\to\ell\ell$ events.

\begin{table}
\begin{center}
\caption{\label{tbl:tightleps}Loose (L), loose and isolated (L$^\dag$), loose, isolated and passing the non-prompt BDT (L*), tight (T) and very tight (T*) light-lepton definitions. Selections for the tighter leptons are applied in addition to the looser ones. For the muons, the L*, T and T* lepton definitions are identical.}
 \begin{tabular}{l|cccccccc}
 \hline\hline
 & \multicolumn{5}{c|}{$e$} & \multicolumn{3}{c}{$\mu$} \\
 & L & L$^\dag$ & L* & T & \multicolumn{1}{c|}{T*} & L & L$^\dag$ & L*/T/T* \\
  \hline

  Isolation        &  No & \multicolumn{4}{|c|}{Yes} & No & \multicolumn{2}{|c}{Yes} \\
  \hline
  Non-prompt lepton BDT   &  \multicolumn{2}{c|}{No} & \multicolumn{3}{c|}{Yes} & \multicolumn{2}{c|}{No} & \multicolumn{1}{c}{Yes} \\
  \hline
  Identification  & \multicolumn{3}{c|}{Loose} & \multicolumn{2}{c|}{Tight} & \multicolumn{3}{c}{Loose} \\
  \hline
  Charge misassignment veto BDT &  \multicolumn{4}{c|}{No} & \multicolumn{1}{c|}{Yes} & \multicolumn{3}{c}{No} \\
  \hline
  Transverse impact parameter significance, $|d_0|/\sigma_{d_0}$  &  \multicolumn{5}{c|}{$<5$} & \multicolumn{3}{c}{$<3$ } \\
  \hline
  Longitudinal impact parameter, $|z_0 \sin \theta|$&  \multicolumn{8}{c}{$< 0.5$ mm} \\
  \hline\hline
 \end{tabular}
\end{center}
\end{table}

Non-prompt leptons are further rejected using a multivariate discriminant, taking as input the energy deposits and charged-particle tracks (including the lepton track) in a cone around the lepton direction, which is referred to as the non-prompt lepton BDT. The jet reconstruction and $b$-tagging algorithms are run on the track collection, and their output is used to train the algorithm together with isolation variables.
A reconstructed track-jet that is matched to a non-prompt lepton is typically a jet initiated by $b$- or $c$-quarks, and may contain a displaced vertex.
The most discriminating variables are thus found to be the angular distance between the lepton and the reconstructed jet, the outputs of the $b$-tagging algorithms, the calorimetric and track isolation variables of the lepton, the number of tracks within the jet and the ratio of the lepton \pt to the jet \pt. The training is performed separately for electrons and muons on prompt and non-prompt leptons from simulated \ttbar events and validated using data in various control regions.
The efficiency at the chosen working point to select well-identified prompt muons (electrons) is about 70\% (60\%) for $\pt \sim 10$~GeV and reaches a plateau of 98\% (96\%) at $\pt \sim 45$~GeV, as shown in Figure~\ref{fig:lepMVAs}, while the rejection factor against leptons from the decay of $b$-hadrons is about 20.
Simulated events are corrected to account for differences between data and simulation for this prompt-lepton isolation efficiency, as well as for the lepton trigger, reconstruction,
and identification efficiencies. The corrections were determined using a so-called tag-and-probe method as described in Refs.~\cite{Aad:2016jkr,Aaboud:2016vfy} and studied as a function of the number of nearby light- and heavy-flavor jets. This is illustrated in Figure~\ref{fig:lepMVAs}, showing that the corrections for the non-prompt lepton BDT efficiencies are at most 10\% at low transverse momentum and decrease with increasing transverse momentum.
The largest contribution to the associated systematic uncertainties comes from pileup effects.

\begin{figure*}[hbtp]
	 \begin{center}
	\includegraphics[width=0.49\linewidth]{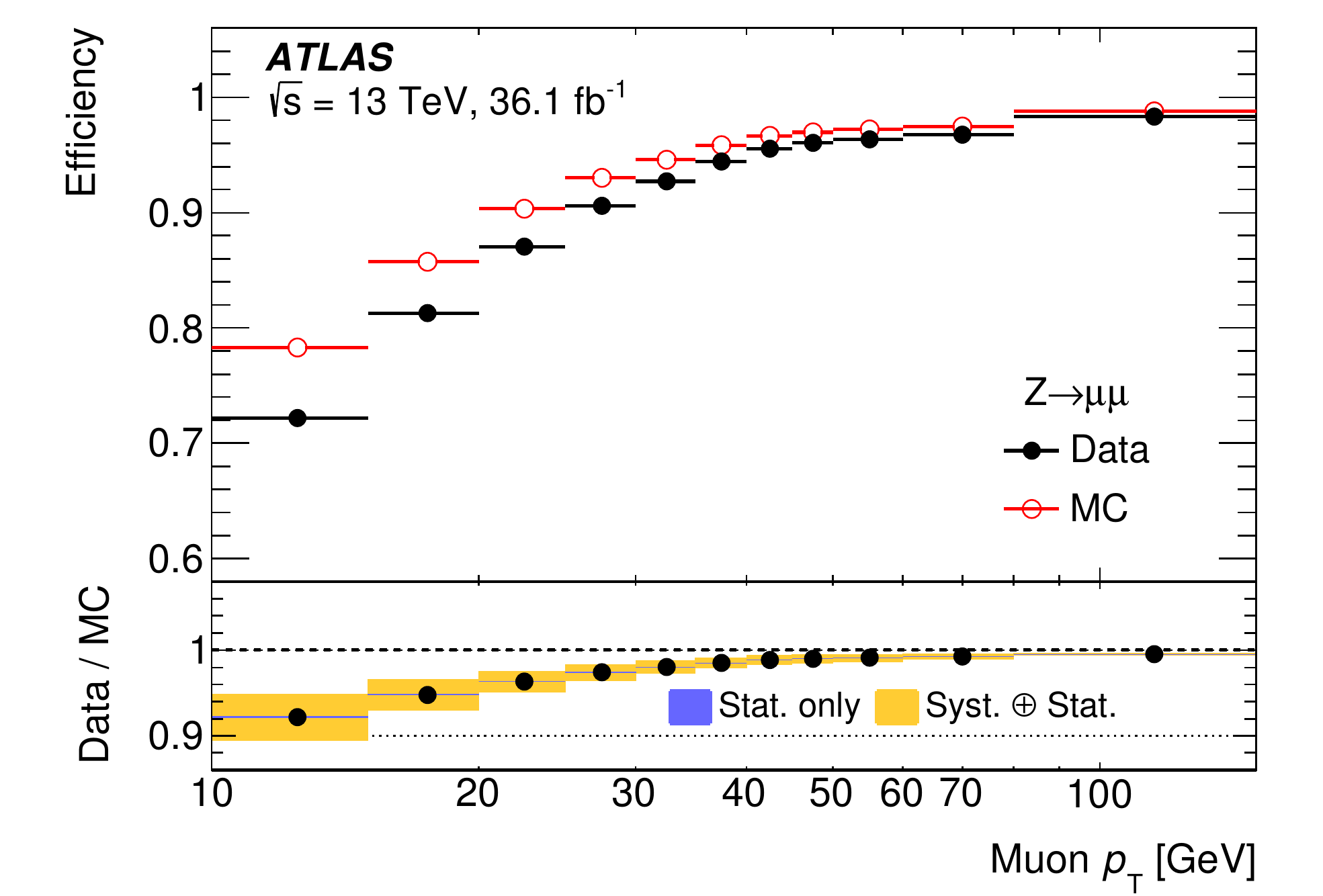}
	\includegraphics[width=0.49\linewidth]{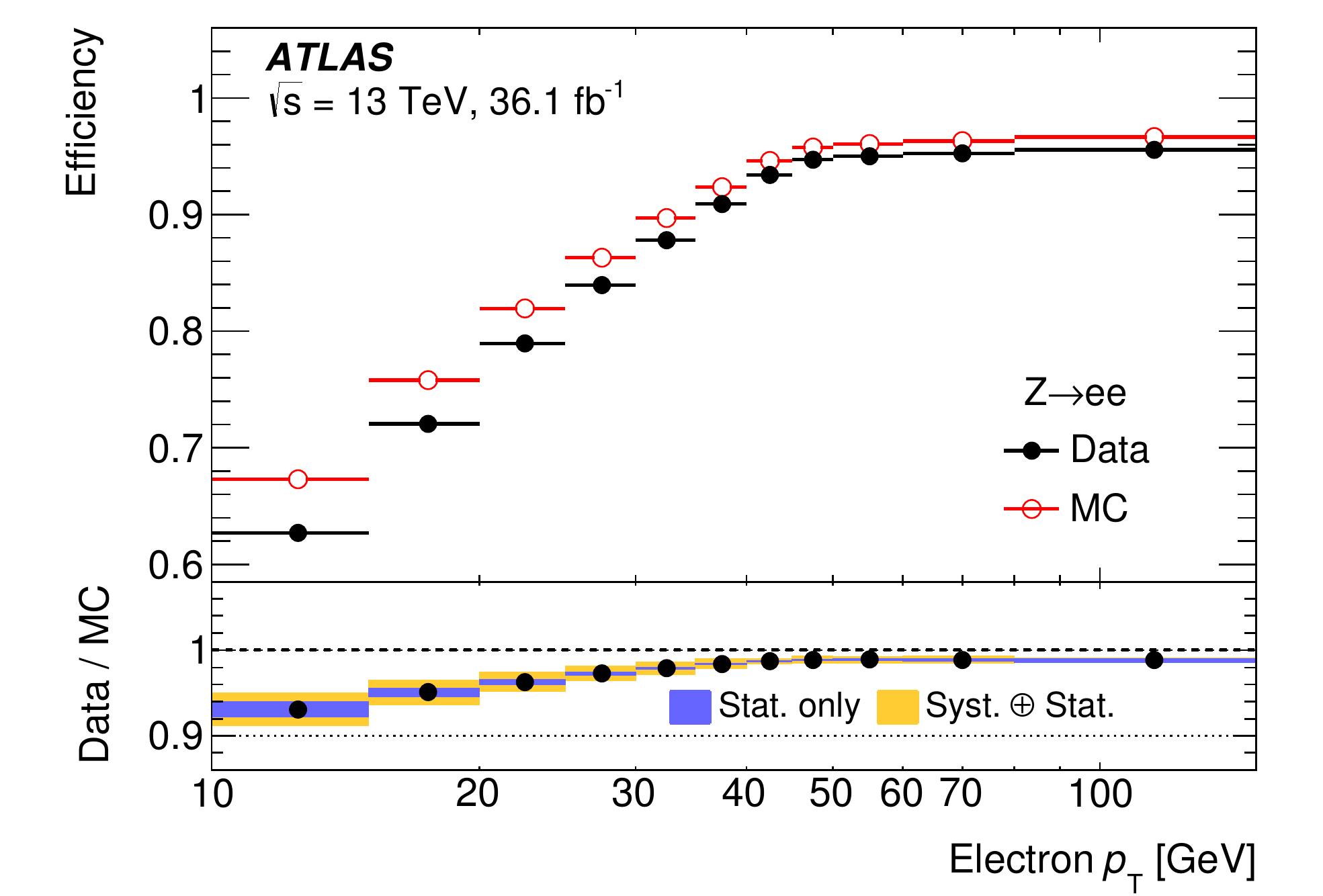}
	\caption{\label{fig:lepMVAs}
The efficiency to select well-identified prompt muons (left) and electrons (right) at the chosen non-prompt lepton BDT working point, as a function of the lepton \pT. The muons are required to pass the loose identification requirements, while the electrons are required to pass the tight identification requirements. The measurements in data (simulation) are shown as  full black (open red) circles. The bottom panel displays the ratio of data to simulation results, with the blue (yellow) band representing the statistical (total) uncertainty. This ratio is the scale factor that is applied to correct the simulation.}
	 \end{center}
\end{figure*}

There is a small, but non-negligible, probability that electrons and positrons are reconstructed with an incorrect charge. This occurs when an electron (positron) emits a hard bremsstrahlung photon; if the photon subsequently converts to an asymmetric electron--positron pair, and the positron (electron) has high momentum and is reconstructed, the lepton charge can be misidentified. Otherwise it occurs when the curvature of a track is poorly estimated, which typically happens at high momentum. The probability for muons to be reconstructed with incorrect charge is small enough that the charge misassignment is negligible.
To reject electrons reconstructed with an incorrect electric charge, a BDT discriminant is built, using the following electron cluster and track properties as input:
the electron's transverse momentum and pseudorapidity, the track curvature
significance (defined as the ratio of the electric charge to the track momentum divided by the estimated uncertainty in the measurement)
and its transverse impact parameter times the electric charge,
the cluster width along the azimuthal direction, and the quality of the matching between the track and the cluster, in terms of both energy/momentum and azimuthal position.
The chosen working point achieves a rejection factor of $\sim$17 for electrons passing the tight identification requirements with a wrong charge
assignment while providing an efficiency of 95\% for electrons with correct charge reconstruction. This requirement is only applied to the very tight electrons.
Correction factors to account for differences in the selection efficiency between data and simulation, which are within a few percent for $|\eta|<2.4$ but larger in the forward region, $2.4 < |\eta| < 2.47$, were applied to the selected electrons in the simulation.

The missing transverse momentum $\overrightarrow{\pT}^{\mathrm{miss}}$ (with magnitude \met) is defined as the negative vector sum of the transverse momenta of all identified and calibrated leptons and jets and remaining unclustered energy, the latter of which is estimated from low-\pt tracks associated with the primary vertex but not assigned to any lepton or jet candidate~\cite{ATL-PHYS-PUB-2015-027, ATL-PHYS-PUB-2015-023}.

\section{Event selection and classification}
\label{sec:event}

The analysis is primarily sensitive to decays of the Higgs boson to $WW^*$ or $\tau\tau$ with a small additional contribution from $H \to ZZ^*$. If the Higgs boson decays to either $WW^*$ or $\tau\tau$, the \ttH events typically contain either $WWWWbb$ or $\tau\tau WWbb$.
In order to reduce the \ttbar background, characterized by a final state of $WWbb$, final states including three or more charged leptons, or two same-charge light leptons, are selected.
Seven final states are analyzed, categorized by the number and flavor of charged-lepton candidates after the preselection requirements, as illustrated in Figure~\ref{fig:categories}. Each of the seven final states is called a ``channel'' and certain channels are further split into categories to gain in significance. Categories include signal and control regions. Additional control regions used for the estimates of the non-prompt backgrounds are discussed in Section~\ref{sec:bkg}.

The seven channels are:
\begin{itemize}
\item two same-charge light leptons and no hadronically decaying $\tau$ lepton candidates (\ll);
\item three light leptons and no hadronically decaying $\tau$ lepton candidates (\lll);
\item four light leptons (\llll);
\item one light lepton and two opposite-charge hadronically decaying $\tau$ lepton candidates (\ltwotau);
\item two same-charge light leptons and one hadronically decaying $\tau$ lepton candidate (\lltau);
\item two opposite-charge light leptons and one hadronically decaying $\tau$ lepton candidate (\OSlltau);
\item three light leptons and one hadronically decaying $\tau$ lepton candidate (\llltau).
\end{itemize}

\begin{figure}[!htbp]
\centering
\includegraphics[width=0.8\textwidth]{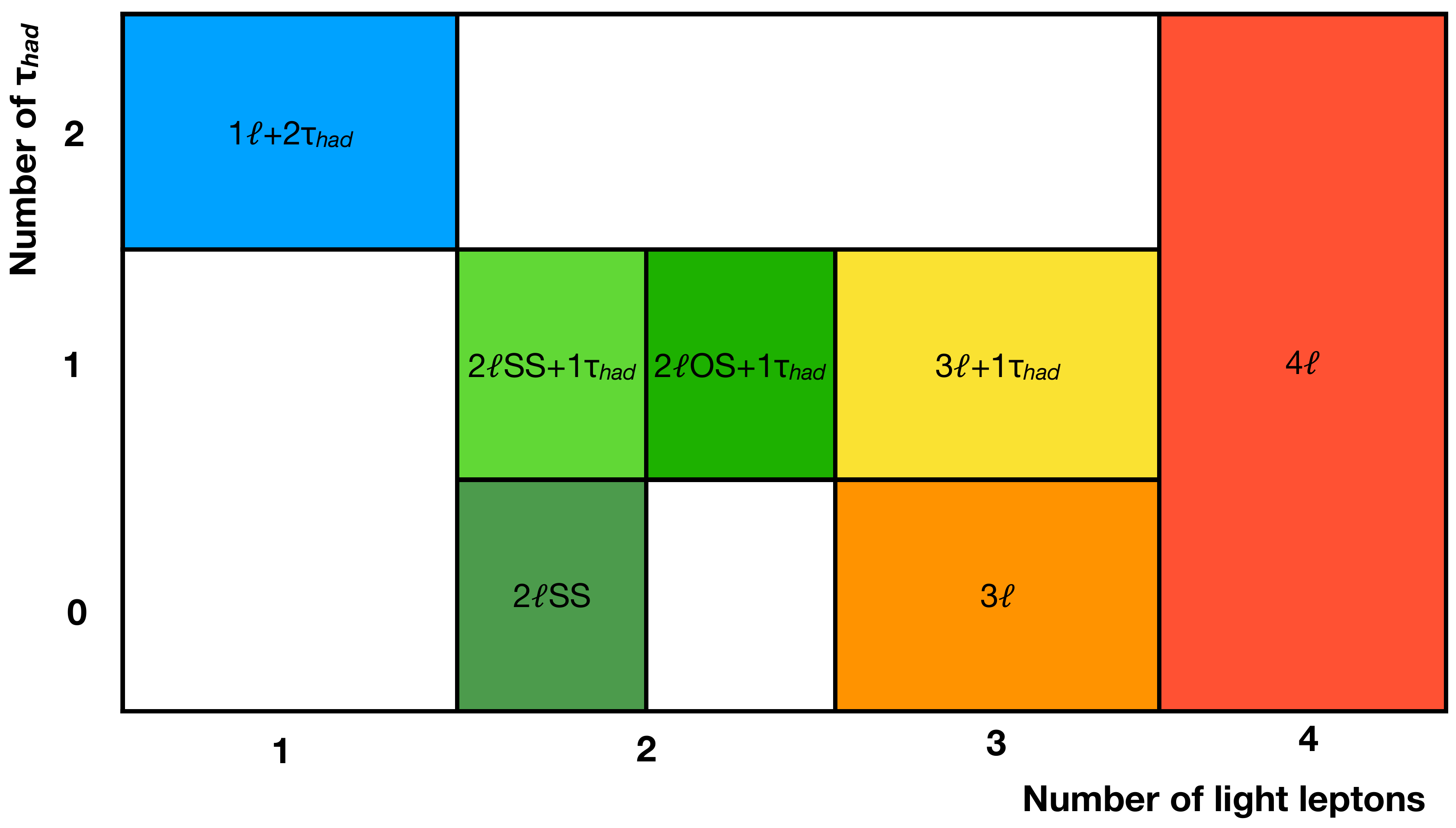}
  \caption{The channels used in the analysis organized according to the number of selected light leptons and \tauh candidates. The selection requirements for each channel are in Table~\ref{tbl:srvr}.\label{fig:categories}}
\end{figure}

The selection criteria are designed to be orthogonal to ensure that each event only contributes to a single channel. Channels are made orthogonal through the requirements on the number of loose light leptons and medium $\tauh$ candidates. A veto on events containing medium $\tauh$ candidates is therefore applied for the \ll and \lll channels, but no veto is applied for the \llll channel because there is no corresponding \tauh channel.
In all channels, the light lepton(s) are required to be matched to the lepton(s) selected by either the single-lepton or dilepton triggers. As the \ltwotau channel has only one light lepton, only single-lepton triggers are used.
In order to reduce the diboson background, all channels also require events to include at least two reconstructed jets and that at least one of these jets must be $b$-tagged.

The detailed criteria for each channel are described below, and summarized in Table~\ref{tbl:srvr}. In addition, Table~\ref{tbl:summary_anaStrat} provides a comparison of the key aspects of the selection used in each channel. After the selection, assuming Standard Model \ttH production, the total expected number of reconstructed signal events summed over all categories is 91, corresponding to 0.50\% of all produced \ttH events. The breakdown in each channel is given in Table~\ref{tbl:acc}.
In total $332\,030$ events are selected in data.
As the background contamination is still large in all channels, except one of the \fourl categories and the \llltau category, further separation of the signal from the background is achieved using multivariate techniques. The TMVA package~\cite{Hocker:2007ht} is used in all channels except for \lll, which uses XGBoost~\cite{xgboost}. Independent cross-check analyses using a simpler cut-and-count categorization were developed for the most sensitive \ll, \threel and \lltau channels.

\begin{table}
 \centering
 \caption{Selection criteria applied in the different channels. Same-flavor, opposite-charge lepton pairs are referred to as SFOC pairs. The common selection criteria for all channels are listed in the first line under the title ``Common''.}
 \begin{tabular}{ll}
  \hline\hline
  Channel & Selection criteria\\
  \hline \hline
  Common & $\njets \ge 2$ and $\nbjets \ge 1$\\
\hline
  \ll & Two very tight light leptons with $\pt > 20$~GeV\\
            & Same-charge light leptons \\
            & Zero medium \tauh candidates \\
            & $\njets \ge 4$ and $\nbjets < 3$\\
  \hline
   \lll & Three light leptons with $\pt > 10$~GeV; sum of light-lepton charges $\pm 1$ \\
              & Two same-charge leptons must be very tight and have $\pt > 15$ GeV\\
              & The opposite-charge lepton must be loose, isolated and pass the non-prompt BDT\\
              & Zero medium \tauh candidates \\
              & $m(\ell^+\ell^-) > 12$~GeV and $|m(\ell^+\ell^-)-91.2\textrm{ GeV}| > 10$~\GeV{} for all SFOC pairs\\
              & $|m(3\ell)-91.2\textrm{ GeV}| > 10$~\GeV{}\\

  \hline
   \llll   & Four light leptons; sum of light-lepton charges 0 \\
              & Third and fourth leading leptons must be tight\\
              & $m(\ell^+\ell^-) > 12$~GeV and $|m(\ell^+\ell^-)-91.2\textrm{ GeV}| > 10$~GeV for all SFOC  pairs \\
              & $|m(4\ell) - 125\textrm{ GeV}| > 5$~GeV\\
              & Split 2 categories: $Z$-depleted (0 SFOC pairs) and $Z$-enriched (2 or 4 SFOC pairs)\\
\hline
   \ltwotau & One tight light lepton with $\pt > 27$~GeV \\
               & Two medium \tauh candidates of opposite charge, at least one being tight \\
              & $\njets \ge 3$ \\
  \hline
   \lltau & Two very tight light leptons with $\pt > 15$~GeV\\
               & Same-charge light leptons \\
               & One medium \tauh candidate, with charge opposite to that of the light leptons\\
               & $\njets \ge 4$\\
& $|m(ee)-91.2\textrm{ GeV}| > 10$ GeV for $ee$ events\\
  \hline
   \OSlltau & Two loose and isolated light leptons with $\pt > 25$, 15~GeV \\
              & One medium \tauh candidate\\
               & Opposite-charge light leptons \\
               & One medium \tauh candidate \\
               & $m(\ell^+\ell^-) > 12$~GeV and $|m(\ell^+\ell^-)-91.2\textrm{ GeV}| > 10$ GeV for the SFOC pair \\
               & $\njets \ge 3$ \\
  \hline
   \llltau & \lll selection, except: \\
               & One medium \tauh candidate, with charge opposite to the total charge of the light leptons \\
	       & The two same-charge light leptons must be tight and have $\pt > 10$ GeV\\
	       & The opposite-charge light lepton must be loose and isolated\\
  \hline\hline
 \end{tabular}
\label{tbl:srvr}
\end{table}

\begin{table}[h!]
\begin{center}
\caption{\label{tbl:summary_anaStrat} Summary of the basic characteristics  of the seven analysis channels. The lepton selection follows the definition in Table~\ref{tbl:tightleps} and is labeled as loose (L), loose and isolated (L$^\dag$), loose, isolated and passing the non-prompt BDT (L*), tight (T) and very tight (T*), respectively. The \tauh selection is labeled as medium (M) and tight (T). }
\begin{tabular}{l|ccccccc}
\hline\hline
                               & \ll      & \lll    & \llll    & \ltwotau    & \lltau     & \OSlltau  & \llltau    \\
\hline
Light lepton          &  2T* & 1L*, 2T* &  2L, 2T &  1T    &  2T*  &  2L$^\dag$ & 1L$^\dag$, 2T     \\
\tauh                 &  0M  & 0M       & --      & 1T, 1M &  1M   &  1M & 1M         \\
\njets, \nbjets     & $\ge 4$, $=1,2$ & $\ge 2$, $ \ge 1$ & $\ge 2$, $\ge 1$ & $\ge 3$, $\ge 1$ & $\ge 4$, $\ge 1$ & $\ge 3$, $\ge 1$ & $\ge 2$, $\ge 1$ \\
\hline\hline
\end{tabular}
\end{center}
\end{table}

\begin{table}
\begin{center}
 \caption{\label{tbl:acc}Acceptance times efficiency ($A\times\epsilon$) for \ttH signal in each analysis channel. This includes Higgs boson and top quark branching fractions, detector acceptance, and reconstruction and selection efficiency, and is computed relative to inclusive \ttH production considering all Higgs boson and top decays. In the \fourl channel, the two numbers correspond to the $Z$-enriched and the $Z$-depleted categories.}
{\small
 \begin{tabular}{lcccccccc}
 \hline\hline
                                         & \ll  & \threel & \fourl & \ltwotau & \lltau & \OSlltau & \llltau & Total \\
 \hline
 $A\times \epsilon$ $[10^{-4}]$   & 23   & 13  &  0.6+0.1   & 2.3      &   1.7  &   7.8   &  0.8    &  50 \\
\hline\hline

 \end{tabular}
}
\end{center}
\end{table}

\subsection{\ll channel}
\label{subsec:twol}

Selected events are required to include exactly two reconstructed light leptons with the same electric charge.
To reduce the background from fake and non-prompt leptons as well as electrons reconstructed with incorrect electric charge, the very tight selection requirements described in Section~\ref{sec:object} are applied
and the leptons are required to satisfy $\pt > 20$~GeV. Events must include at least four reconstructed jets to suppress \ttbar and $\ttW$ backgrounds, among which either one or two are required to be $b$-tagged.
A slight disagreement is observed between the Standard Model prediction and the data for events containing two same-charge light leptons and three or more $b$-jets. To avoid any potential systematic bias, these events are vetoed, at no expense in sensitivity.

Two independent BDTs are trained using the selected events. The first aims to separate the signal from the non-prompt and fake background, while the second aims to separate the signal from the \ttV background. The data-driven estimate of the non-prompt and fake background described in Section~\ref{subsubsec:fakestwothreel} is used in the training, which is performed for both BDTs with the nine variables listed in Table~\ref{tbl:mvavar}. The outputs of the two BDT classifiers are combined to maximize the signal significance.

A cross-check is provided by an independent cut-and-count analysis using twelve categories, which places requirements on the jet multiplicity, $b$-tagged jet multiplicity and the lepton flavor.

\subsection{\threel channel}
\label{subsec:threel}

Selected events are required to include exactly three reconstructed light leptons with the total charge equal to $\pm 1$. The lepton of opposite charge to the other two is found to be prompt in 97\% of the selected events in \ttbar simulated samples and therefore only required to be loose, isolated and pass the non-prompt BDT selection requirements, as described in Section~\ref{sec:object}. To reduce the background from fake and non-prompt leptons, leptons in the same-charge pair are required to be very tight and to satisfy $\pt > 15$~GeV.
Events containing a same-flavor opposite-charge lepton pair with an invariant mass below 12~GeV are removed to suppress background from resonances that decay to light lepton pairs.
A $Z$-veto is applied, excluding events containing an same-flavor opposite-charge lepton pair with an invariant mass within 10~GeV of the $Z$ mass to suppress the \ttZ background. Finally, to eliminate  potential backgrounds with $Z$ decays to $\ell\ell\gamma^{(*)} \to \ell\ell \ell'(\ell')$, where one lepton has very low momentum and is not reconstructed, the three-lepton invariant mass must satisfy $|m(3\ell) - 91.2\textrm{ GeV}| > 10$ GeV.

Selected events are classified using a five-dimensional multinomial boosted decision tree. The five classification targets used in the training are: \ttH, \ttW, \ttZ, \ttbar and diboson.
In total, 28 variables based on topological aspects of the events as listed in Table~\ref{tbl:mvavar} are used in the training.
The output discriminants are mapped into the five categories to maximize the signal significance using a variable multidimensional binning procedure~\cite{Dannheim2009717}, while accounting for the uncertainties in the background estimates: \ttH, \ttW, \ttbar, \ttZ and diboson. The \ttH category is the signal region and the remaining four categories are control regions. Events not explicitly assigned to any category are found to largely contain non-prompt or fake leptons and hence are included in the \ttbar category.
The $Z$-veto is removed during the categorization process and then applied in the \ttH, \ttW and \ttbar categories because this was found to decrease the \ttZ background in the signal region.
The data-driven estimate of the non-prompt and fake background described in Section~\ref{subsubsec:fakestwothreel} is used for the categorization process, while the simulation is used for the training due to the small size of the sample used in the non-prompt estimate. The \ttH discriminant is used in the signal region.

A cross-check is provided by an independent cut-and-count analysis using twelve categories, which places requirements on the jet multiplicity, $b$-tagged jet multiplicity, the lepton flavor and the invariant mass of the opposite-charge pair of leptons with the smallest $\Delta R$ separation.

\subsection{\fourl channel}
\label{subsec:fourl}

Selected events are required to include exactly four loose light leptons with the total charge equal to zero.
To reduce the background from fake and non-prompt leptons, the third and fourth leptons ordered by decreasing transverse momentum are required to satisfy tight selection requirements described
in Section~\ref{sec:object}.
No requirements are applied to the number of $\tauh$ candidates and any jets also reconstructed as \tauh candidates are treated only as jets.
To further suppress the $\ttZ$ background, the $Z$-veto described for the \lll channel in Section~\ref{subsec:threel} is applied.
To suppress background from resonances that decay to light leptons, events containing a same-flavor opposite-charge lepton pair with an invariant mass below 12~GeV are also removed.
To reduce contamination from other Higgs boson production processes and to ensure minimal overlap
with the dedicated search for \ttH production with $H \to ZZ^* \to 4\ell$~\cite{4l} decay, a $H \rightarrow 4\ell$ veto $|m(4\ell) - 125\textrm{ GeV}| > 5$~GeV is applied.

Selected events are separated by the presence or absence of a same-flavor, opposite-charge lepton pair into two categories, referred to respectively as the $Z$-enriched and $Z$-depleted categories.
Background events in the $Z$-enriched category can arise from off-shell $Z^*$ and $\gamma^* \to \ell^+\ell^-$ processes while in the $Z$-depleted category these backgrounds are absent. Therefore, a BDT is trained in the $Z$-enriched category to further discriminate the signal from the $\ttZ$ background.
Seven variables listed in Table~\ref{tbl:mvavar} are used in the training, including a pseudo-matrix-element discriminator exploiting partially reconstructed resonances ($t$, $H$ and $Z$)~\cite{TOPQ-2014-14}. A requirement on the BDT discriminant is then imposed to define the $Z$-enriched signal region.

\subsection{\ltwotau channel}

Selected events are required to include exactly one tight light lepton and exactly two medium \tauh candidates of opposite charge.
At least one of the \tauh candidates is required to be tight.
In order to suppress the \ttbar and $\ttV$ backgrounds, events must include at least three reconstructed jets.
A BDT is trained to further reduce the main \ttbar background, in which events had one or two fake $\tauh$ candidates.
Seven variables listed in Table~\ref{tbl:mvavar} are used in the training, including the invariant mass of the visible decay products of the $\tauh\tauh$ system.

\subsection{\lltau channel}
Selected events are required to contain exactly one medium \tauh candidate but otherwise to meet the requirements for the \ll channel discussed in Section~\ref{subsec:twol},
except that the light-lepton \pt threshold is lowered from 20 to 15~GeV and that events with 3 or more $b$-jets are included.
The reconstructed charge of the $\tauh$ candidate must be opposite to that of the light leptons.
The $Z$-veto is applied to dielectron events to suppress $Z$+jets events with a misassigned charge.
A BDT is trained using the 13 variables listed in Table~\ref{tbl:mvavar} on events with relaxed selection requirements: the light leptons are required to be loose instead of tight and the requirement on the number of jets is reduced to two. This BDT is used to further reduce the \ttbar background.

A cross-check is provided by an independent cut-and-count analysis using three categories, which places requirements on the maximum $|\eta|$ of the two light leptons and the \pT of the subleading jet.

\subsection{\OSlltau channel}

Selected events are required to include exactly two reconstructed loose and isolated leptons of opposite charge with leading (subleading) $\pt >$ 25~(15) \gev, and exactly one medium \tauh candidate.
In order to reduce the \ttbar, $Z$+jets and $\ttV$ backgrounds, events must include at least three reconstructed jets.
The $Z$-veto is applied to same-flavor lepton pairs to suppress the $Z$+jets background with a fake \tauh candidate.
To suppress background from resonances that decay to light leptons, events containing a same-flavor lepton pair with an invariant mass below 12~GeV are also removed.
A BDT is trained using the 13 variables listed in Table~\ref{tbl:mvavar} on the selected events, with the aim of further reducing the main \ttbar background with a fake $\tauh$ candidate.

\subsection{\llltau channel}
Selected events are required to contain exactly one medium \tauh candidate but otherwise to meet the requirements for the \threel channel discussed in Section~\ref{subsec:threel},
except that the two same-charge leptons must be tight and have $\pt > 10$ GeV and the opposite-charge lepton must be loose and isolated. The reconstructed charge of the $\tauh$ candidate must be opposite to the total charge of the light leptons. Due to the high purity of the signal, no further selection is required and only the event yields are used in the fit.

\begin{table}
\begin{center}
 \caption{\label{tbl:mvavar}Variables used in the multivariate analysis (denoted by $\times$) for the \ll, \threel, \fourl ($Z$-enriched category),
 \ltwotau, \lltau and \OSlltau channels. For \ll and \lltau, lepton~0 and lepton~1 are the leading and subleading leptons, respectively. For \threel, lepton~0 is the lepton with charge opposite to that of the same-charge pair, while the same-charge leptons are labeled with increasing index (lepton~1 and lepton~2) as \pT decreases.  The best $Z$-candidate dilepton invariant mass is the mass of the dilepton pair closest to the $Z$ boson mass. The variables also used in the cross-check analyses are indicated by a $*$.}
{\small \begin{tabular}{clcccccc}
 \hline\hline
  & Variable                                       & \ll      & \threel    & \fourl    & \ltwotau    & \lltau     & \OSlltau      \\
 \hline
 \parbox[t]{5mm}{\multirow{13}{*}{\rotatebox[origin=c]{90}{Lepton properties}}} & Leading lepton \pt                              &          & $\times$\phantom{*}   &           &             &            &               \\
 & Second leading lepton \pt                       & $\times$\phantom{*} & $\times$\phantom{*}   &           &             &  $\times$\phantom{*}  &               \\
 & Third lepton \pt                                &          & $\times$\phantom{*}   &           &             &            &               \\
 & Dilepton invariant mass (all combinations)     & $\times$\phantom{*} & $\times${$*$}   &           &             &            &  $\times$\phantom{*}     \\
 & Three-lepton invariant mass                     &          & $\times$\phantom{*}   &           &             &            &               \\
 & Four-lepton invariant mass                      &          &            & $\times$\phantom{*}  &             &            &               \\
 & Best $Z$-candidate dilepton invariant mass     &          &            & $\times$\phantom{*}  &             &            &               \\
 & Other $Z$-candidate dilepton invariant mass    &          &            & $\times$\phantom{*}  &             &            &               \\
 & Scalar sum of all leptons \pt                   &          &            & $\times$\phantom{*}  &             &            &  $\times$\phantom{*}     \\
 & Second leading lepton track isolation           &          &            &           &             &  $\times$\phantom{*}  &               \\
 & Maximum $|\eta|$ (lepton 0, lepton 1)     & $\times$\phantom{*} &            &           &             &  $\times${$*$}  &               \\
 & Lepton flavor                                  & $\times${$*$}  & $\times${$*$}   &           &             &            &               \\
 & Lepton charge                                   &          & $\times$\phantom{*}   &           &             &            &               \\
 \hline
 \parbox[t]{5mm}{\multirow{12}{*}{\rotatebox[origin=c]{90}{Jet properties}}} & Number of jets                                  & $\times${$*$}  & $\times${$*$}    &           &  $\times$\phantom{*}   &  $\times$\phantom{*}  &  $\times$\phantom{*}     \\
 & Number of $b$-tagged jets                         & $\times${$*$}  & $\times${$*$}    &           &  $\times$\phantom{*}   &  $\times$\phantom{*}  &  $\times$\phantom{*}     \\
 & Leading jet \pt                                 &          &            &           &             &             &  $\times$\phantom{*}     \\
 & Second leading jet \pt                          &          & $\times$\phantom{*}   &           &             &  $\times${$*$}  &               \\
 & Leading $b$-tagged jet \pt                      &          & $\times$\phantom{*}   &           &             &            &               \\
 & Scalar sum of all jets \pt                      &          & $\times$\phantom{*}   &           &  $\times$\phantom{*}   &  $\times$\phantom{*}  &  $\times$\phantom{*}     \\
 & Scalar sum of all $b$-tagged jets \pt           &          &            &           &             &            &  $\times$\phantom{*}     \\
 & Has leading jet highest $b$-tagging weight?&          & $\times$\phantom{*}   &           &             &            &               \\
 & $b$-tagging weight of leading jet        &          & $\times$\phantom{*}           &           &             &   &               \\
 & $b$-tagging weight of second leading jet        &          & $\times$\phantom{*}           &           &             &  $\times$\phantom{*}  &               \\
 & $b$-tagging weight of third leading jet         &          &            &           &             &  $\times$\phantom{*}  &               \\
 & Pseudorapidity of fourth leading jet            &          &            &           &             &  $\times$\phantom{*}  &               \\
 \hline
 \parbox[t]{2mm}{\multirow{4}{*}{\rotatebox[origin=c]{90}{\tauh}}} & Leading \tauh \pt                               &          &            &           &  $\times$\phantom{*}   &            &  $\times$\phantom{*}     \\
 & Second leading \tauh \pt                        &          &            &           &  $\times$\phantom{*}   &            &               \\
 & Di-\tauh invariant mass                           &          &            &           &  $\times$\phantom{*}   &            &               \\
 & Invariant mass \tauh--furthest lepton           &          &            &           &             &  $\times$\phantom{*}  &               \\
 \hline
 \parbox[t]{5mm}{\multirow{14}{*}{\rotatebox[origin=c]{90}{Angular distances}}} & $\Delta R$(lepton~0, lepton~1)                &          & $\times$\phantom{*}   &           &             &            &               \\
 & $\Delta R$(lepton~0, lepton~2)                &          & $\times$\phantom{*}   &           &             &            &               \\
 & $\Delta R$(lepton~0, closest jet)            & $\times$\phantom{*} & $\times$\phantom{*}   &           &             &            &               \\
 & $\Delta R$(lepton~0, leading jet)            &          & $\times$\phantom{*}   &           &             &  $\times$\phantom{*}  &               \\
 & $\Delta R$(lepton~0, closest $b$-jet)        &          & $\times$\phantom{*}   &           &             &            &               \\
 & $\Delta R$(lepton~1, closest jet)            & $\times$\phantom{*} & $\times$\phantom{*}   &           &             &            &               \\
 & $\Delta R$(lepton~2, closest jet)            &          & $\times$\phantom{*}   &           &             &            &               \\
 & Smallest $\Delta R$(lepton, jet)            &          & $\times$\phantom{*}   &           &             &            &  $\times$\phantom{*}     \\
 & Smallest $\Delta R$(lepton, $b$-tagged jet) &          &            &           &             &            &  $\times$\phantom{*}     \\
 & Smallest $\Delta R$(non-tagged jet, $b$-tagged jet)   &          &            &           &             &            &  $\times$\phantom{*}     \\
 & $\Delta R$(lepton~0, \tauh)                  &          &            &           &             &            &  $\times$\phantom{*}     \\
 & $\Delta R$(lepton~1, \tauh)                  &          &            &           &             &            &  $\times$\phantom{*}     \\
 & Minimum $\Delta R$ between all jets       &          &            &           &  $\times$\phantom{*}   &            &               \\
 & $\Delta R$ between two leading jets       &          &            &           &             &  $\times$\phantom{*}  &               \\
 \hline
 \parbox[t]{5mm}{\multirow{3}{*}{\rotatebox[origin=c]{90}{$\overrightarrow{\pT}^{\mathrm{miss}}$}}} & Missing transverse momentum \met                      & $\times$\phantom{*} &            & $\times$\phantom{*}  &             &            &               \\
 & Azimuthal separation $\Delta\phi$(leading jet, $\overrightarrow{\pT}^{\mathrm{miss}}$)           &          & $\times$\phantom{*}   &           &             &            &               \\
 & Transverse mass leptons ($H/Z$ decay) - $\overrightarrow{\pT}^{\mathrm{miss}}$  &          &            & $\times$\phantom{*}  &             &            &               \\
 \hline
 & Pseudo-Matrix-Element                           &          &            & $\times$\phantom{*}  &             &            &               \\
\hline\hline
 \end{tabular}
}
\end{center}
\end{table}

\subsection{Channel summary}

Twelve categories are defined in the previous subsections: eight signal regions and four control regions (CR) from the \lll channel.
The fraction of the expected signal arising from different Higgs boson decay modes in each signal region  is shown in Figure~\ref{fig:soverb} (left).
The signal-to-background ratio $S/B$ for each signal and control region is shown in Figure~\ref{fig:soverb} (right). This ranges from 0.014 to almost 2.
The ratio $S/\sqrt B$ is also indicated. The acceptance for each channel is shown in Table~\ref{tbl:acc}. The background composition in each region is shown in Figure~\ref{fig:piechart}. The background prediction methods are described in the next section. Multivariate techniques have been applied in most channels to improve the discrimination between the signal and the background. The variables used in each channel are indicated in Table~\ref{tbl:mvavar}. The modeling of each variable was checked and no significant disagreement between data and simulation was found.\\

\begin{figure}[!htbp]
\centering
{\includegraphics[width=0.47\textwidth]{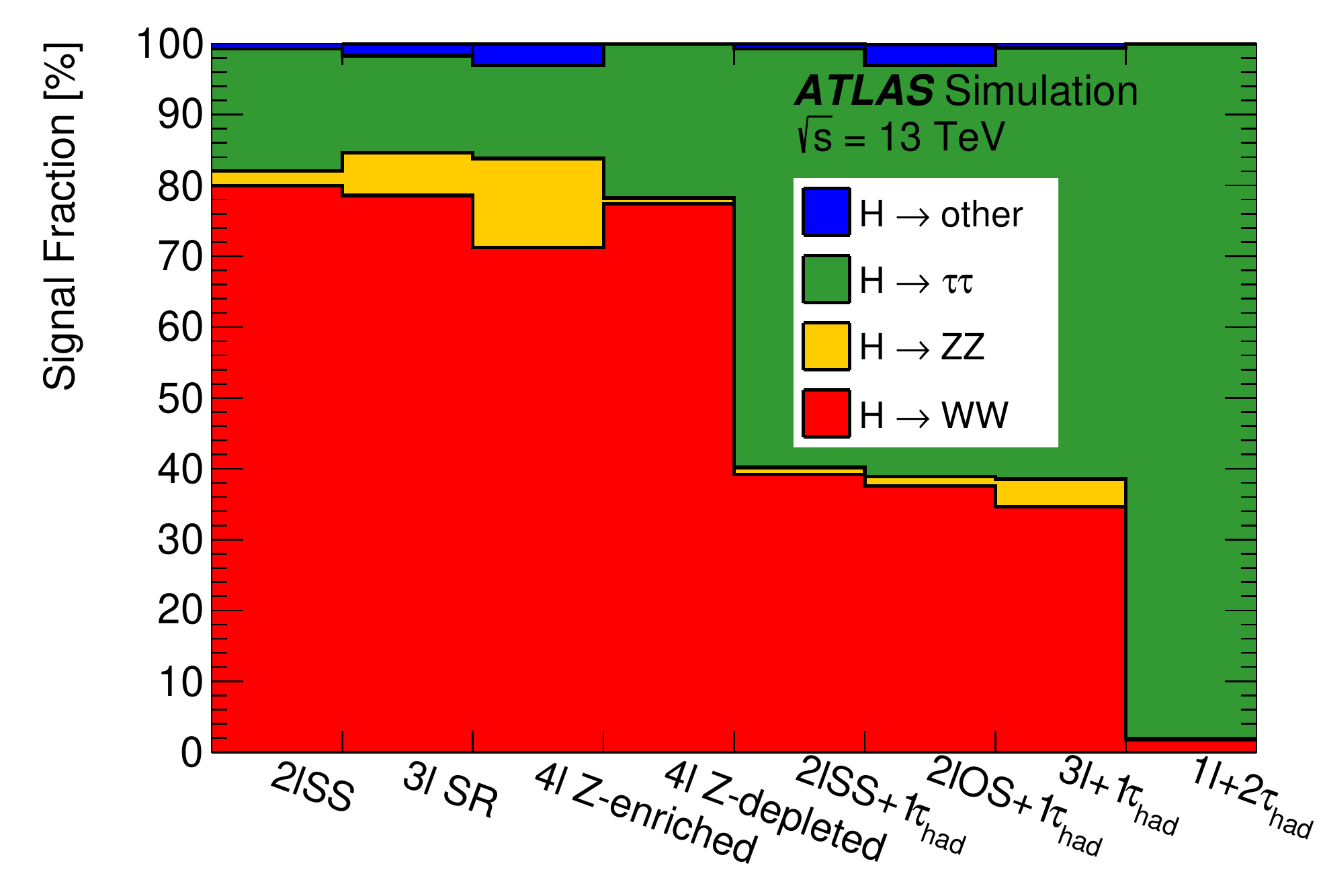}}
{\includegraphics[width=0.47\textwidth]{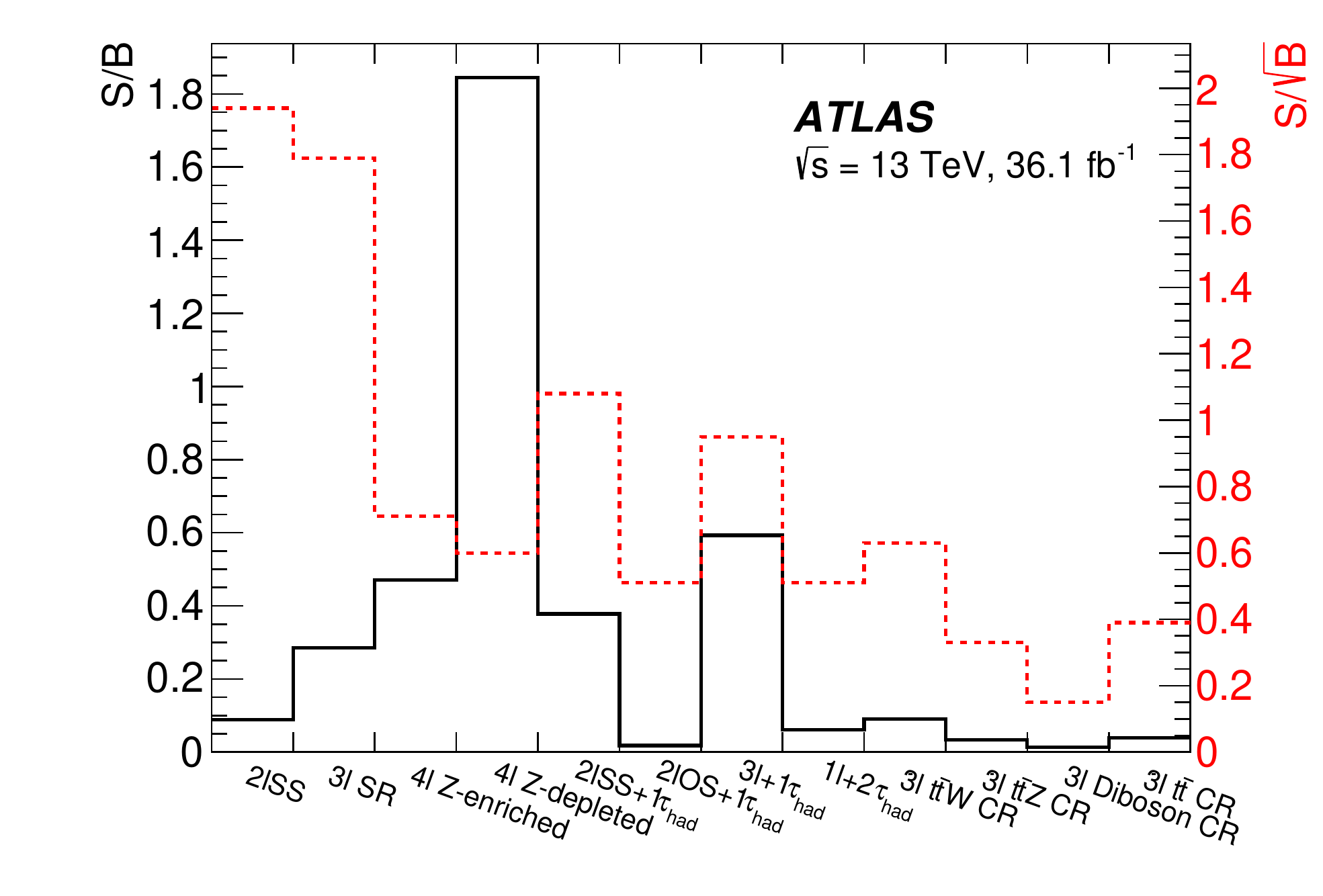}}
  \caption{Left: The fraction of the expected \ttH signal arising from different Higgs boson decay modes in each signal region. The decays labeled as ``other'' are mostly $H\to\mu\mu$ and $H\to b\bar b$.
  Right: Pre-fit $S/B$ (black line) and $S/\sqrt B$ (red dashed line) ratios for each of the twelve analysis categories including the four \lll control regions. The background prediction methods are described in Section~\ref{sec:bkg}.\label{fig:soverb}}
\end{figure}

\begin{figure}[!htbp]
\centering
\includegraphics[width=0.75\textwidth]{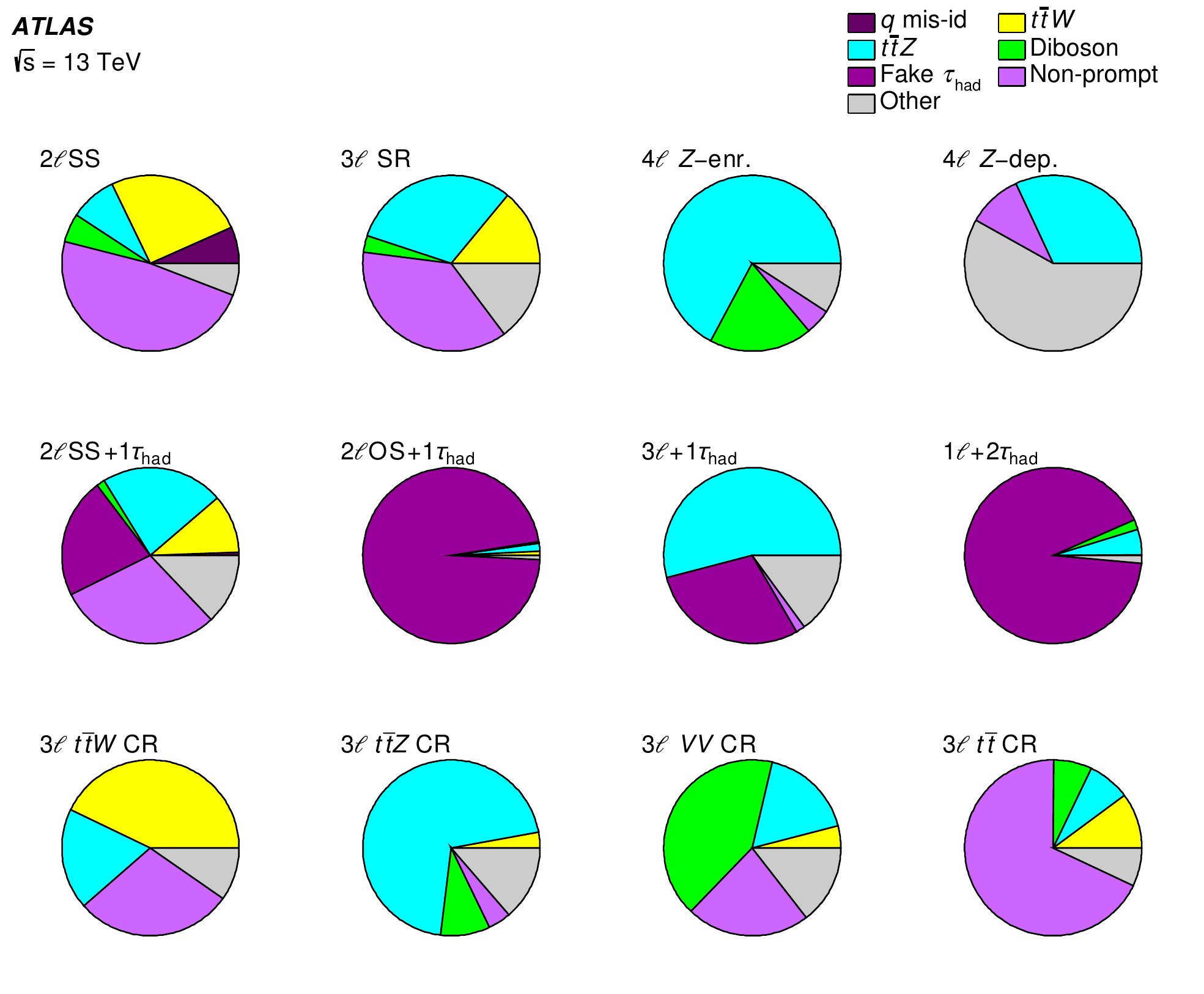}
  \caption{The fractional contributions of the various backgrounds to the total predicted background in each of the twelve analysis categories. The background prediction methods are described in Section~\ref{sec:bkg}:
  ``Non-prompt'', ``Fake \tauh'' and ``$q$ mis-id'' refer to the data-driven background estimates (largely \ttbar but also include other electroweak processes), and rare processes ($tZ$, $tW$, $tWZ$, $\ttbar WW$, triboson production, $t\bar t t$, $t\bar t t \bar t$, $tH$, rare top decay) are labeled as ``Other''. \label{fig:piechart}}
\end{figure}

\clearpage

\section{Background estimation}
\label{sec:bkg}

The irreducible backgrounds all have selected light leptons produced in $W$ or $Z/\gamma^*$ boson decays or leptonic $\tau$ decays (prompt leptons, Section~\ref{subsec:promptbkg}). The reducible backgrounds have at least one lepton arising from another source (Section~\ref{subsec:nonpromptbkg}).
In the latter case, light leptons originate from heavy-flavor hadron decays, photon conversions, improper reconstruction of other particles such as hadronic jets, or prompt leptons whose charge is misassigned.
Such misidentified and non-prompt light leptons are collectively referred to as non-prompt leptons in the following, as this is the dominant source.
The fake \tauh candidates are typically jets, including HF jets.

\subsection{Backgrounds with prompt leptons}
\label{subsec:promptbkg}
Background contributions with prompt leptons originate from a wide range of processes and the relative importance of individual processes varies by channel. The largest backgrounds with prompt leptons are from top production in association with a vector boson, \ttW and $\ttbar (Z/\gamma^*)$, and diboson production, $VV$. These background estimates are a crucial part of the analysis, because their final state and kinematics are similar to the signal. In addition, there are contributions from a number of rare processes:  rare top decay, $tZ$, $tW$, $tWZ$, $\ttbar WW$, $VVV$, $t\bar t t$ and $t\bar t t \bar t$ production.
The associated production of single top quarks with a Higgs boson, which contributes at most 2\% in any signal region, is also considered as a background process.
All other Higgs boson production mechanisms contribute negligibly (<0.2\%) in any signal region.

All these backgrounds are estimated from simulation using the samples described in Section~\ref{sec:mc}. The systematic uncertainties in the modeling of these processes by the simulation are discussed in Section~\ref{sec:syst}. The prompt-lepton estimates were validated in various regions, as illustrated in Figure~\ref{fig:promptvalid} for the \lll \ttZ and \ttW control regions.

\begin{figure*}[htbp]
\centering
{\includegraphics[width=0.47\textwidth]{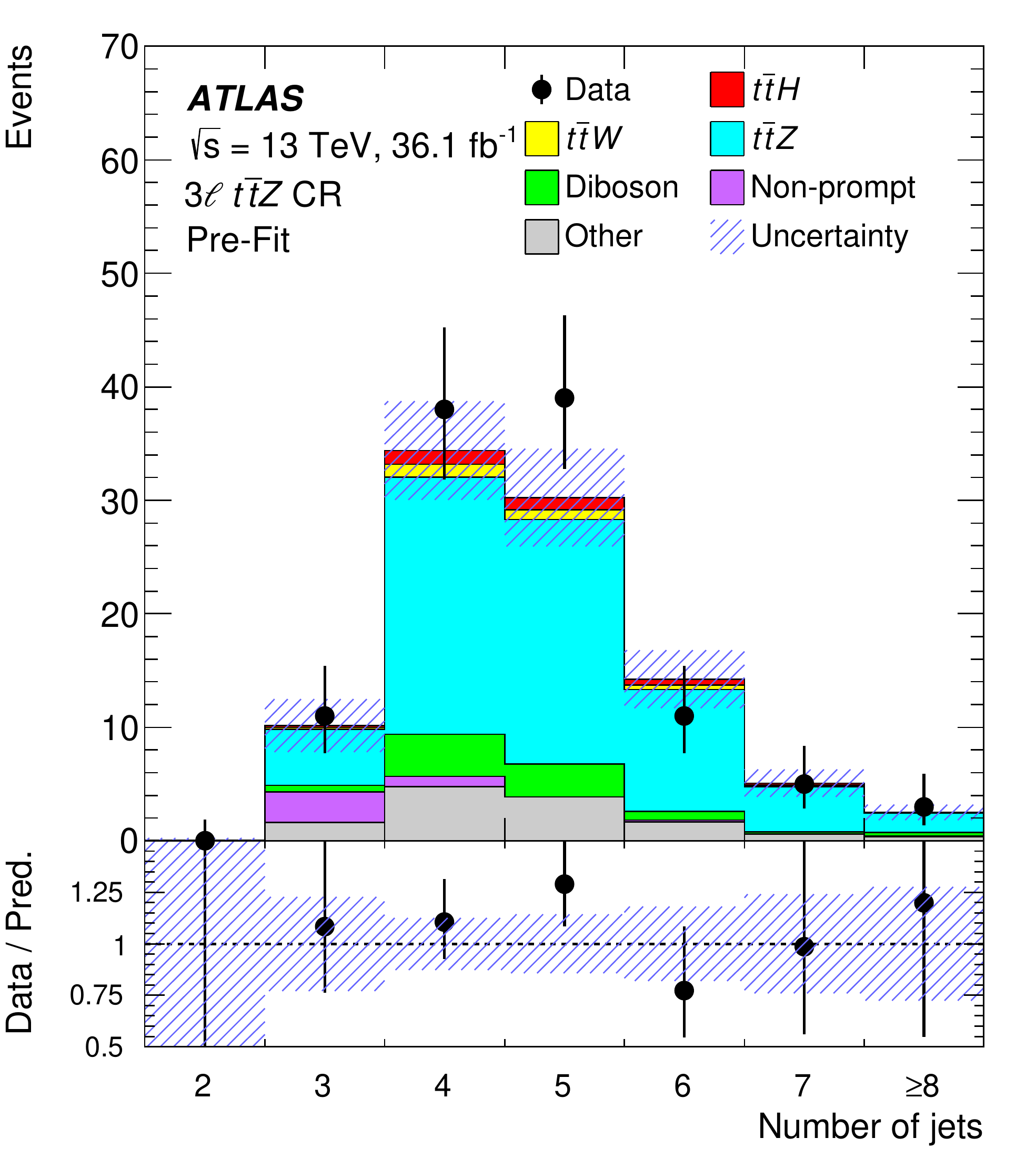}}
{\includegraphics[width=0.47\textwidth]{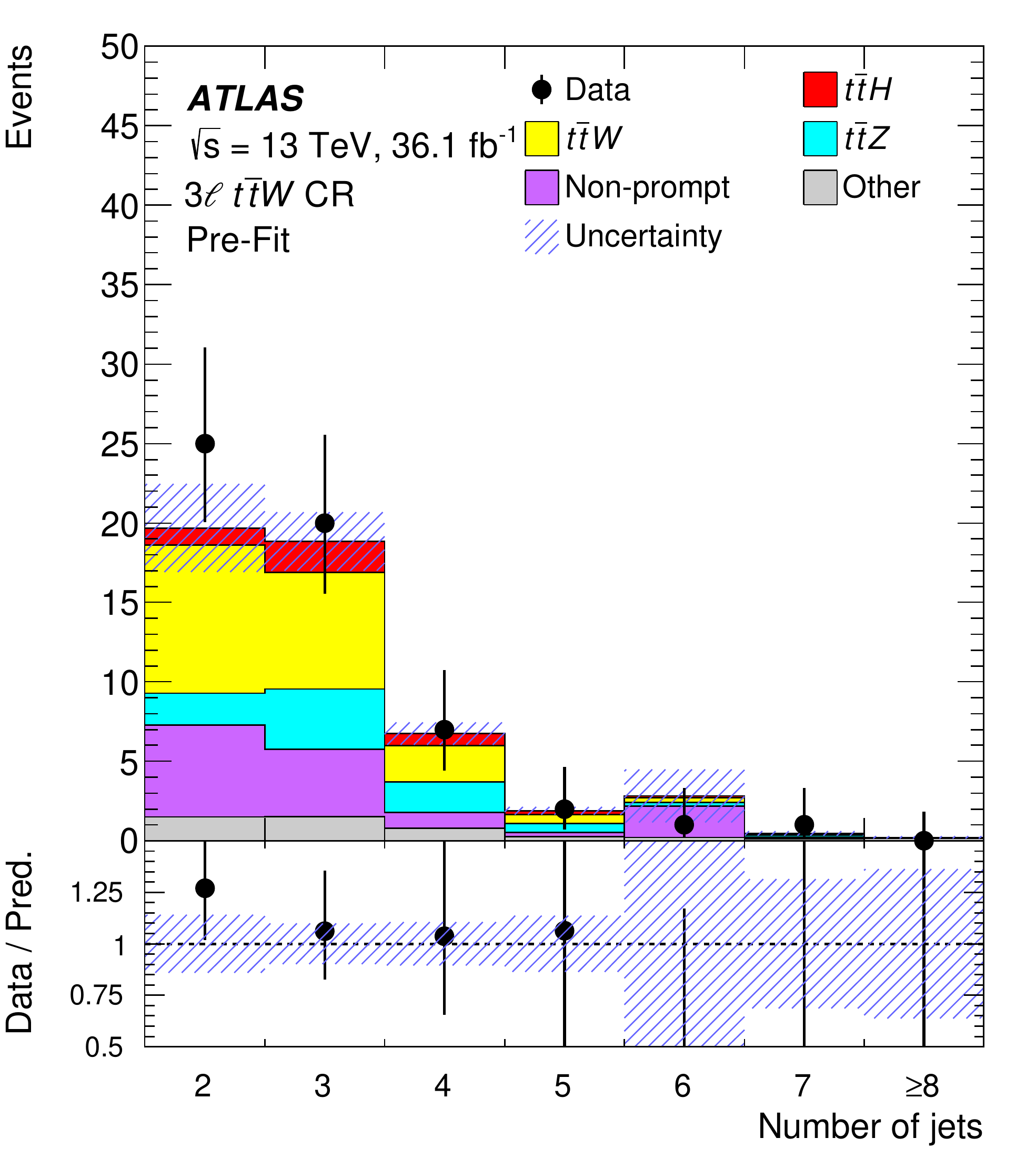}}
  \caption{Comparison of data and prediction of the jet multiplicity in the (left) \lll \ttZ and the (right) \lll \ttW control regions. The last bin in each figure contains the overflow. The bottom panel displays the ratio of data to the total prediction. The hatched area represents the total uncertainty in the background. The background prediction for non-prompt leptons is described in Section~\ref{subsec:nonpromptbkg} and the other backgrounds are normalized according to the predictions from simulation. \label{fig:promptvalid}}
\end{figure*}

\subsection{Backgrounds with non-prompt leptons and fake \tauh candidates}
\label{subsec:nonpromptbkg}
Data-driven methods are used to estimate the backgrounds with non-prompt light leptons and fake \tauh candidates, defining control regions enriched in such backgrounds and extrapolating the observed yields to the signal regions. The control regions used for this purpose are summarized in Table~\ref{tbl:fakecr}. They are orthogonal to the signal regions.
Figure~\ref{fig:fakeComp} summarizes the origin of the non-prompt leptons and fake \tauh candidates in these control regions and some signal regions based on predictions from simulation, where the statistical uncertainties of the absolute fractions can be as large as 7\%.

Table~\ref{tbl:summary_fakeStrat} summarizes the strategies used to estimate the non-prompt lepton and fake \tauh backgrounds in each of the channels, motivated by the different event topologies and the statistical power available in the control regions. The matrix method and fake-factor method are largely similar, but differ in that the fake-factor method estimates the prompt contribution from simulation, while the matrix method uses the measured prompt lepton efficiency from data.

\begin{table}[!h]
 \centering
 \caption{Selection criteria applied to define the control regions used for the non-prompt lepton (top part) and fake \tauh (bottom part) estimates. The \ll CR is used for both the \ll and \lll channels, as indicated by putting \lll in parenthesis. Same-flavor, opposite-charge (same-charge) lepton pairs are referred to as SFOC (SFSC) pairs.}
 \begin{tabular}{lll}
  \hline\hline
  Channel & Region & Selection criteria\\
  \hline \hline
  \ll    &   & $2 \le \njets \le 3 $ and $\nbjets \ge 1$\\
  (\lll)      &   & One very tight, one loose light lepton with $\pt > 20~(15)$~GeV\\
        &   & Zero \tauh candidates \\
        &   $\epsilon_{\mathrm{real}}$ & Opposite charge; opposite flavor\\
        &   $\epsilon_{\mathrm{fake}}$  & Same charge; opposite flavor or $\mu\mu$ \\
  \hline
   \llll   & & 1 $\le \njets \le 2 $ \\
  & & Three loose light leptons; sum of light lepton charges $\pm1$ \\
            &   & Subleading same-charge lepton must be tight\\
            &   & Veto on \lll selection \\
            & Either & One SFOC pair with $|m(\ell^+\ell^-)-91.2\textrm{ GeV}| < 10$~\GeV{} \\
            & & $\met < 50$~GeV, $m_{\mathrm{T}} < 50$~GeV\\
            & or & No SFOC pair \\
            & & Subleading jet $\pT > 30$~GeV \\
  \hline
   \lltau & &  $2 \le \njets \le 3$ and $\nbjets \ge 1$\\
    &   & One very tight, one loose light lepton with $\pt > 15$~GeV\\
           & & A SFSC pair \\
          &       & $|m(ee)-91.2\textrm{ GeV}| > 10$ GeV \\
           &       & Zero or one medium \tauh candidate, opposite in charge to the light leptons \\
\hline\hline
   \ltwotau &&  $\njets \ge 3$ and $\nbjets \ge 1$\\
   & & One tight light lepton, with $\pt > 27$~GeV \\
            &   & Two \tauh candidates of same charge \\
	          & & At least one \tauh candidate has to satisfy tight identification criteria \\
  \hline
  \OSlltau  & & Two loose and isolated light leptons, with $\pt > 25$, 15~GeV \\
         & & One loose \tauh candidate \\
            & & $|m(\ell^+\ell^-)-91.2\textrm{ GeV}| > 10$ GeV and $m(\ell^+\ell^-) > 12$~GeV \\
              & & $\njets \ge 3 $ and $\nbjets = 0$\\
  \hline\hline
 \end{tabular}
\label{tbl:fakecr}
\end{table}

\begin{figure}[!htbp]
\centering
\subfigure[Fake and non-prompt lepton composition]{\includegraphics[width=0.47\textwidth]{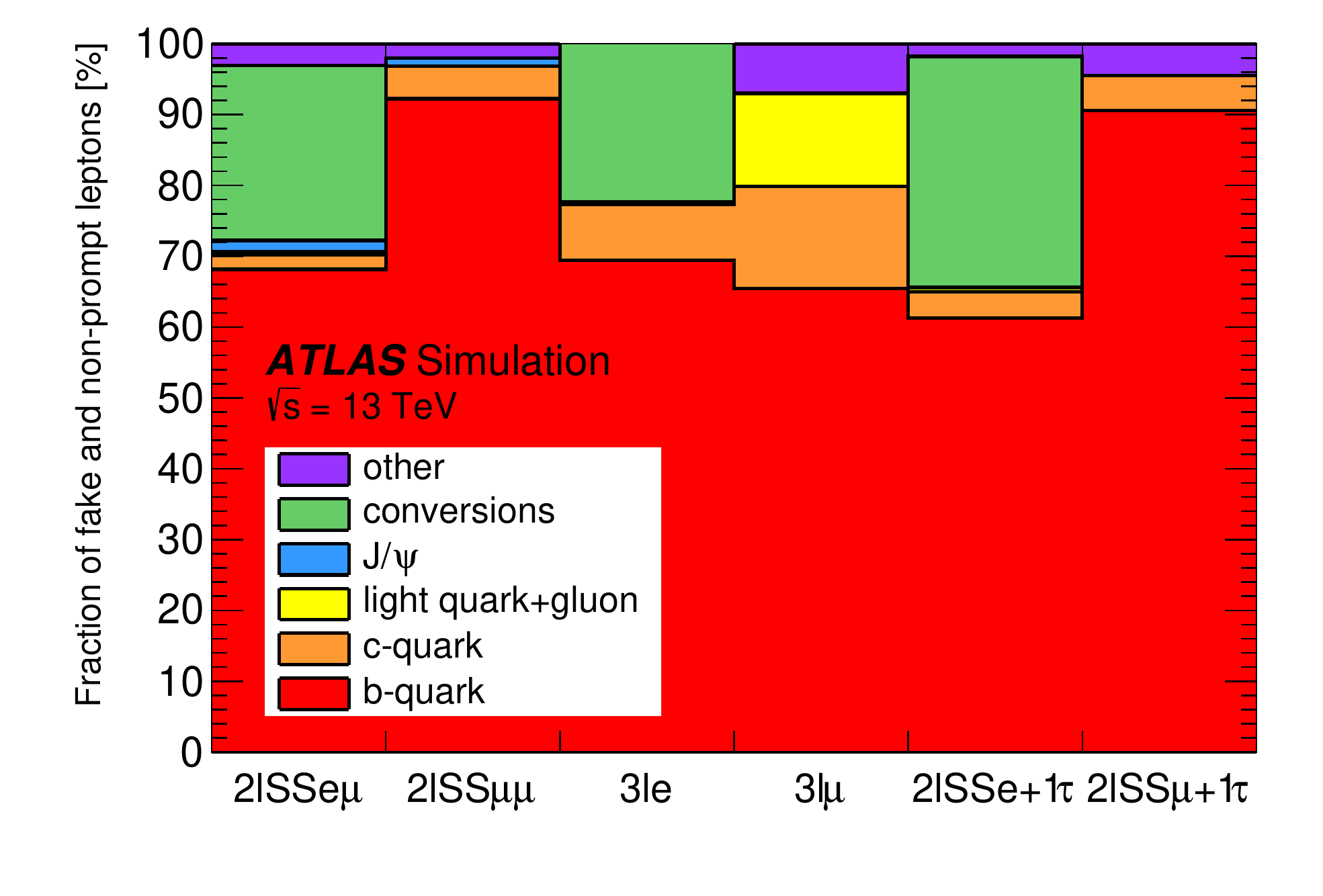}}
\subfigure[Fake \tauh composition]{\includegraphics[width=0.47\textwidth]{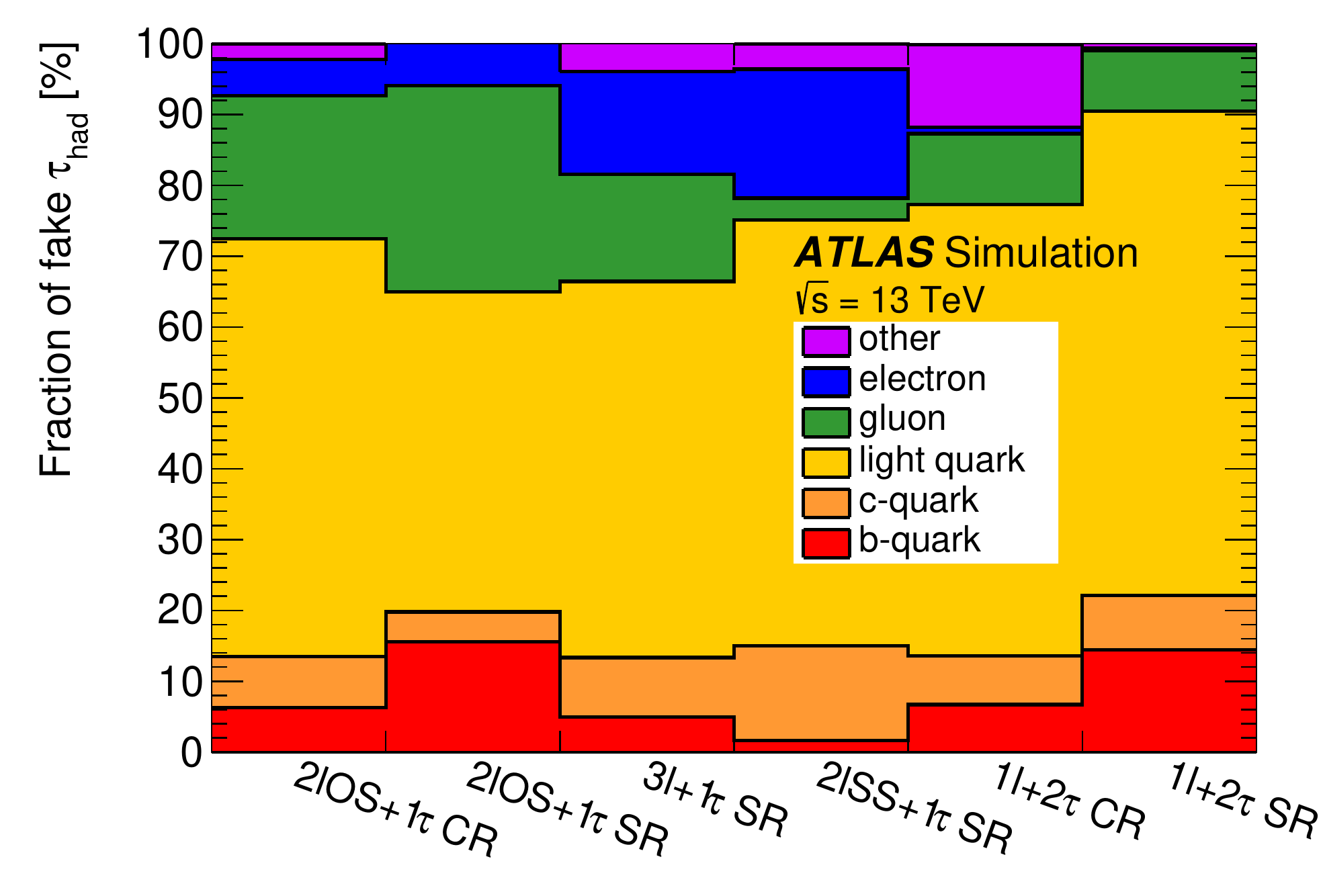}}
  \caption{The composition from simulation of (left) the fake and non-prompt light leptons and (right) the fake \tauh in selected analysis regions. The light-lepton composition is shown separately depending on the lepton flavor in the regions used in the estimate of the non-prompt contribution. The control regions labeled `2lSSxx' are used for the \ll and \lll channels; those labeled `3lx' are used for the \llll channel where x denotes the flavor of the lowest-\pT lepton and those labeled `2lSSx+1$\tau$' are used for the \lltau channel. The non-prompt lepton background has been separated into the components from $b$-jets, $c$-jets, other jets, $J/\psi$, photon conversions and other contributions. The latter includes pion, kaon and non-prompt tau decays and cases where reconstructed leptons cannot be assigned unambiguously to a particular source.
The \tauh composition is shown both in the control regions used in the estimates and in the signal regions of each channel. The \tauh background has been separated into the components from $b$-jets, $c$-jets, light-quark jets, gluon jets, electrons and other contributions. The latter includes muons, hadrons and cases where reconstructed leptons cannot be assigned unambiguously to a particular source.\label{fig:fakeComp}}
\end{figure}

\begin{table}[h!]
\begin{center}
\caption{\label{tbl:summary_fakeStrat} Summary of the non-prompt lepton and fake \tauh background estimate strategies of the seven analysis channels. DD means data-driven background estimates and the techniques used are the matrix method (MM) and the fake-factor method (FF). The scale factor method (SF), which scales the estimate from simulation by a correction factor measured in data, is partially data-driven. The lower half of the table lists the selection requirements used to define the control regions. The lepton selection follows the same convention as in Table~\ref{tbl:tightleps} and is labeled as loose (L), loose and isolated (L$^\dag$), loose, isolated and passing the non-prompt BDT (L*), tight (T) and very tight (T*), respectively. Analogously, the \tauh selection is labeled as medium (M) and tight (T).}
\resizebox{\textwidth}{!}{
\begin{tabular}{l|ccccccc}
\hline\hline
                               & \ll      & \lll    & \llll    & \ltwotau    & \lltau     & \OSlltau  & \llltau    \\
\hline
Non-prompt lepton strategy             &  DD  & DD  & semi-DD & MC & DD  & MC & MC \\
        &  (MM) & (MM) & (SF)   &    &  (FF)  &    &    \\
Fake \tauh strategy                      & -- & -- & -- & DD  & semi-DD  & DD & semi-DD \\
            &   &    &   & (SS data) & (SF)  & (FF)   & (SF)   \\
\hline
\multicolumn{8}{c}{Control Region Selection} \\
\hline
Light lepton          &  \multicolumn{2}{c|}{1T*, 1L} & 3L &  1T  &  1T*, 1L  &  2L$^\dag$ & --    \\
\tauh                 & \multicolumn{3}{c|}{0M} & 1T, 1M & $\le1$M & 1L & -- \\
\njets     &
\multicolumn{2}{c|}{$2\le  \njets \le 3$}
& $1 \le \njets \le 2$
& $\ge 3$
& $2 \le \njets \le 3$
& $\ge 3$
& -- \\
\nbjets & \multicolumn{5}{c|}{$\ge 1$} & $=0$ & -- \\
\hline\hline
\end{tabular}
}
\end{center}
\end{table}

\subsubsection{Non-prompt leptons in the \ll and \lll channels}
\label{subsubsec:fakestwothreel}
The non-prompt lepton background in the \ll and \lll channels is a mixture of leptons from semileptonic HF decays and conversions.
These backgrounds are estimated using a matrix method similar to that described in Refs.~\cite{TOPQ-2010-01,TOPQ-2011-01}. The matrix method estimates the number of non-prompt leptons in the signal region by selecting events passing all selection requirements except the tight-lepton requirements and splitting the events into four categories. The four categories contain exactly two tight leptons, one tight and one  loose-but-not-tight lepton, one loose-but-not-tight and one tight lepton, and two loose-but-not-tight leptons (where the leptons are ordered according to their \pt). The probabilities for both the loose prompt and non-prompt leptons to be tight are measured in control regions independent from the signal regions. These are used to estimate the number of non-prompt events in the signal regions via the following formula: $f_\mathrm{SR} = w_\mathrm{TT}N^\mathrm{TT} + w_{\bar{\mathrm{T}}\mathrm{T}}N^{\bar{\mathrm{T}}\mathrm{T}} + w_{\mathrm{T}\bar{\mathrm{T}}}N^{\mathrm{T}\bar{\mathrm{T}}} +w_{\bar{\mathrm{T}}\bar{\mathrm{T}}}N^{\bar{\mathrm{T}}\bar{\mathrm{T}}}$. The $w$ weights depend on the measured prompt and non-prompt lepton efficiencies, T and $\bar{\mathrm{T}}$ denote leptons passing the tight and loose-but-not-tight lepton selections respectively.

In the \ll channel, the method allows either of the candidate leptons to be non-prompt, while in the \lll channel, the opposite-charge lepton is assumed to always be prompt, as is seen in the simulation for 97\% of the cases. The efficiencies are measured separately for electrons and muons.

The control regions used to measure the prompt ($\epsilon_{\mathrm{real}}$) and non-prompt ($\epsilon_{\mathrm{fake}}$) lepton efficiencies are defined in Table~\ref{tbl:fakecr}. They have lower jet multiplicity than the signal regions.
The lepton efficiencies are parameterized as a function of \pt. The non-prompt electron efficiency is additionally parameterized as a function of the number of $b$-jets in the events to account for changes in the composition of fakes. The non-prompt muon efficiency is additionally parameterized as a function of the angular distance between the lepton and the closest jet to account for effects of nearby jets.
The residual prompt background in the control regions is subtracted using the prediction from simulation, while the background from charge misassignment is subtracted using the estimate described in Section~\ref{subsubsec:chargemisrec}.

The efficiency for electrons from conversions is significantly higher than that for electrons from HF decays; therefore the change in the fraction of conversions when going from the control to the signal regions is estimated from simulation and used to correct $\epsilon_{\mathrm{fake}}$. Systematic uncertainties in this correction are estimated to be 40\%. They include a 15\% uncertainty in the modeling of conversions in the simulation~\cite{Aaboud:2017pjd}, a 20\% uncertainty from a measurement of $\ttbar\gamma$~\cite{Aaboud:2017era2}, a 50\% uncertainty in the modeling of semileptonic $b$-decays and the uncertainties in the non-prompt lepton efficiencies.

The performance of the matrix method was tested in simulation using a closure test by comparing the prediction from the method to the results from the simulation. Closure tests were performed for each channel using \ttbar simulation and the level of the non-closure is found to be at most ($11 \pm 8$)\% and ($9 \pm 18$)\% for the \ll and \lll channels, respectively, which
is accounted for as a systematic uncertainty. Additional systematic uncertainties due to the subtraction of the prompt backgrounds in the control regions are included. The total uncertainty in the non-prompt lepton estimate varies from 20\% for $e^\pm\mu^\pm$ to 30\% for \lll. The ratio for the non-prompt background yield in data to the predictions from simulation is found to be $2.0\pm0.5$ for $ee$, $1.5\pm0.5$ for $\mu\mu$ and $1.7\pm0.4$ for $e\mu$ in the \ll signal region. It is $1.8\pm0.8$ for \lll in the signal region and $2.2 \pm 0.5$ in the \ttbar control region.
The non-prompt lepton estimates were validated in various regions, as illustrated in Figure~\ref{fig:fakevalid}(a) and~\ref{fig:fakevalid}(b) in a region identical to the \ll signal region except for being orthogonal in the \njets requirement (low multiplicity \njets$=2,3$).

\begin{figure*}[htbp]
\centering
\subfigure[]{\includegraphics[width=0.47\textwidth]{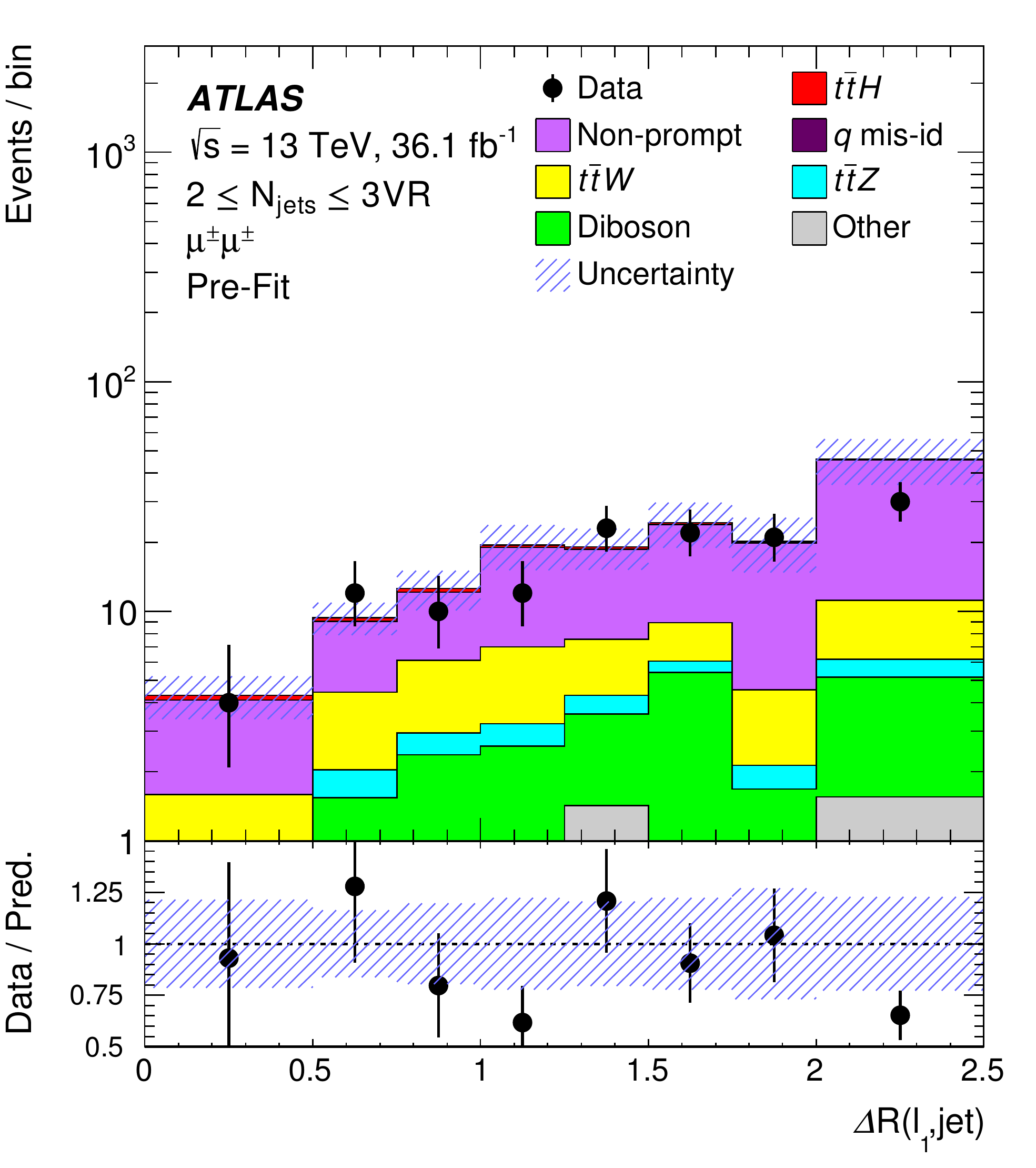}}
\subfigure[]{\includegraphics[width=0.47\textwidth]{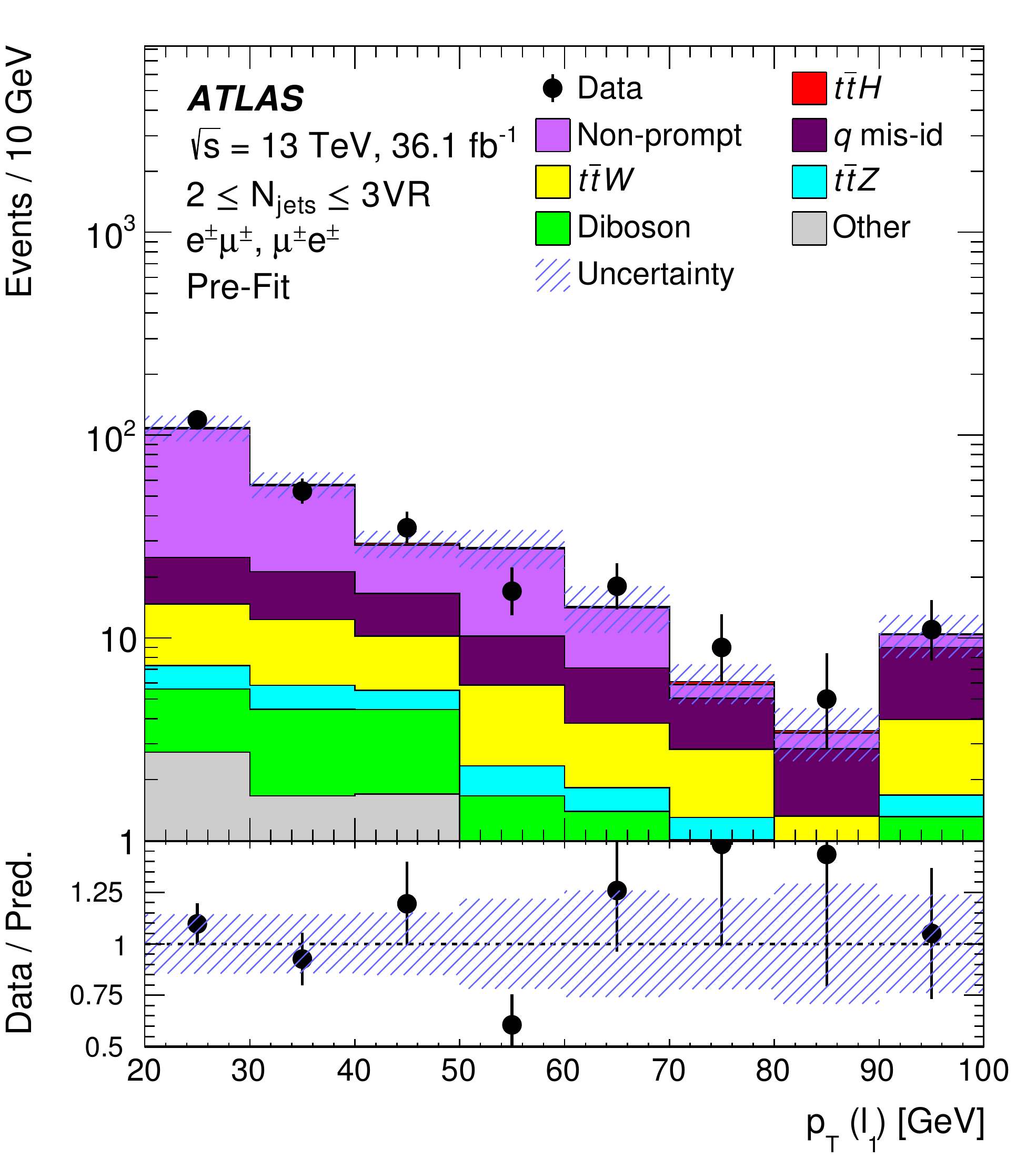}}
\subfigure[]{\includegraphics[width=0.47\textwidth]{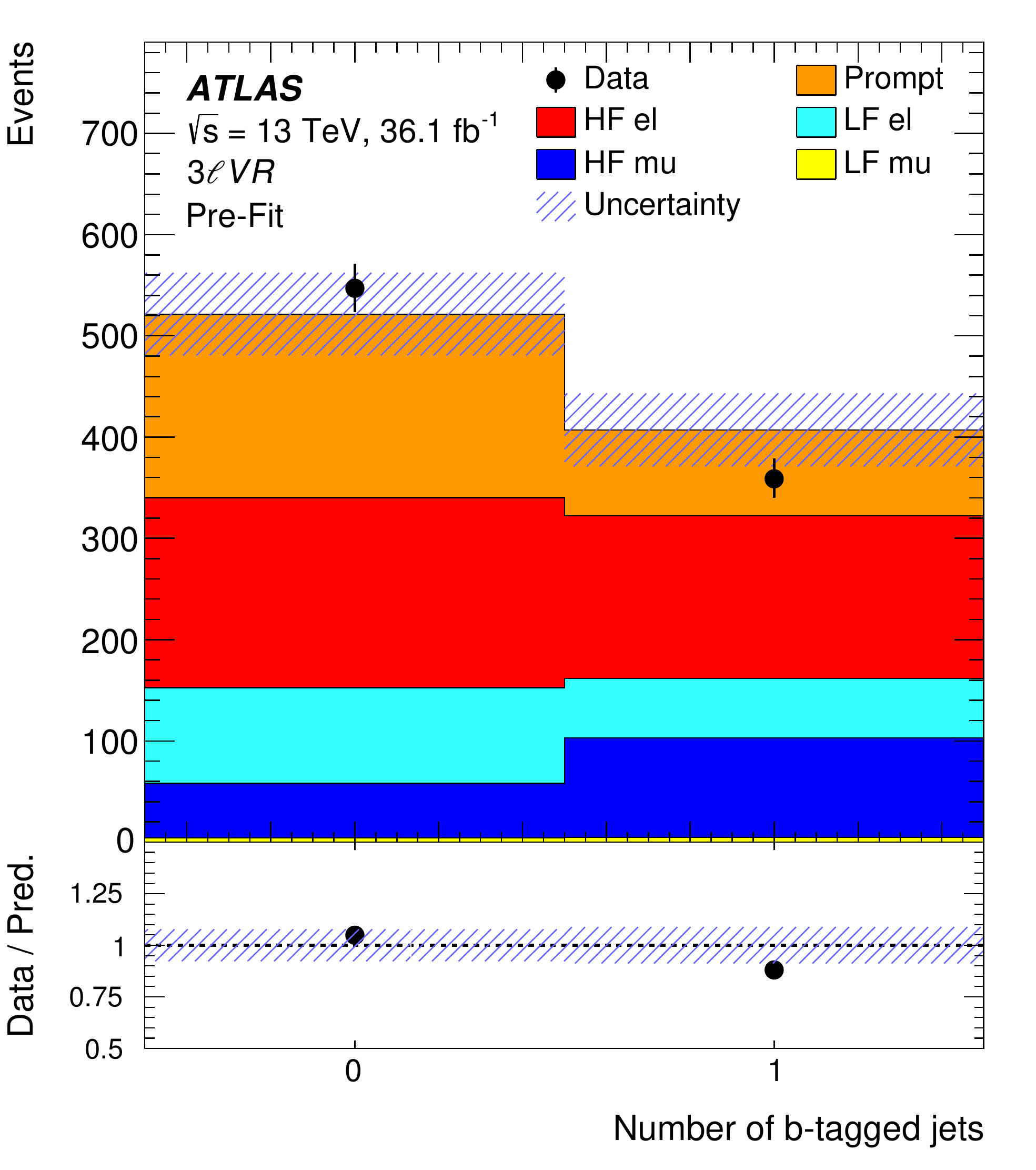}}
\subfigure[]{\includegraphics[width=0.47\textwidth]{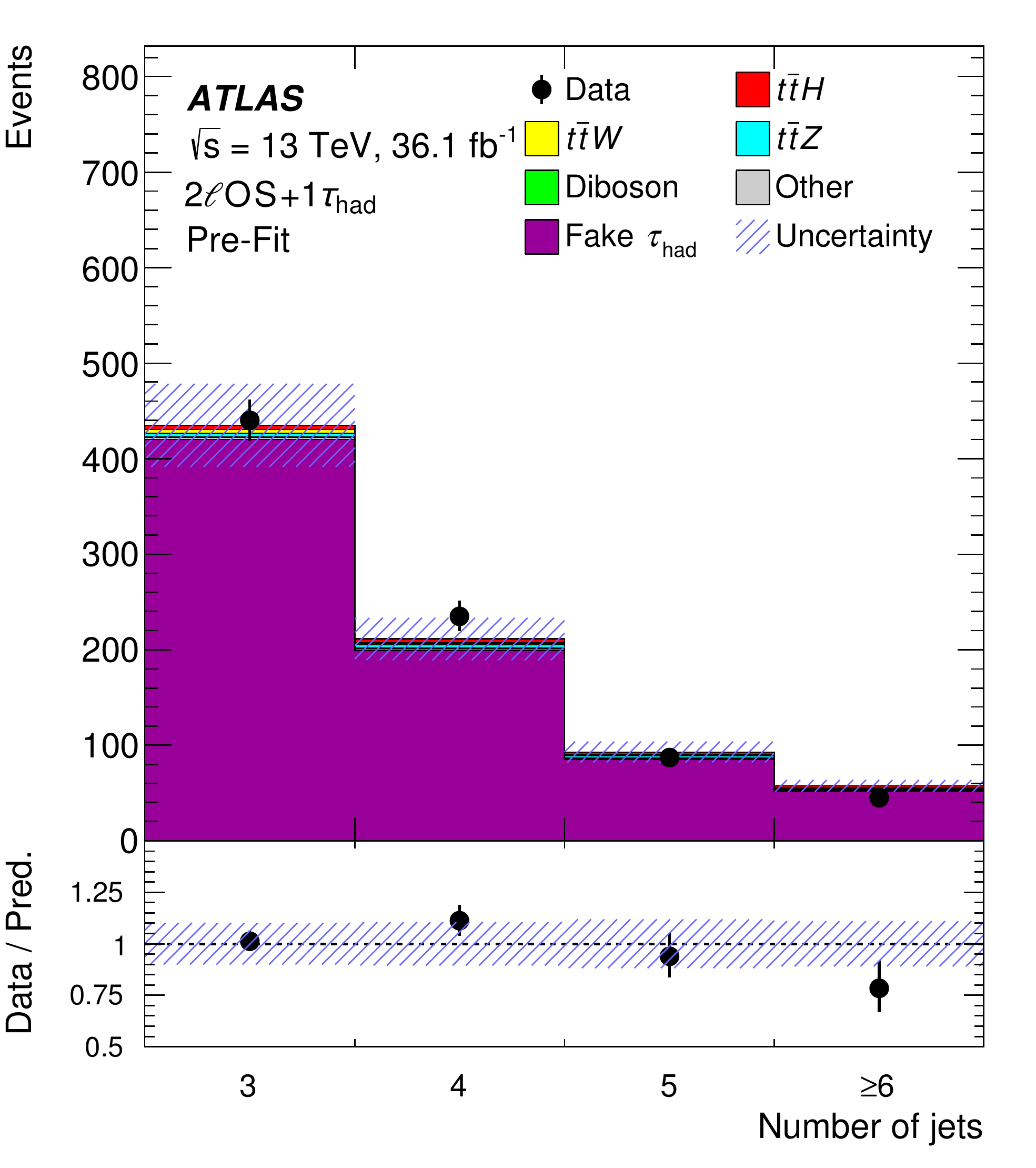}}
  \caption{Comparison of data and prediction of (a) the angular distance between the subleading lepton and the closest jet in the $\mu\mu$ channel and (b) the subleading lepton \pt in the opposite-flavor channel, in a \ll low-\njets validation region (VR); (c) the $b$-tagged jet multiplicity in a validation region similar to the control region used in the \fourl channel but at higher \njets multiplicity (called $3\ell$~VR), with the leptons categorized according to their origin: prompt, heavy-flavor (HF) and light-flavor (LF), see text; (d) the jet multiplicity in the \OSlltau category. The last bin in each figure contains the overflow. The bottom panel displays the ratio of data to the total prediction. The hatched area represents the total uncertainty in the background. \label{fig:fakevalid}}
\end{figure*}

\subsubsection{Non-prompt leptons in the \fourl channel}
A semi-data-driven estimate of the non-prompt leptons is used in the \fourl channel. Leptons are separated according to their origin: prompt, heavy-flavor and light-flavor, with the latter designation including leptons from photon conversions. As the rate of non-prompt muons originating from light-flavor hadrons is extremely low, the muons of heavy- and light-flavor origin are treated together. The control region defined in Table~\ref{tbl:fakecr} for the non-prompt lepton estimate in the \fourl channel, where three light leptons are required, is used. It is composed of roughly 50\% $Z$+jets, 30\% diboson and 20\% \ttbar events.
The control region is separated into four categories according to the flavor of the leptons ($eee$, $ee\mu$, $e\mu\mu$ and $\mu\mu\mu$) and a fit to the leading jet \pT distribution is performed to extract three normalization factors: $\lambda_{\textrm{heavy}}^{e} = 1.48\pm 0.22$, $\lambda_{\textrm{light}}^{e}=0.72 \pm 0.53$ and $\lambda^{\mu}=0.66 \pm 0.19$, where
the errors are statistical. The normalization factors are applied to all events containing non-prompt leptons to correct the yields from the simulation in each category to data. The composition of the non-prompt leptons in the control region is shown in Figure~\ref{fig:fakeComp}(a). The systematic uncertainty in each normalization factor is estimated to be 30\% by varying the \pT requirements on the leptons.
The non-prompt lepton estimates were validated in various regions, as illustrated in Figure~\ref{fig:fakevalid}(c) in a region identical to the control region used to extract the normalization factors except for being orthogonal in the \njets requirement (higher multiplicity \njets$>2$).

\subsubsection{Non-prompt leptons and fake \tauh candidates in other channels}
In the \llltau, \OSlltau and \ltwotau channels, the background from non-prompt light leptons is a few percent and is estimated from simulation, but the fake \tauh background, mainly arising from \ttbar and \ttV, is estimated from data. In the \lltau channel, both backgrounds are significant and hence are estimated from data.

In the \OSlltau channel, the fake-factor method is used to estimate the background from events containing a fake \tauh candidate. The method assumes that the real contribution is described well by simulation. The fake factors are estimated using the control region defined in Table~\ref{tbl:fakecr}, which applies the nominal \OSlltau selection but requires at least three jets and vetoes events containing $b$-jets. The fake factors are parameterized as a function of $\pT^{\tauh}$ and no significant dependence on other key event properties was found. Systematic uncertainties include the statistical uncertainty in the control regions, differences in the fake composition between the control and signal regions and the variation in the fake factors between different control regions. The total systematic uncertainty in the fake \tauh background estimate in this channel is 11\%.
Figure~\ref{fig:fakevalid}(d) illustrates a validation of this estimate in the  \OSlltau selection region, which is largely dominated by events with a fake \tauh.

As the origin of the \tauh fakes is very similar between the channels, as demonstrated in Figure~\ref{fig:fakeComp}(b), an extrapolation is made to the \lltau and \llltau channels. The fake factors derived in the \OSlltau channel are converted into a scale factor to correct the simulation of fake \tauh candidates coming from jets in order to better describe the data. The scale factor is derived in the \OSlltau control region and then applied in the respective signal regions. Its dependence on \pT was found to be negligible. Uncertainties in the scale factor are derived by comparing the value in the nominal control region to those obtained in control regions enriched in \ttbar and $Z$ boson events, respectively. The final scale factor is $1.36 \pm 0.16$ including statistical and systematic uncertainties.

In the \lltau channel, this scale factor is applied only to backgrounds containing prompt leptons and fake \tauh candidates. An additional fake-factor method is used to estimate the background from events containing non-prompt light leptons. This fake factor is derived in a control region defined in Table~\ref{tbl:fakecr}, which differs from the signal region by looser lepton requirements and lower jet multiplicity.
As in the \ll and \lll non-prompt lepton estimates, the change in the fraction of conversions from the control to the signal region is taken into account, with the same associated uncertainties.
The total systematic uncertainty in the non-prompt lepton estimate in this channel is 55\%, dominated by the statistical uncertainty in the closure test of the method found in simulation.

The dominant background in the \ltwotau signal region is \ttbar production where one or two \tauh are fakes from \ttbar decays. As there is equal probability for a jet to be reconstructed as a positively or negatively charged \tauh, the fakes are estimated from a control region identical to the signal region except that the \tauh candidates are required to have the same charge, as shown in Table~\ref{tbl:fakecr}. This region contains almost entirely fakes from \ttbar decays. The estimate is extrapolated to the signal region after using simulation to subtract the contribution from real \tauh in the control region. Using simulation, the non-closure of this method was found to be below 30\%, which is included as a systematic uncertainty.

\subsubsection{Charge misassignment}
\label{subsubsec:chargemisrec}
The electron charge misassignment rate is measured in data, and the corresponding background is taken into account in the \ll, \lltau channels and, indirectly, in the \lll channel via the non-prompt background estimate, by scaling opposite-charge data events by this rate. The measurement is performed within a sample of $Z \rightarrow ee$ events reconstructed as same-charge pairs and as opposite-charge pairs. Six bins in $|\eta|$ and four bins in \pT are used. The bins were chosen in accord with the size of the event sample and the variation of the rate with $|\eta|$ and \pT. The background is subtracted using a sideband method. The charge misassignment rate varies from $5 \times 10^{-5}$ for low-\pt electrons (\pt$\approx10$~GeV) at small $|\eta|$ to $10^{-2}$ for high-\pt electrons (\pt$\gtrapprox100$~GeV) with $|\eta| > 2$.

The electron charge misassignment measurement is validated by a closure test in simulation using same-charge pairs, with the observed difference between measured and predicted rates being taken as the systematic uncertainty. An additional validation is performed in data by comparing the measured and estimated numbers of same-charge events. The results are found to agree within uncertainties. Additional systematic uncertainties applied to the estimate include the statistical uncertainty from the data and the variation in the rates when the $Z$-peak range definition is varied.  The total systematic uncertainty in the charge misassignment background estimate is about 30\%, with the dominant contribution at low \pT from the closure tests and at high \pT from the statistical uncertainty.

\clearpage

\section{Systematic uncertainties}
\label{sec:syst}

The sources of systematic uncertainty considered in this analysis are summarized in Table~\ref{tab:SystSummary}.
They impact the estimated signal and background rates, the migration of events between categories and/or the shape of the BDT discriminants used in the final fit. Systematic uncertainties are implemented in the fit as normalization factors that affect the normalization of a process in a given analysis category or as a shape variation that only affects the distribution of a discriminant in a given category but not its normalization. The impact of all these systematic uncertainties on the measured signal strength is discussed quantitatively in Section~\ref{sec:result}.

The uncertainty in the combined 2015+2016 integrated luminosity is 2.1\%.
It is derived, following a methodology similar to that detailed in Ref.~\cite{DAPR-2013-01}, from a calibration of the luminosity scale using $x$--$y$ beam-separation scans performed in August 2015 and May 2016.

The experimental uncertainties are related to the reconstruction and identification of light leptons and hadronically decaying  $\tau$-leptons, to the reconstruction and $b$-tagging of jets, and to the reconstruction of \met. The sources that contribute to the uncertainty in the jet energy scale~\cite{PERF-2012-01,PERF-2011-04} are decomposed into uncorrelated components and treated as independent sources in the analysis. The total jet uncertainty varies from 1.0\% to 5.5\% depending on the jet \pT. The largest impact of experimental uncertainties on the signal strength $\mu = \sigma_{\ttH, \mathrm{obs}}/\sigma_{\ttH, \mathrm{SM}}$ arises from the jet energy scale, in particular contributions from the different responses to quark and gluon jets, pileup subtraction, and in situ calibration in data~\cite{PERF-2016-04}.

The uncertainties in the $b$-tagging efficiencies measured in dedicated calibration analyses~\cite{PERF-2012-04_corr} are also decomposed into uncorrelated components. The large number of components for $b$-tagging is due to the calibration of the distribution of the BDT discriminant. The approximate relative size of the $b$-tagging efficiency uncertainty is 2\% for $b$-jets, 10\% for $c$-jets and $\tau$s, and 30\% for light jets. The impact of the tagging uncertainty for jets containing either $c$-hadrons or \tauh is significant and, due to the calibration procedure applied, is taken as fully correlated between the two jet flavors.

Uncertainties in light-lepton reconstruction, identification, isolation and trigger efficiencies have negligible impact. The uncertainty in the identification efficiency for \tauh is ~6\%~\cite{ATLAS-CONF-2017-029}.

The systematic uncertainties associated with the estimation of the fake and non-prompt lepton backgrounds, as well as electron charge misassignment, are discussed in Section~\ref{sec:bkg}.
They have large effects on the background estimates in all channels.

The systematic uncertainties associated with the generation of signal and background processes are due to uncertainties in the assumed cross sections and acceptance modeling for each process, and they are assessed in each category. The former are evaluated by varying the cross section of each process within its uncertainty, as described in Section~\ref{sec:mc}.
The latter are estimated by comparing the results with those obtained using alternative simulated samples detailed in Section~\ref{sec:mc}.
The most important uncertainty arising from theoretical predictions is in the assumed SM cross sections and the modeling of the acceptance for \ttH, \ttZ and \ttW production. The uncertainty in the shape of the simulated \ttW and \ttZ backgrounds due to the choice of event generator varies by at most 10\% between bins. The uncertainties for $\ttbar\gamma$, $t Z$, $t W Z$, and $VV (\to \ell \ell XX)$ include extrapolation uncertainties into the analysis phase space.

\begin{table}
\centering
\caption{Sources of systematic uncertainty considered in the analysis.
``N'' means that the uncertainty is taken as normalization-only for all processes and channels affected,
whereas ``S'' denotes uncertainties that are considered shape-only in all processes and channels.
``SN'' means that the uncertainty applies to both shape and normalization.
Some of the systematic uncertainties are split into several components, as indicated by the number in the rightmost column.}
\begin{tabular}{lcr}
\hline\hline
Systematic uncertainty                          & Type  & Components  	      \\
\hline
\hline
Luminosity                                      &  N	& 1		      \\
Pileup reweighting                             &  SN	& 1		      \\
\textbf {Physics Objects}                       &	&		      \\
\ \ Electron                                        & SN	& 6		      \\
\ \ Muon                                            & SN	& 15		      \\
\ \ \tauh                                             & SN	& 10		      \\
\ \ Jet energy scale and resolution                 & SN	& 28                  \\
\ \ Jet vertex fraction                             & SN	& 1		      \\
\ \ Jet flavor tagging                             & SN	& 126		      \\
\ \ \met                                            & SN	& 3		      \\
\hline
Total (Experimental)                            & --    & 191		     \\
\hline
\hline
\textbf {Data-driven non-prompt/fake leptons and charge misassignment}                              &	&		     \\
\ \ Control region statistics     			& SN	& 38    		 \\
\ \ Light-lepton efficiencies & SN    & 22			 \\
\ \ Non-prompt light-lepton estimates: non-closure  &  N    & 5			 \\
\ \ $\gamma$-conversion fraction                    &  N    & 5          		 \\
\ \ Fake \tauh estimates                            &  N/SN & 12          		 \\
\ \ Electron charge misassignment                  & SN	& 1		         \\
\hline
Total (Data-driven reducible background)        & --    & 83		         \\
\hline
\hline
\textbf {\ttH modeling}                        &	&		      \\
\ \ Cross section                        &  N	& 2		      \\
\ \ Renormalization and factorization scales        &  S	& 3		      \\
\ \ Parton shower and hadronization model           &  SN	& 1		      \\
\ \ Higgs boson branching fraction                           &  N	& 4		      \\
\ \ Shower tune				 	&  SN	& 1		      \\
\textbf {\ttW modeling}                        &	&		      \\
\ \ Cross section                       &  N	& 2		      \\
\ \ Renormalization and factorization scales                         &  S	& 3		      \\
\ \ Matrix-element MC event generator                     &  SN	& 1		      \\
\ \ Shower tune				 	&  SN	& 1		      \\
\textbf {\ttZ modeling}                        &	&		      \\
\ \ Cross section                        &  N	& 2		     \\
\ \ Renormalization and factorization scales                                       &  S	& 3		      \\
\ \ Matrix-element MC event generator                    &  SN	& 1		      \\
\ \ Shower tune				 	&  SN	& 1		      \\
\textbf {Other background modeling}            &	&		      \\
\ \ Cross section                                   &  N	& 15		      \\
\ \ Shower tune       				&  SN	& 1		      \\
\hline
Total (Signal and background modeling)         & --    & 41		     \\
\hline\hline
Total (Overall)                                 & --    & 315	      \\
\hline\hline
\end{tabular}
\label{tab:SystSummary}
\end{table}

\section{Statistical model and results}
\label{sec:result}

Table~\ref{tbl:yields} (top part) shows a comparison of the predicted yields to data in the eight signal and four control regions defined in Section~\ref{sec:event}.

A maximum-likelihood fit is performed on all these twelve categories simultaneously to extract the \ttH signal cross section normalized to the prediction from the SM ($\mu$) with the signal acceptance in the different regions derived assuming the SM. The statistical analysis of the data uses a binned likelihood function $\mathcal{L}(\mu, \vec{\theta})$, which is constructed from a product of Poisson probability terms to estimate $\mu$. The Higgs boson branching fractions and the cross section for associated production of a Higgs boson and a single top quark, which is treated as background, are set to their SM expectations with appropriate theoretical uncertainties. As mentioned in Section~\ref{sec:event} and summarized in Table~\ref{tbl:summary_channOverview}, a BDT shape is used as the final discriminant in five of the eight signal regions. The exceptions are the \fourl $Z$-enriched (defined after placing a requirement on a BDT discriminant), the \fourl $Z$-depleted and the \llltau categories, which use a single bin because there are few events. A single bin is also used in the four control regions from the \lll channel. The total number of bins used in the fit is 32 and the details of each category are presented in Table~\ref{tbl:summary_channOverview}.

The impact of systematic uncertainties on the signal and background expectations is described by nuisance parameters (NPs), $\vec{\theta}$, which are constrained by Gaussian or log-normal probability density functions. The latter are used for normalization factors to ensure that they are always positive. The expected numbers of signal and background events are functions of $\vec{\theta}$. The prior for each NP is added as a penalty term to the likelihood, $\mathcal{L}(\mu, \vec{\theta})$, to decrease it when $\theta$ is shifted away from its nominal value. The statistical uncertainties in the simulated background predictions and the control regions used for the non-prompt and fake estimates are included as bin-by-bin NPs using the Beeston--Barlow technique~\cite{Barlow:1993dm}.

The test statistic, $q_\mu$, is constructed from the profile log-likelihood ratio: \newline
$q_\mu = -2 \ln{\Lambda_\mu} = -2 \ln{\mathcal{L}(\mu,\hat{\hat{\vec{\theta}}})/\mathcal{L}(\hat{\mu},\hat{\vec{\theta}})}$, where $\hat{\mu}$ and $\hat{\vec{\theta}}$ are the parameters that maximize the likelihood and $\hat{\hat{\vec{\theta}}}$ are the NPs that maximize the likelihood for a given $\mu$. The test statistic is used to quantify how well the observed data agrees with the background-only hypothesis.

The fitted $\hat{\mu}$ value is obtained by maximizing the likelihood function with respect to all parameters and the total uncertainty, $\sigma_\mu$, is obtained from the variation of $-2 \ln{\Lambda_\mu}$ by one unit from its minimum.
Systematic uncertainties are found by subtracting in quadrature the statistical uncertainty, determined by fixing all NPs to their best-fit values, from the total uncertainty.
The expected results are obtained in the same way as the observed results by replacing the data in each input bin by the prediction from simulation and the data-driven fake and non-prompt estimates with all NPs set to their best-fit values obtained from the fit to data.  The significance is obtained from the test statistic in the asymptotic limit~\cite{asym}. As the \llll channel has few events, the validity of this assumption was verified using pseudo-experiments.

\begin{landscape}
\begin{table}[htbp]
\begin{center}
\caption{\label{tbl:yields} Background, signal and observed yields in the twelve analysis categories in 36.1 \ifb\ of data at $\sqrt{s} =$ 13~TeV. Uncertainties in the background estimates due to systematic effects and to limited simulation sample size are shown.  ``Non-prompt'', ``Fake \tauh'' and ``$q$ mis-id'' refer to the data-driven background estimates described in Section~\ref{sec:bkg}. Rare processes ($tZ$, $tW$, $tWZ$, $\ttbar WW$, triboson production, $t\bar t t$, $t\bar t t \bar t$, $tH$, rare top decay) are labeled as ``Other''. In the top part, the pre-fit values are quoted, i.e. using the initial values of background systematic uncertainty nuisance parameters and the signal expected from the SM. In the bottom part, the corresponding post-fit values are quoted. In the post-fit case, the prediction and uncertainties for \ttH reflect the best-fit production rate of 1.6~$^{+0.5}_{-0.4}$ times the Standard Model prediction and the uncertainty in the total background estimate is smaller than for the pre-fit values due to anticorrelations between the nuisance parameters obtained in the fit.}
 \resizebox{1.3\textwidth}{!}{
 \begin{tabular}{lcccccccccr}
\hline\hline
Category         & Non-prompt	    & Fake \tauh      &  $q$ mis-id     & \ttW	         & \ttZ		   & Diboson	     & Other         & Total Bkgd.	 & \ttH 	   & Observed	\\
\hline
\multicolumn{11}{c}{Pre-fit yields} \\
\hline
 \ll             & 233\phantom{.}  $\pm$ 39\phantom{0.} &	 --		                  & 33\phantom{0.} $\pm$ 11\phantom{0.} & 123\phantom{.} $\pm$ 18\phantom{0.}  & 41.4\phantom{} $\pm$ 5.6\phantom{0}     & 25\phantom{0.} $\pm$ 15\phantom{0.}                     & 28.4\phantom{} $\pm$ 5.9\phantom{0}     & 484\phantom{.} $\pm$ 38\phantom{0.} & 42.6\phantom{} $\pm$ 4.2\phantom{0}  & 514    \\
 \lll SR         & 14.5\phantom{}  $\pm$ 4.3\phantom{0} &	--		                  &     --	 		        &  5.5\phantom{0} $\pm$ 1.2\phantom{0} & 12.0\phantom{} $\pm$ 1.8\phantom{0}     & 1.2\phantom{0} $\pm$ 1.2\phantom{0}                     & 5.8\phantom{0} $\pm$ 1.4\phantom{0}     & 39.1\phantom{} $\pm$ 5.2\phantom{0} & 11.2\phantom{} $\pm$ 1.6\phantom{0}  &  61    \\
 \lll \ttW~CR    & 13.3\phantom{}  $\pm$ 4.3\phantom{0} &	--		                  &     --	                        & 19.9\phantom{} $\pm$ 3.1\phantom{0}  &  8.7\phantom{0} $\pm$ 1.1\phantom{0}    &   $<0.2$                                                 & 4.53\phantom{} $\pm$ 0.92\phantom{}     & 46.5\phantom{} $\pm$ 5.4\phantom{0} & 4.18\phantom{} $\pm$ 0.46\phantom{}  &  56    \\
 \lll \ttZ~CR    &  3.9\phantom{0} $\pm$ 2.5\phantom{0} &	--		                  &     --	                        &  2.71\phantom{} $\pm$ 0.56\phantom{} &   66\phantom{0.} $\pm$ 11\phantom{0.}   & 8.4\phantom{0} $\pm$ 5.3\phantom{0}                     & 12.9\phantom{} $\pm$ 4.2\phantom{0}     &  93\phantom{0.} $\pm$ 13\phantom{0.}& 3.17\phantom{} $\pm$ 0.41\phantom{}  & 107    \\
 \lll VV CR      & 27.7\phantom{}  $\pm$ 8.7\phantom{0} &	--		                  &     --	                        & 4.9\phantom{0} $\pm$ 1.0\phantom{0}  & 21.3\phantom{} $\pm$ 3.4\phantom{0}     & 51\phantom{0.}  $\pm$ 30\phantom{0.}                    & 17.9\phantom{} $\pm$ 6.1\phantom{0}     & 123\phantom{.} $\pm$ 32\phantom{0.} & 1.67\phantom{} $\pm$ 0.25\phantom{}  & 109    \\
 \lll \ttbar CR  & 70\phantom{.0}  $\pm$ 17\phantom{0.} &	--		                  &     --	                        & 10.5\phantom{} $\pm$ 1.5\phantom{0}  &  7.9\phantom{0} $\pm$ 1.1\phantom{0}    & 7.2\phantom{0} $\pm$ 4.8\phantom{0}                     & 7.3\phantom{0} $\pm$ 1.9\phantom{0}     & 103\phantom{.} $\pm$ 17\phantom{0.} & 4.00\phantom{} $\pm$ 0.49\phantom{}  &  85    \\
 \fourl $Z$-enr. & 0.11\phantom{}  $\pm$ 0.07\phantom{} &	 --		                  &     --	                        &  $<0.01$	                       & 1.52\phantom{} $\pm$ 0.23\phantom{}   & 0.43\phantom{} $\pm$ 0.23\phantom{}                     & 0.21\phantom{} $\pm$ 0.09\phantom{}     & 2.26\phantom{} $\pm$ 0.34\phantom{} & 1.06\phantom{} $\pm$ 0.14\phantom{}  &   2    \\
 \fourl $Z$-dep. & 0.01\phantom{}  $\pm$ 0.01\phantom{} &	 --		                  &     --	                        &  $<0.01$	                       & 0.04\phantom{} $\pm$ 0.02\phantom{}   &   $<0.01$                                               & 0.06\phantom{} $\pm$ 0.03\phantom{}     & 0.11\phantom{} $\pm$ 0.03\phantom{} & 0.20\phantom{} $\pm$ 0.03\phantom{}  &   0    \\
 \ltwotau        &      --                              & 65\phantom{0.} $\pm$ 21\phantom{0.}     &     --  	                        & 0.09\phantom{} $\pm$ 0.09\phantom{}  & 3.3\phantom{0} $\pm$ 1.0\phantom{0}    & 1.3\phantom{0} $\pm$ 1.0\phantom{0}                     & 0.98\phantom{} $\pm$ 0.35\phantom{}     & 71\phantom{0.} $\pm$ 21\phantom{0.}	& 4.3\phantom{0} $\pm$ 1.0\phantom{0}  &  67    \\
 \lltau          & 2.4\phantom{0}  $\pm$ 1.4\phantom{0} & 1.80\phantom{} $\pm$ 0.30\phantom{}     & 0.05\phantom{} $\pm$ 0.02\phantom{} & 0.88\phantom{} $\pm$ 0.24\phantom{}  & 1.83\phantom{} $\pm$ 0.37\phantom{}       & 0.12\phantom{} $\pm$ 0.18\phantom{}                     & 1.06\phantom{} $\pm$ 0.24\phantom{}     & 8.2\phantom{0} $\pm$ 1.6\phantom{0} & 3.09\phantom{} $\pm$ 0.46\phantom{}  &  18    \\
 \OSlltau        &      --                              & 756\phantom{.} $\pm$ 80\phantom{0.}     &     --  	                        & 6.5\phantom{0} $\pm$ 1.3\phantom{0}  & 11.4\phantom{} $\pm$ 1.9\phantom{0}     & 2.0\phantom{0} $\pm$ 1.3\phantom{0}                     & 5.8\phantom{0} $\pm$ 1.5\phantom{0}     & 782\phantom{.} $\pm$ 81\phantom{0.}	& 14.2\phantom{} $\pm$ 2.0\phantom{0}  & 807    \\
 \llltau         &      --                              & 0.75\phantom{} $\pm$ 0.15\phantom{}     &     --  	                        & 0.04\phantom{} $\pm$ 0.04\phantom{}  & 1.38\phantom{} $\pm$ 0.24\phantom{}  & 0.002 $\pm$ 0.002                                       & 0.38\phantom{} $\pm$ 0.10\phantom{}     & 2.55\phantom{} $\pm$ 0.32\phantom{} & 1.51\phantom{} $\pm$ 0.23\phantom{}  &   5    \\
\hline
\multicolumn{11}{c}{Post-fit yields} \\
\hline
 \ll             & 211\phantom{.} $\pm$ 26\phantom{0.}  &	  --	                          & 28.3\phantom{} $\pm$ 9.4\phantom{0} & 127\phantom{.} $\pm$ 18\phantom{0.}  & 42.9\phantom{} $\pm$ 5.4\phantom{0}    & 20.0\phantom{} $\pm$ 6.3\phantom{0}                     & 28.5\phantom{} $\pm$ 5.7\phantom{0}     & 459\phantom{0.} $\pm$ 24\phantom{0.} & 67\phantom{0.} $\pm$ 18\phantom{0.}  & 514    \\
 \lll SR         & 13.2\phantom{} $\pm$ 3.1\phantom{0}  &	 --	                          &	  --	                        &  5.8\phantom{0} $\pm$ 1.2\phantom{0} & 12.9\phantom{} $\pm$ 1.6\phantom{0}& 1.2\phantom{0} $\pm$ 1.1\phantom{0}                     & 5.9\phantom{0} $\pm$ 1.3\phantom{0}     & 39.0\phantom{0} $\pm$ 4.0\phantom{0} & 17.7\phantom{} $\pm$ 4.9\phantom{0}  &  61    \\
 \lll \ttW~CR    & 11.7\phantom{} $\pm$ 3.0\phantom{0}  &	 --	                          &	  --	                        & 20.4\phantom{} $\pm$ 3.0\phantom{0}  &  8.9\phantom{0} $\pm$ 1.0\phantom{0}  &    $<0.2$	                                        & 4.54\phantom{} $\pm$ 0.88\phantom{}     & 45.6\phantom{0} $\pm$ 4.0\phantom{0} & 6.6\phantom{0} $\pm$ 1.9\phantom{0}  &  56    \\
 \lll \ttZ~CR    & 3.5\phantom{0} $\pm$ 2.1\phantom{0}  &	 --	                          &	  --	                        &  2.82\phantom{} $\pm$ 0.56\phantom{} &  70.4\phantom{} $\pm$ 8.6\phantom{0}  & 7.1\phantom{0} $\pm$ 3.0\phantom{0}                  & 13.6\phantom{} $\pm$ 4.2\phantom{0}     &  97.4\phantom{0} $\pm$ 8.6\phantom{0}& 5.1\phantom{0} $\pm$ 1.4\phantom{0}  & 107    \\
 \lll VV CR      & 22.4\phantom{} $\pm$ 5.7\phantom{0}  &	 --	                          &	  --	                        & 5.05\phantom{} $\pm$ 0.94\phantom{}  & 22.0\phantom{} $\pm$ 3.0\phantom{0}  & 39\phantom{0.}  $\pm$ 11\phantom{0.}                  & 18.1\phantom{} $\pm$ 5.9\phantom{0}     & 106.8\phantom{} $\pm$ 9.4\phantom{0} & 2.61\phantom{} $\pm$ 0.82\phantom{}  & 109    \\
 \lll \ttbar CR  & 56.0\phantom{} $\pm$ 8.1\phantom{0}  &	 --	                          &	  --	                        & 10.7\phantom{} $\pm$ 1.4\phantom{0}  &  8.1\phantom{0} $\pm$ 1.0\phantom{0}  & 5.9\phantom{0} $\pm$ 2.7\phantom{0}                  & 7.1\phantom{0} $\pm$ 1.8\phantom{0}     & 87.8\phantom{0} $\pm$ 7.9\phantom{0} & 6.3\phantom{0} $\pm$ 1.8\phantom{0}  &  85    \\
 \fourl $Z$-enr. & 0.10\phantom{} $\pm$ 0.07\phantom{}  &      --  	                          &     --  	                        &   $<0.01$	                       & 1.60\phantom{} $\pm$ 0.22\phantom{} & 0.37\phantom{} $\pm$ 0.15\phantom{}                     & 0.22\phantom{} $\pm$ 0.10\phantom{}     & 2.29\phantom{0} $\pm$ 0.28\phantom{} & 1.65\phantom{} $\pm$ 0.47\phantom{}  &   2    \\
 \fourl $Z$-dep. & 0.01\phantom{} $\pm$ 0.01\phantom{}  &      --  	                          &     --  	                        &   $<0.01$	                       & 0.04\phantom{} $\pm$ 0.02\phantom{} &	$<0.01$                                                 & 0.07\phantom{} $\pm$ 0.03\phantom{}     & 0.11\phantom{0} $\pm$ 0.03\phantom{} & 0.32\phantom{} $\pm$ 0.09\phantom{}  &   0    \\
 \ltwotau        &      --                              & 58.0\phantom{} $\pm$ 6.8\phantom{0}     &     --  	                        & 0.11\phantom{} $\pm$ 0.11\phantom{}  & 3.31\phantom{} $\pm$ 0.90\phantom{}    & 0.98\phantom{} $\pm$ 0.75\phantom{}                     & 0.98\phantom{} $\pm$ 0.33\phantom{}     & 63.4\phantom{0} $\pm$ 6.7\phantom{0} &  6.5\phantom{0} $\pm$ 2.0\phantom{0} &  67    \\
 \lltau          & 1.86\phantom{} $\pm$ 0.91\phantom{}  & 1.86\phantom{} $\pm$ 0.27\phantom{}     & 0.05\phantom{} $\pm$ 0.02\phantom{} & 0.97\phantom{} $\pm$ 0.26\phantom{}  & 1.96\phantom{} $\pm$ 0.37\phantom{}       & 0.15\phantom{} $\pm$ 0.20\phantom{}                     & 1.09\phantom{} $\pm$ 0.24\phantom{}     & 7.9\phantom{00} $\pm$ 1.2\phantom{0} & 5.1\phantom{0} $\pm$ 1.3\phantom{0}  &  18    \\
 \OSlltau        &      --                              & 756\phantom{.} $\pm$ 28\phantom{0.}     &     --  	                        & 6.6\phantom{0} $\pm$ 1.3\phantom{0}  & 11.5\phantom{} $\pm$ 1.7\phantom{0}     & 1.64\phantom{} $\pm$ 0.92\phantom{}                     & 6.1\phantom{0} $\pm$ 1.5\phantom{0}     & 782\phantom{0.} $\pm$ 27\phantom{0.}	& 21.7\phantom{} $\pm$ 5.9\phantom{0}  & 807    \\
 \llltau         &      --                              & 0.75\phantom{} $\pm$ 0.14\phantom{}     &     --  	                        & 0.04\phantom{} $\pm$ 0.04\phantom{}  & 1.42\phantom{} $\pm$ 0.22\phantom{}   & 0.002 $\pm$ 0.002                                       & 0.40\phantom{} $\pm$ 0.10\phantom{}     & 2.61\phantom{0} $\pm$ 0.30\phantom{} & 2.41\phantom{} $\pm$ 0.68\phantom{}  &   5    \\
\hline\hline
 \end{tabular}
 }
\end{center}
\end{table}
\end{landscape}

\begin{table}[h!]
\begin{center}
\caption{\label{tbl:summary_channOverview} Summary of the basic characteristics and analysis strategies of all channels. In the \fourl channel, the two entries correspond to the $Z$-enriched and the $Z$-depleted categories; 1D and 5D refer to one- and five-dimensional BDTs respectively.}
\resizebox{\textwidth}{!}{
\begin{tabular}{l|ccccccc}
\hline\hline
                               & \ll      & \lll    & \llll    & \ltwotau    & \lltau     & \OSlltau  & \llltau    \\
\hline
BDT trained against            & Fakes and \ttV & \ttbar, \ttW, \ttZ, VV & \ttZ~/ - & \ttbar & all & \ttbar & - \\
Discriminant                   & 2$\times$1D BDT    & 5D BDT    & Event count    &  BDT    &   BDT    &   BDT    & Event count \\
Number of bins                 & 6 & 5 & 1 / 1 & 2 & 2 & 10& 1\\
Control regions                & - & 4 &  -  & - & - & - & - \\
\hline\hline
\end{tabular}
}
\end{center}
\end{table}

As described in Section~\ref{sec:syst}, a large number of systematic uncertainties, whose effects are accounted for using NPs, affect the final results. In total 315~NPs are considered, most having experimental origin. The experimental uncertainties are fully correlated across categories, with the exception of those related to the quark/gluon jet composition and some uncertainties associated with the fake and non-prompt lepton background determinations, which are specific to the different categories, as detailed in Section~\ref{sec:bkg}.
As the residual prompt (mainly \ttW and $VV$) background contribution is subtracted from the control regions to extract the fake and non-prompt leptons, the associated nuisance parameters are taken as fully correlated with the theoretical cross-section systematic uncertainties. The same treatment is used for the uncertainty associated to the measurement of the background from charge misassignment, which is also subtracted from the control regions.

The fit uses templates constructed from the predicted yields for the signal and the various backgrounds in the bins of the input distribution in each region. The systematic uncertainties are encoded in templates of variations relative to the nominal template for each upward or downward ($\pm \sigma$) variation. A smoothing procedure is applied to remove large local fluctuations in the templates for some background processes in certain regions. Systematic uncertainties that have a negligible impact on the final results are removed to improve the speed of the fit: a normalization or a shape uncertainty is not applied if the associated variation is below 1\% in all bins; this reduces the number of nuisance parameters to 230. Most of the neglected nuisance parameters are those related to flavor tagging.

The behavior of the global fit is studied by performing a number of checks including evaluating how much each NP is pulled from its nominal value, how much its uncertainty decreases from the nominal uncertainty and which correlations develop between initially uncorrelated systematic uncertainties. The stability of the results was tested by performing fits for each channel independently and in combination.

The impact of each systematic uncertainty on the final result is assessed by performing the fit with the parameter fixed to its fitted value varied up or down by its fitted uncertainty, with all the other parameters allowed to vary and calculating the $\Delta\mu$ to the baseline fit. The ranking obtained for those nuisance parameters with the largest contribution to the uncertainty in the signal strength is shown in Figure~\ref{fig:ranking}.
The NP with the largest pull from its nominal value is the uncertainty in the non-prompt lepton estimate due to the non-closure in the \lll channel. This is mainly due to the slight deficit observed in the \lll \ttbar control region relative to the background prediction. As the fit includes bins with high purity of non-prompt light leptons and fake \tauh backgrounds, the precision of these estimates is increased, as is shown in Table~\ref{tbl:yields}.
The correlations between the nuisance parameters were checked and no unexpected correlations were observed.
The impact of the most important groups of systematic uncertainties on the measured value of $\mu$ is shown in Table~\ref{tab:systs}. The uncertainties with the largest impact are those associated with the signal modeling, the jet energy scale and the non-prompt light-lepton estimate. The signal uncertainty is separated into two components to show the uncertainty due to the acceptance and the one due to the cross section. The uncertainties in the non-prompt light-lepton estimates, the fake \tauh estimates and the charge misassignment have large statistical components due to the small data sample size. The large impact of the luminosity uncertainty is due to its effect on both the signal and simulated background predictions. Although the individual groups are initially largely uncorrelated, a small correlation is introduced by the fit to data.

\begin{figure}[tbp]
\centering
\includegraphics[width=0.8\textwidth]{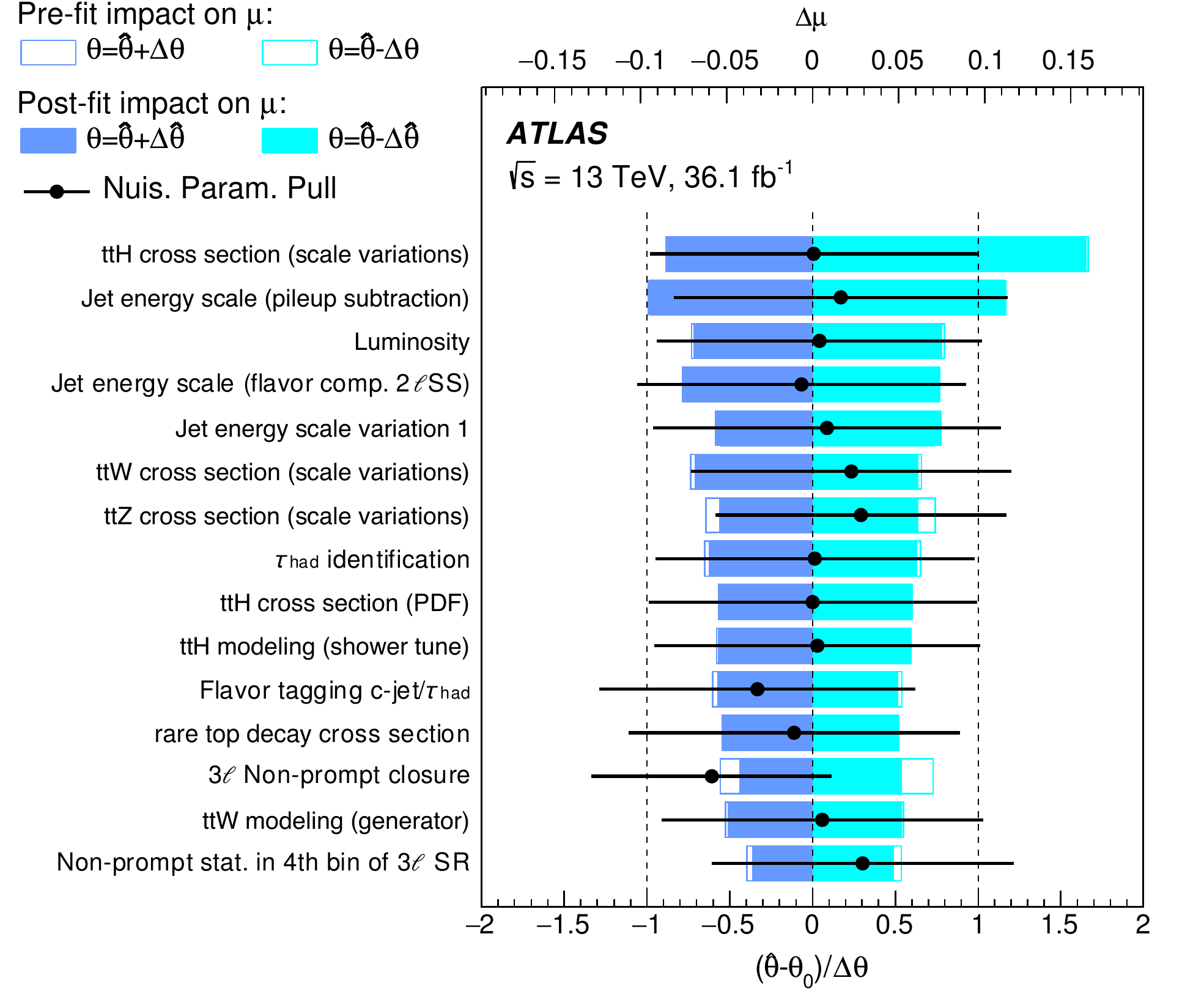}
  \caption{The impact of systematic uncertainties on the fitted signal-strength parameter \muhat~for the combined fit of all channels. The systematic uncertainties are listed in decreasing order of their impact on \muhat~on the $y$-axis, and only the 15 most important ones are displayed. The filled blue boxes show the variations of \muhat~from the central value, $\Delta\mu$, referring to the upper $x$-axis, when fixing the corresponding individual nuisance parameter, $\theta$, to its post-fit value $\hat{\theta}$ modified upwards or downwards by its post-fit uncertainty, and repeating the fit. The empty blue boxes represent the corresponding pre-fit impact. The black points, which refer to the lower $x$-axis, show the fitted values and uncertainties of the nuisance parameters, relative to their pre-fit values, $\theta_0$, and uncertainties, $\Delta\theta$. The black lines show the post-fit uncertainties of the nuisance parameters, relative to their nominal uncertainties, which are indicated by the dashed line.\label{fig:ranking}}
\end{figure}

\begin{table}[htbp]
\centering
\caption{\label{tab:systs} Summary of the effects of the most important groups of systematic uncertainties on $\mu$. Due to rounding effects and small correlations between the different sources of uncertainty, the total systematic uncertainty is different from the sum in quadrature of the individual sources.}
\begin{tabular}{lcc}
\hline\hline
Uncertainty Source                       &   \multicolumn{2}{c}{$\Delta\mu$} \\
\hline
\ttH modeling (cross section)              &   ${+0.20}$ & ${-0.09}$    \\
Jet energy scale and resolution          &   ${+0.18}$ & ${-0.15}$    \\
Non-prompt light-lepton estimates       &   ${+0.15}$ & ${-0.13}$    \\
Jet flavor tagging and \tauh identification &   ${+0.11}$ & ${-0.09}$    \\
\ttW modeling                            &   ${+0.10}$ & ${-0.09}$    \\
\ttZ modeling                            &   ${+0.08}$ & ${-0.07}$    \\
Other background modeling                &   ${+0.08}$ & ${-0.07}$    \\
Luminosity                               &   ${+0.08}$ & ${-0.06}$    \\
\ttH modeling (acceptance)                 &   ${+0.08}$ & ${-0.04}$    \\
Fake \tauh estimates                     &   ${+0.07}$ & ${-0.07}$    \\
Other experimental uncertainties         &   ${+0.05}$ & ${-0.04}$    \\
Simulation sample size           &   ${+0.04}$ & ${-0.04}$    \\
Charge misassignment                     &   ${+0.01}$ & ${-0.01}$    \\
\hline
Total systematic uncertainty             &   ${+0.39}$ & ${-0.30}$   \\
\hline\hline
\end{tabular}
\end{table}

Figure~\ref{fig:summary_postfit} and Table~\ref{tbl:yields} (bottom part) compare the data to the yields after the predictions were adjusted by the fit in the twelve signal and control regions.
Figures~\ref{fig:NonTauPlots} and~\ref{fig:TauPlots} show the distributions of the discriminating variables used by the fit in the eight signal regions. Distributions are shown both before and after the fit to the data. An excess of events over the expected Standard Model background is found with an observed (expected) significance of 4.1 (2.8) standard deviations. The observed (expected) best-fit value of $\mu$ is 1.6~$^{+0.3}_{-0.3}$ (stat.) $^{+0.4}_{-0.3}$ (syst.) $=$1.6~$^{+0.5}_{-0.4}$ (1.00~$^{+0.3}_{-0.3}$ (stat.) $^{+0.3}_{-0.3}$ (syst.) $=$1.00~$^{+0.4}_{-0.4}$). The best-fit value of $\mu$ for each individual channel and the combination of all channels are shown in Figure~\ref{fig:mu} and Table~\ref{tab:mu}. The individual channel results are extracted from the full fit but with a separate parameter of interest for each channel. The probability that the fitted signal strengths in the seven channels are compatible is 34\%. When assuming that the observed signal is due to the SM Higgs boson, the excess over the SM signal-plus-background hypothesis has a significance of 1.4$\sigma$. A model-dependent extrapolation is made to the inclusive phase space, and the measured \ttH production cross section is $\sigma(\ttH) = 790^{+150}_{-150} \mathrm{(stat.)}^{+170}_{-150}\mathrm{(syst.)}$~fb $ = 790^{+230}_{-210}$~fb. The predicted cross section is $\sigma(\ttH) = 507^{+35}_{-50}$~fb.

For the \fourl, \OSlltau and \llltau channels, the uncertainties in $\mu$ are mainly statistical, while the statistical and systematic uncertainties are of comparable size for the \ll, \lll, \lltau and \ltwotau channels.
Figure~\ref{fig:SoverB} shows the data, background and signal yields, where the final-discriminant bins in all signal regions are combined into bins of log($S/B$), $S$ being the expected signal yield and $B$ the fitted background yield.

The most sensitive \ll, \threel and \lltau analyses were cross-checked with simpler cut-and-count analyses with reduced sensitivity. The observed significance relative to the background-only hypothesis is 1.2$\sigma$, 2.3$\sigma$ and 2.3$\sigma$, respectively. The observed signal strengths in the cross-check analyses are found to be statistically compatible with those from the nominal analyses.

An alternative fit where \ttW and \ttZ normalizations were left free together with $\mu$ was performed as a cross-check. The expected sensitivity to $\mu$ is 15\% worse than with the nominal fit. The observed best-fit value of $\mu$ is 1.6~$^{+0.6}_{-0.5}$, in agreement with the result obtained with the nominal fit. The fitted \ttW and \ttZ cross-section modifiers are $0.92\pm0.32$ and 1.17~$^{+0.25}_{-0.22}$, respectively, in agreement with the SM predictions.

\begin{figure}[!htbp]
\centering
\includegraphics[width=0.8\textwidth]{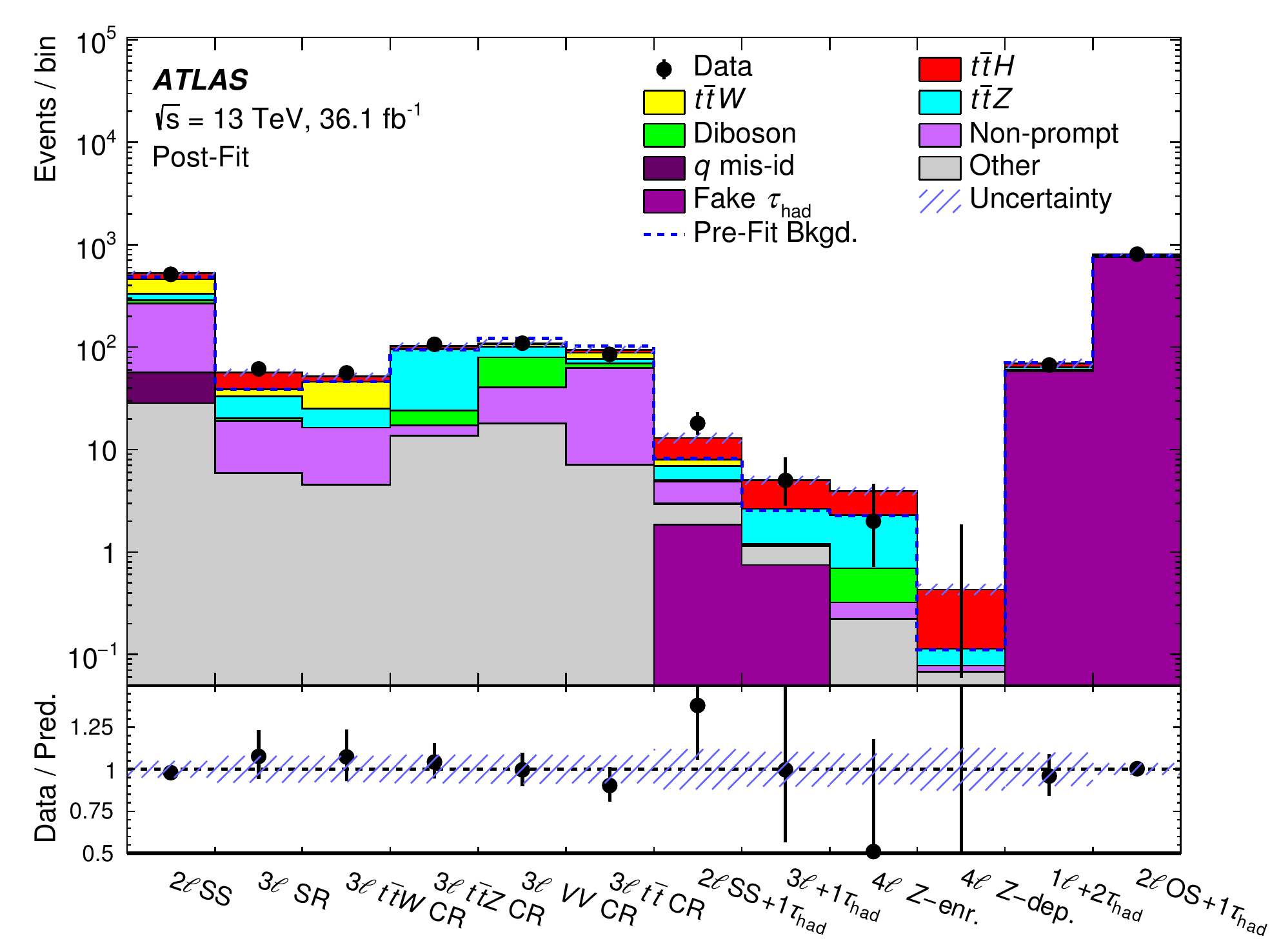}
  \caption{Comparison of prediction to data after the fit in the eight signal and four control regions.
The background contributions after the global fit are shown as filled histograms.
The total background before the fit is shown as a dashed blue histogram.
The Higgs boson signal (\mH = 125~GeV), scaled according to the results of the fit, is shown as a filled red histogram superimposed on the fitted backgrounds.
The size of the combined statistical and systematic uncertainty in the sum of the signal
and fitted background is indicated by the blue hatched band. The ratio of the data to the sum of the
signal and fitted background is shown in the lower panel. The yields in each region are shown in Table~\ref{tbl:yields}.\label{fig:summary_postfit}}
\end{figure}

\begin{figure}[!htbp]
\centering
\subfigure[]{\includegraphics[width=0.47\textwidth]{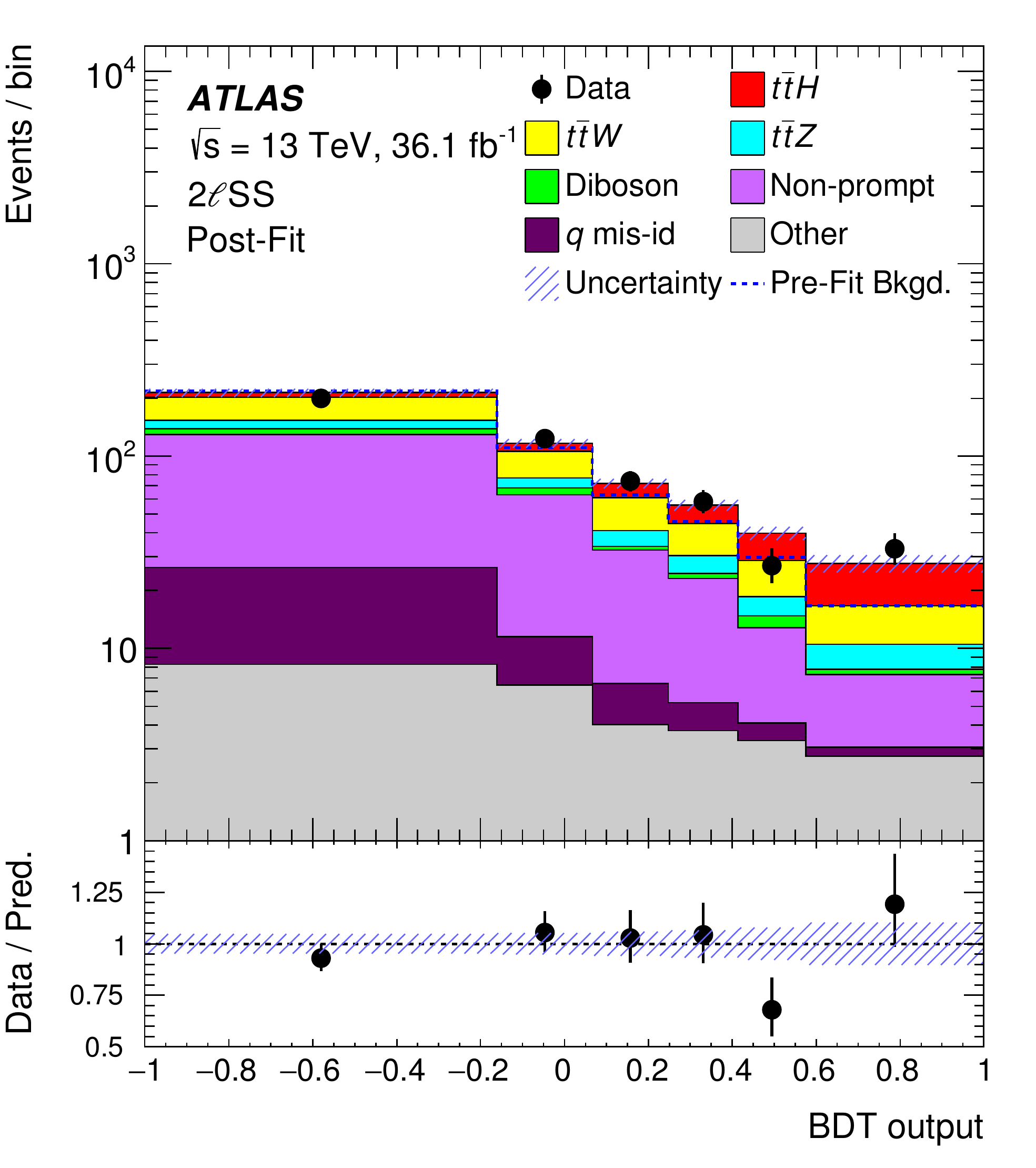}}
\subfigure[]{\includegraphics[width=0.47\textwidth]{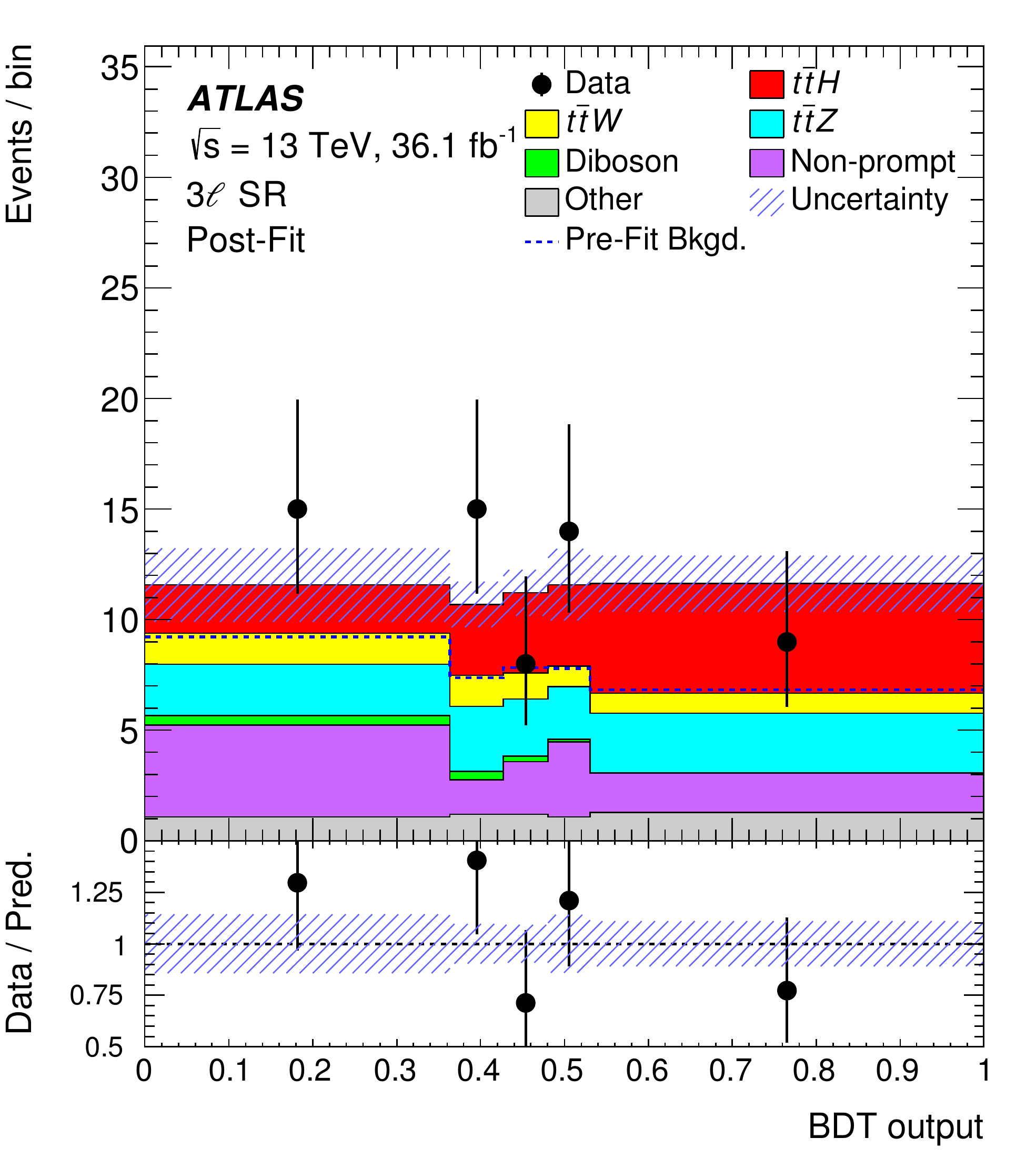}}
\subfigure[]{\includegraphics[width=0.47\textwidth]{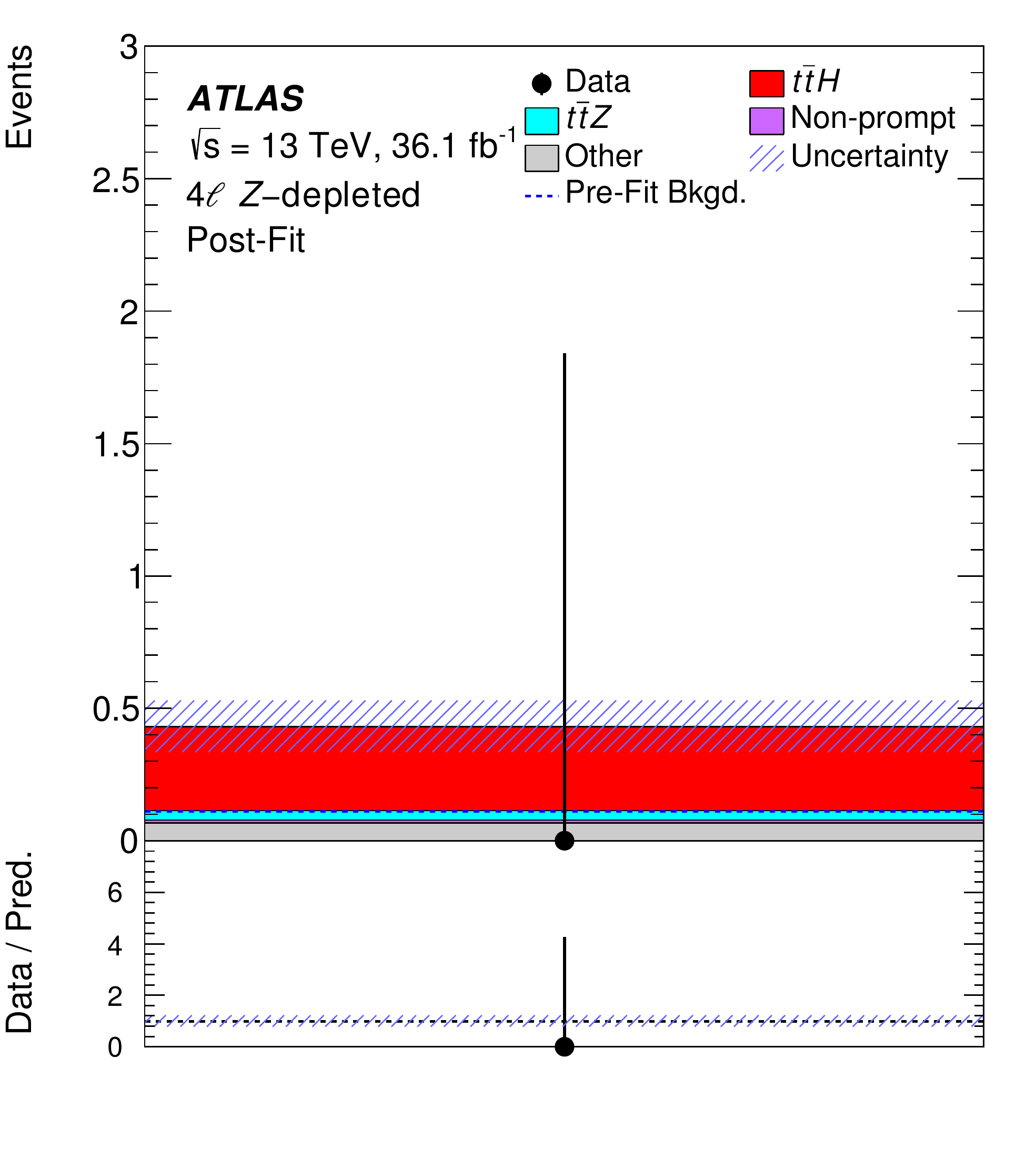}}
\subfigure[]{\includegraphics[width=0.47\textwidth]{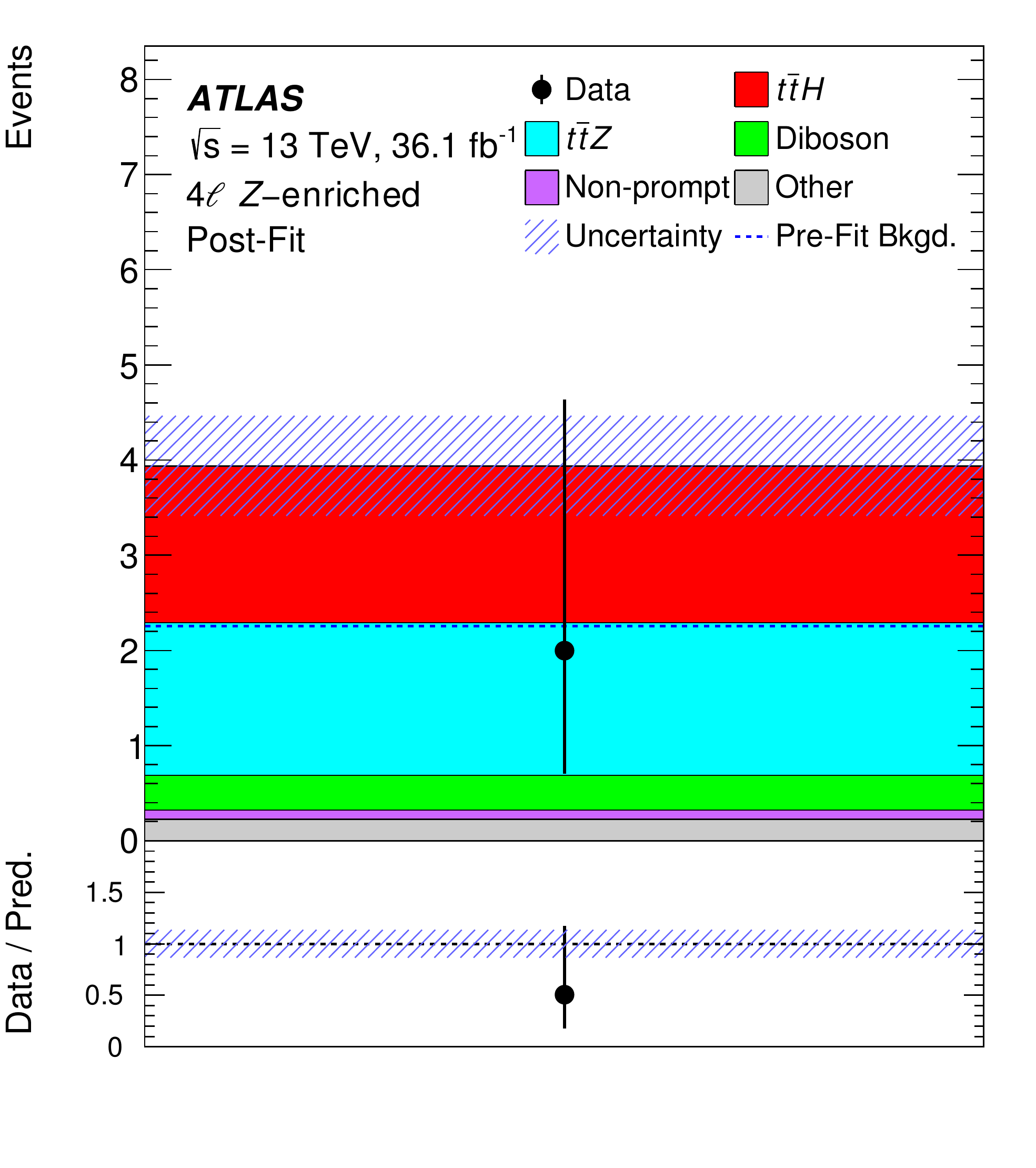}}
  \caption{The distribution of the discriminating variables observed in data (points with bars indicating the statistical errors) and expected
(histograms) in the (a) \ll,  (b) \lll, (c) \llll ($Z$-enriched) and  (d) \llll ($Z$-depleted) signal regions.
The background contributions after the global fit are shown as filled histograms.
The total background before the fit is shown as a dashed blue histogram.
The Higgs boson signal (\mH = 125 GeV), scaled according to the results of the fit, is shown as a filled red histogram superimposed on the fitted backgrounds.
The size of the combined statistical and systematic uncertainty in the sum of the signal
and fitted background is indicated by the blue hatched band. The ratio of the data to the sum of the
signal and fitted background is shown in the lower panel.\label{fig:NonTauPlots}}
\end{figure}

\begin{figure}[!htbp]
\centering
\subfigure[]{\includegraphics[width=0.47\textwidth]{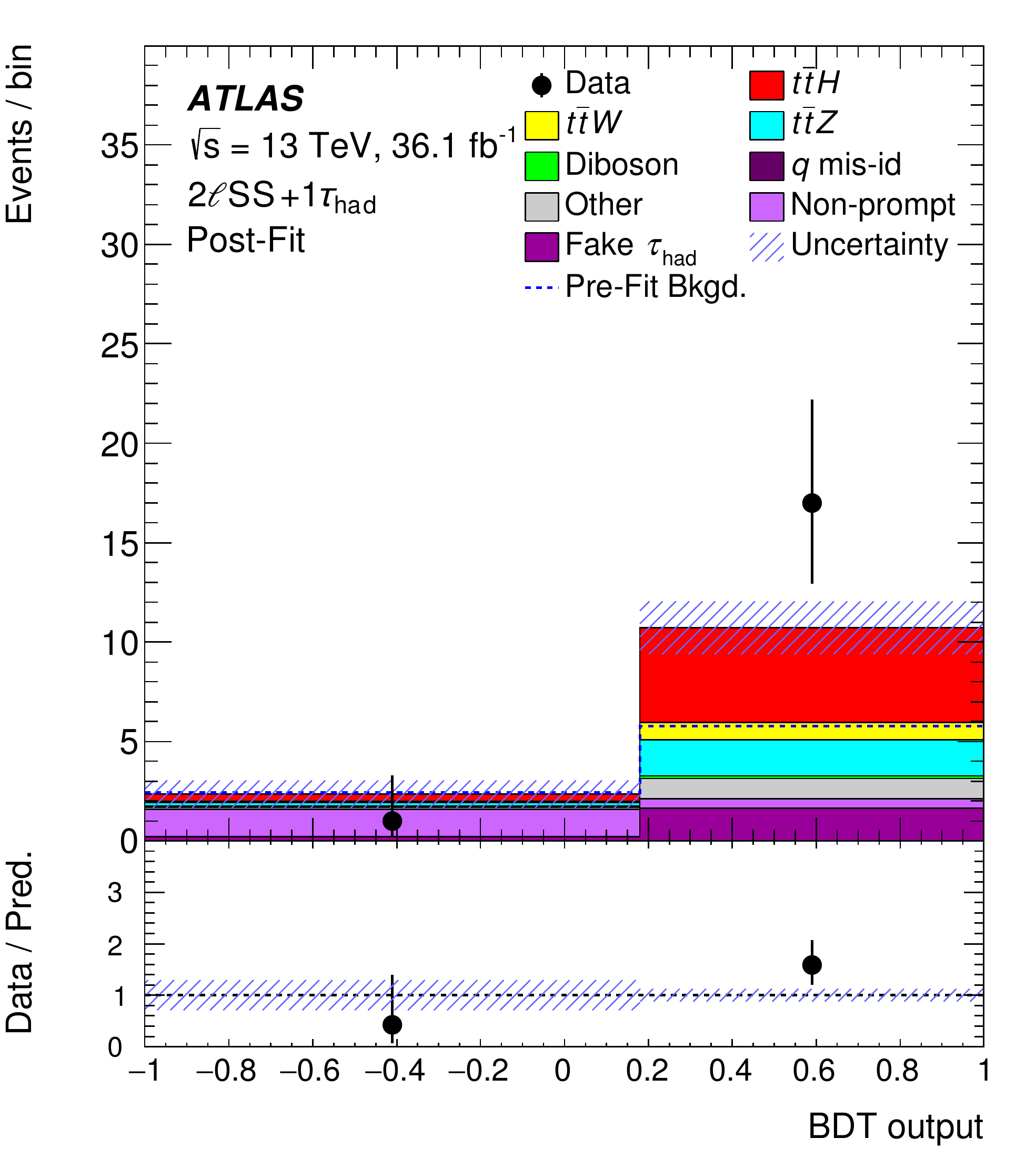}}
\subfigure[]{\includegraphics[width=0.47\textwidth]{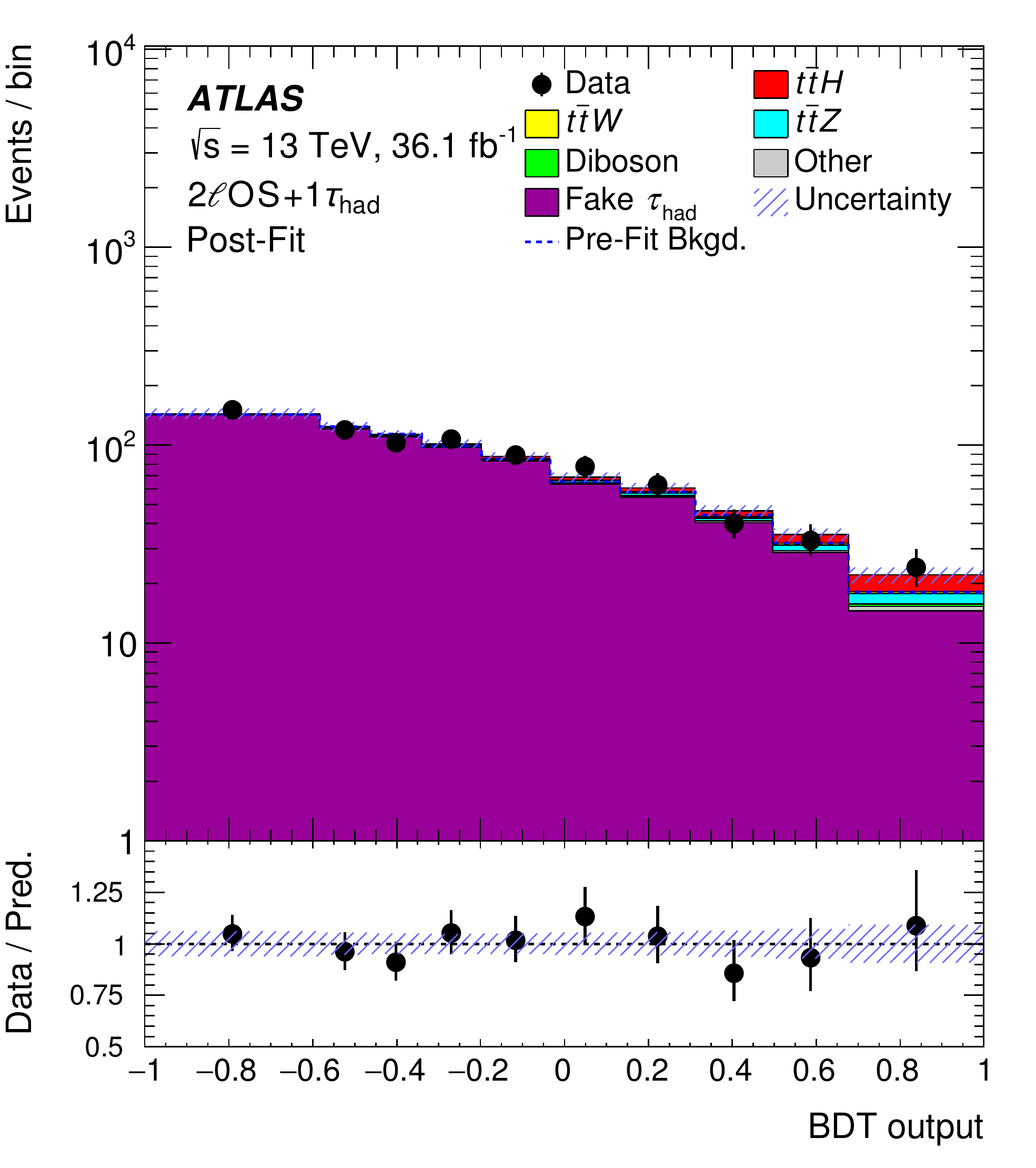}}
\subfigure[]{\includegraphics[width=0.47\textwidth]{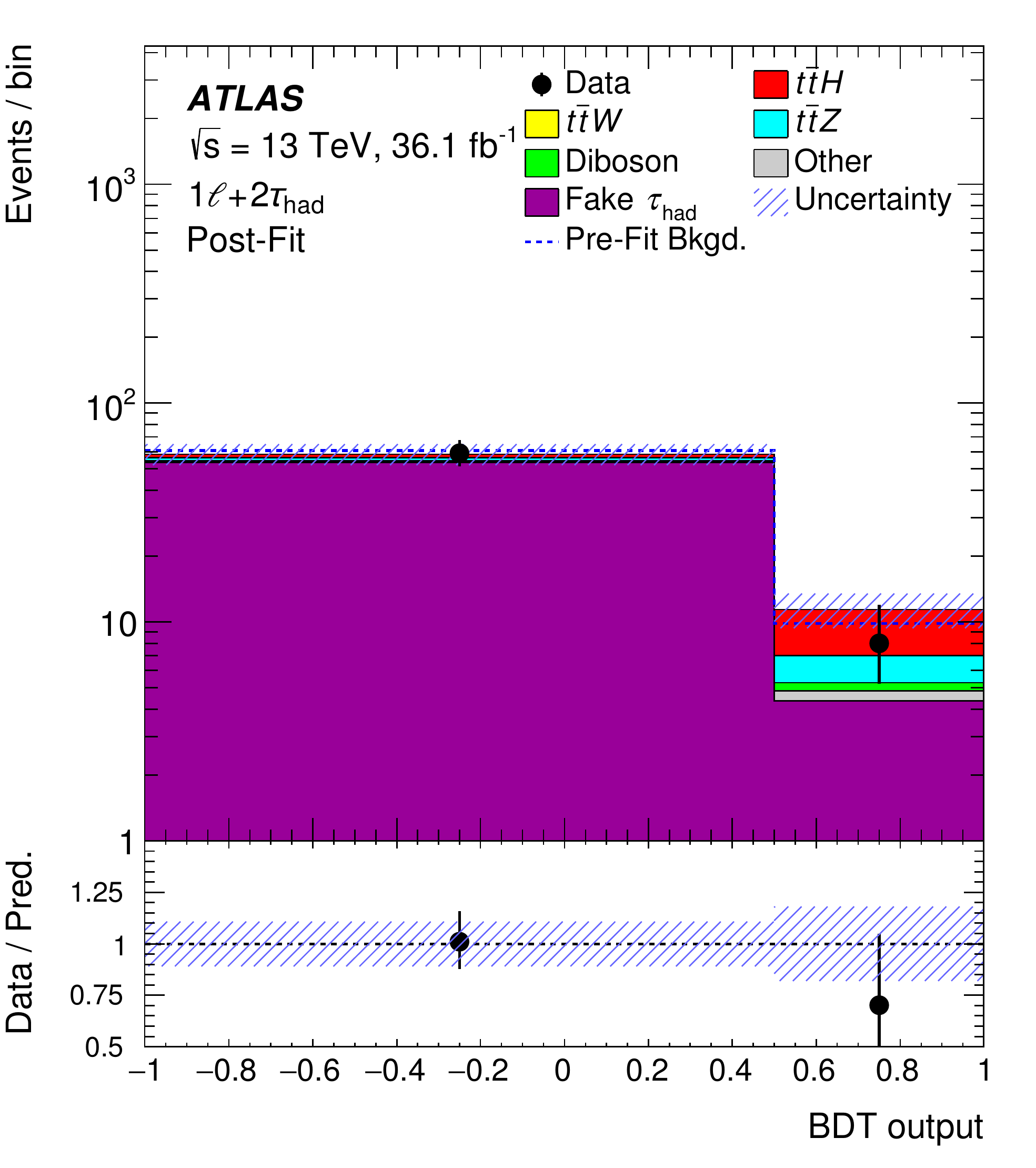}}
\subfigure[]{\includegraphics[width=0.47\textwidth]{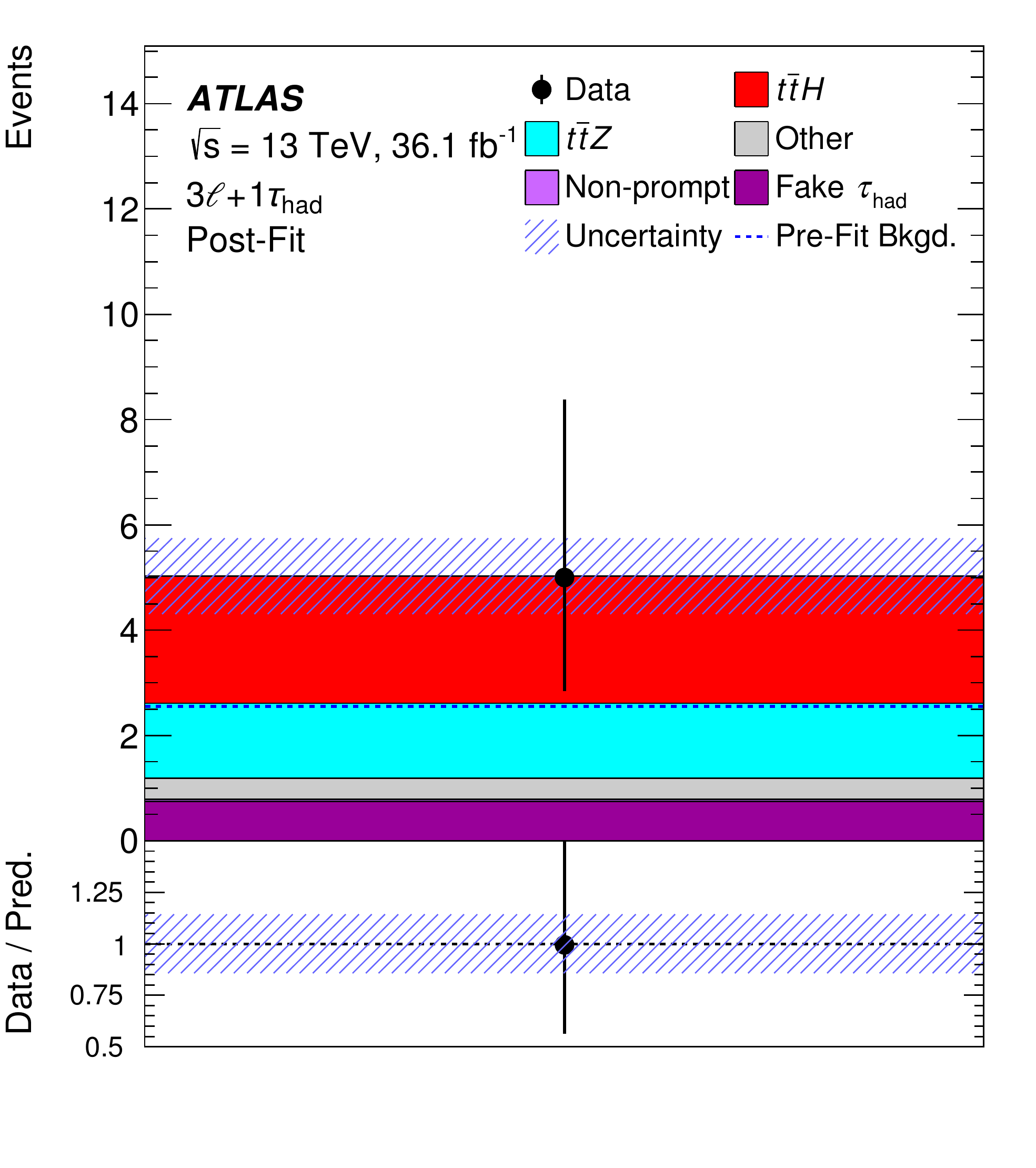}}
  \caption{The distribution of the discriminating variables observed in data (points with bars indicating the statistical errors) and expected
(histograms) in the (a) \lltau, (b) \OSlltau, (c) \ltwotau and (d) \llltau signal regions.
The background contributions after the global fit are shown as filled histograms.
The total background before the fit is shown as a dashed blue histogram.
The Higgs boson signal (\mH = 125 GeV), scaled according to the results of the fit, is shown as a filled red histogram superimposed on the fitted backgrounds.
The size of the combined statistical and systematic uncertainty in the sum of the signal
and fitted background is indicated by the blue hatched band. The ratio of the data to the sum of the
signal and fitted background is shown in the lower panel.\label{fig:TauPlots}}
\end{figure}

\begin{figure}[!htbp]
\centering
\includegraphics[width=0.9\textwidth]{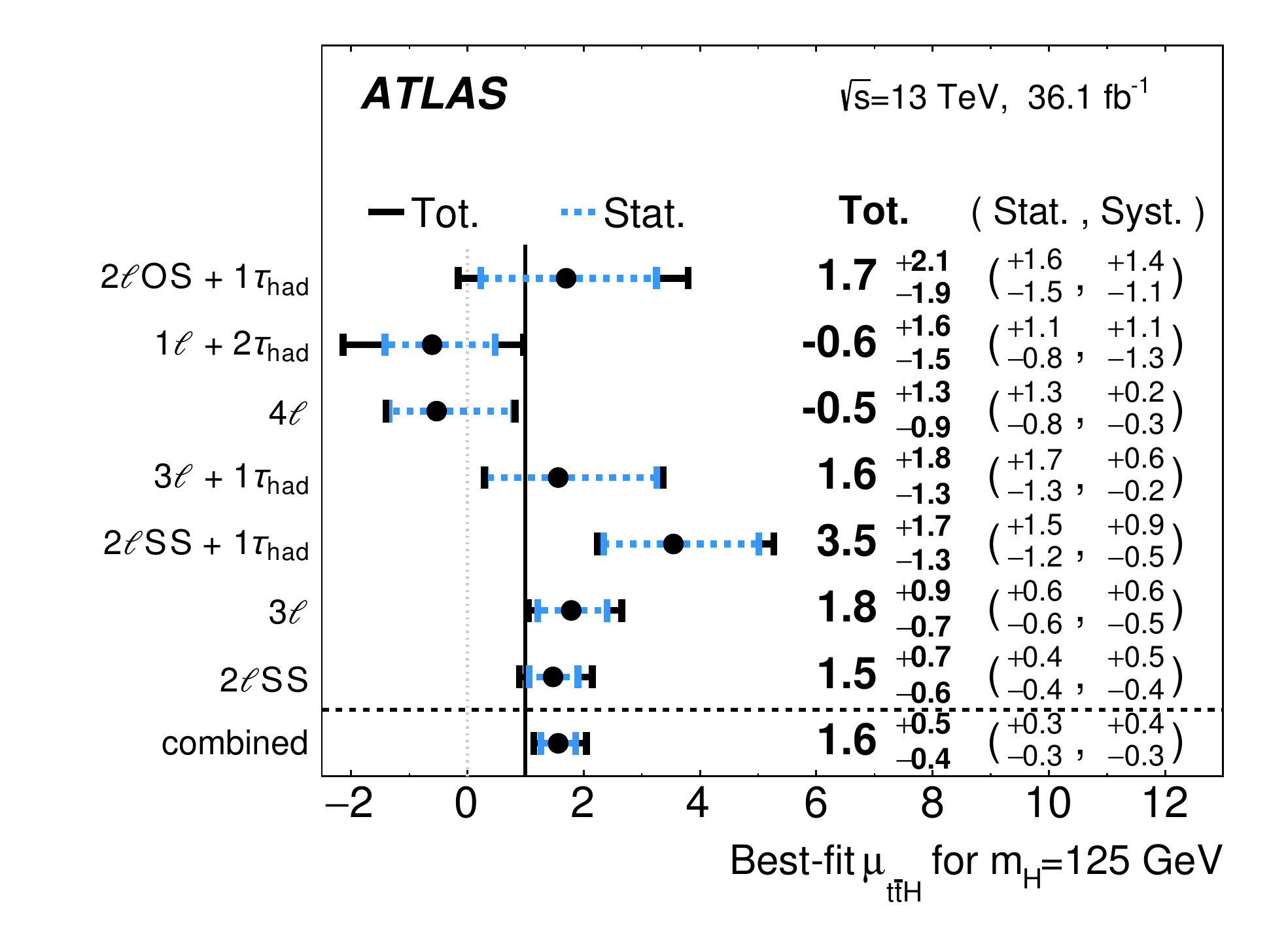}
  \caption{The observed best-fit values of the \ttH signal strength $\mu$ and their uncertainties by final-state category and combined. The individual $\mu$ values for the channels are obtained from a simultaneous fit with the signal-strength parameter for each channel floating independently. The SM prediction is $\mu=1$.\label{fig:mu}}
\end{figure}

\begin{figure}[!htbp]
\centering
\includegraphics[width=0.8\textwidth]{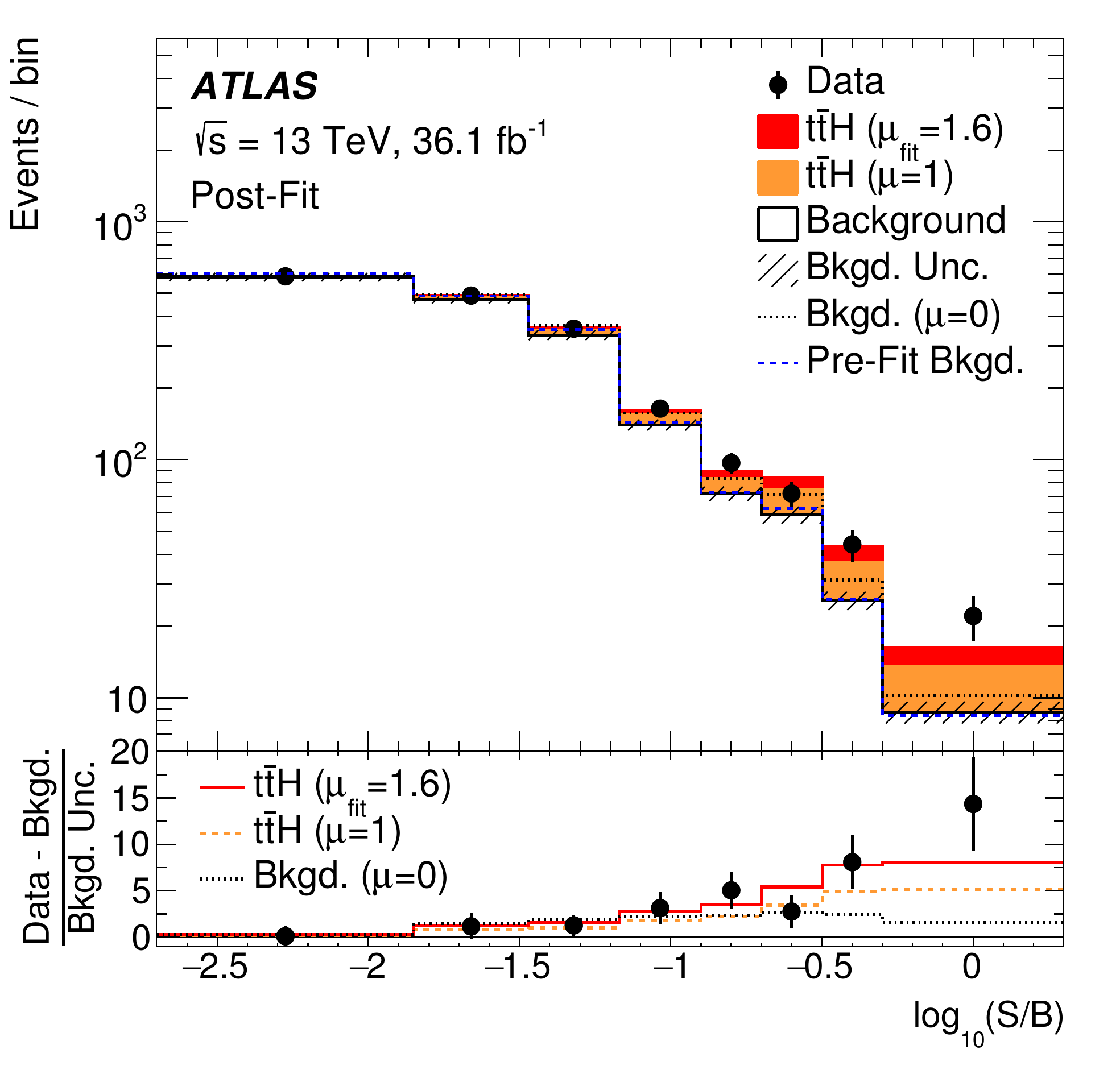}
  \caption{Event yields as a function of log$_{10}(S/B)$ for data, background and a Higgs boson signal with $m_\mathrm{H} = 125$~GeV. The discriminant bins in all signal regions are combined into bins of log$_{10}(S/B)$, where $S$ is the expected signal yield and $B$ the background yield from the unconditional fit. The background yields are shown as the fitted values, while the signal yields are shown for the fitted value ($\mu$=1.6) and the SM prediction ($\mu$=1). The total background before the fit is shown as a dashed blue histogram.
The pull (residual divided by its uncertainty) of the data relative to the background-only prediction is shown in the lower panel, where the full red line (dashed orange line) indicates the pull of the prediction for signal with $\mu$=1.6 ($\mu$=1) and background relative to the background-only prediction.
The background is also shown after the fit to data assuming zero signal contribution as well as its pull (dotted black line) relative to the background from the nominal fit.
\label{fig:SoverB}}
\end{figure}

\begin{table}
 \begin{center}
 \caption{\label{tab:mu} Observed and expected best-fit values of the signal strength $\mu$ and associated significance under the SM background-only hypothesis. The expected values are shown for the pre-fit background estimates. The observed significance is indicated with a $-$ for the channels where $\mu$ is negative.}
  \begin{tabular}{lcccc}
  \hline\hline
   Channel  & \multicolumn{2}{c}{Best-fit $\mu$}       &  \multicolumn{2}{c}{Significance}\\
            & Observed              & Expected            & Observed              & Expected \\
   \hline\noalign{\vskip 1mm}
   \OSlltau  &  \phantom{$-$}1.7 $^{+1.6}_{-1.5}$ (stat.) $^{+1.4}_{-1.1}$ (syst.) & 1.0 $^{+1.5}_{-1.4}$ (stat.) $^{+1.2}_{-1.1}$ (syst.) & 0.9$\sigma$ & 0.5$\sigma$\\ \noalign{\vskip 1mm}
   \ltwotau  & $-0.6$ $^{+1.1}_{-0.8}$ (stat.) $^{+1.1}_{-1.3}$ (syst.) & 1.0 $^{+1.1}_{-0.9}$ (stat.) $^{+1.2}_{-1.1}$ (syst.) & $-$          & 0.6$\sigma$\\  \noalign{\vskip 1mm}
   \fourl    & $-0.5$ $^{+1.3}_{-0.8}$ (stat.) $^{+0.2}_{-0.3}$ (syst.) & 1.0 $^{+1.7}_{-1.2}$ (stat.) $^{+0.4}_{-0.2}$ (syst.) & $-$          & 0.8$\sigma$\\  \noalign{\vskip 1mm}
   \llltau   &  \phantom{$-$}1.6 $^{+1.7}_{-1.3}$ (stat.) $^{+0.6}_{-0.2}$ (syst.) & 1.0 $^{+1.5}_{-1.1}$ (stat.) $^{+0.4}_{-0.2}$ (syst.) & 1.3$\sigma$ & 0.9$\sigma$\\ \noalign{\vskip 1mm}
   \lltau    &  \phantom{$-$}3.5 $^{+1.5}_{-1.2}$ (stat.) $^{+0.9}_{-0.5}$ (syst.) & 1.0 $^{+1.1}_{-0.8}$ (stat.) $^{+0.5}_{-0.3}$ (syst.) & 3.4$\sigma$ & 1.1$\sigma$\\  \noalign{\vskip 1mm}
   \threel   &  \phantom{$-$}1.8 $^{+0.6}_{-0.6}$ (stat.) $^{+0.6}_{-0.5}$ (syst.) & 1.0 $^{+0.6}_{-0.5}$ (stat.) $^{+0.5}_{-0.4}$ (syst.) & 2.4$\sigma$ & 1.5$\sigma$\\ \noalign{\vskip 1mm}
   \ll       &  \phantom{$-$}1.5 $^{+0.4}_{-0.4}$ (stat.) $^{+0.5}_{-0.4}$ (syst.) & 1.0 $^{+0.4}_{-0.4}$ (stat.) $^{+0.4}_{-0.4}$ (syst.) & 2.7$\sigma$ & 1.9$\sigma$\\ \noalign{\vskip 1mm}
   \hline \noalign{\vskip 1mm}
   Combined  &  \phantom{$-$}1.6 $^{+0.3}_{-0.3}$ (stat.) $^{+0.4}_{-0.3}$ (syst.) & 1.0 $^{+0.3}_{-0.3}$ (stat.) $^{+0.3}_{-0.3}$ (syst.) & 4.1$\sigma$ & 2.8$\sigma$\\  \noalign{\vskip 1mm}
   \hline\hline
  \end{tabular}
 \end{center}
\end{table}

\clearpage

\section{Combination of ATLAS \ttH searches}
\label{sec:combo}

In addition to the results reported in Section~\ref{sec:result} (referred to
hereafter as the multilepton analysis), the ATLAS Collaboration has
carried out searches for \ttH production at $\sqrt{s} = 13$ TeV using other
Higgs boson decay modes:

\begin{itemize}
\item \Htobb, in the lepton+jets and dileptonic \ttbar final states \cite{ttHbb}.
\item \Htoyy, in lepton+jets/dileptonic and all-hadronic \ttbar decay channels
  \cite{yy}. In addition, specialized categories sensitive to $tHqb/WtH$
  production also have significant \ttH acceptance and are included.
  \item $\HtoZZ \to 4\ell$ (hereafter $H \to 4\ell$), in a single category including all \ttbar decay
    channels \cite{4l}.
\end{itemize}

All analyses use the same 36.1~\ifb\ of data.  The overlap between the signal and control regions of all the analyses was checked and found to be negligible.  All analyses use the same Monte
Carlo event generators
for \ttH production, and use nominal Higgs boson decay branching fractions from
Ref.~\cite{deFlorian:2016spz} assuming $m_{H} = 125$~GeV.

For the extraction of the \ttH signal strength $\mu$,
the single top quark and Higgs boson associated production processes $tHqb$ and $WtH$ are considered
backgrounds and fixed to their SM predictions with appropriate theoretical
uncertainties.  All other Higgs boson production mechanisms contribute
negligibly to the multilepton and \Htobb analyses and are ignored.
The searches for \ttH production in \Htoyy and $H \to 4\ell$ reported in
Refs.~\cite{yy} and \cite{4l} both utilize categories targeting \ttH production in global analyses
of all Higgs boson production; in the following result, only
the \ttH-enhanced categories from those results are considered.  These
categories have non-negligible
contamination from other production mechanisms (4--21\% in
the \ttH categories with
\Htoyy, 21--64\% in the $tHqb/WtH$ categories with \Htoyy, 23\% in $H\to
4\ell$).  The best-fit values for $\mu$ obtained in those analyses result from
multiple-parameter-of-interest fits that allow other Higgs boson production mode
signal strengths to
take on non-SM values.  In the following discussion, non-\ttH Higgs boson
production mechanism cross sections and all Higgs boson branching fractions are set to SM
expectations with theoretical errors considered as systematic
uncertainties \cite{deFlorian:2016spz}.  This results in slightly different $\mu$ values than reported in
the standalone analyses. Details of the modeling and simulation of non-\ttH
production modes can be found in Refs.~\cite{yy,4l}.

The combined likelihood function \likeli is obtained from the product of likelihood
functions of the individual analyses.  The nuisance parameters associated with
the same sources in the different analyses are treated as follows:
\begin{itemize}
  \item \textbf{Higgs boson production and decay:} all analyses use the same nominal production
    cross sections and decay branching fractions.  All theoretical
    uncertainties associated with these parameters are fully
    correlated between analyses.
  \item \textbf{Background uncertainties:} The cross-section and modeling
    uncertainties for MC-estimated \ttZ, \ttW, $tZqb/WtZ$, $WZ/ZZ$, $Wt$, $t\bar t t\bar
    t$, and $t\bar t WW$ production are correlated between the \Htobb and
    multilepton analyses.  The modeling systematic uncertainties of the dominant
    background of \ttbar in the \Htobb analyses are not applied to any
    other channels, as the relevant regions of phase space are not similar and
    other channels have independent methods of estimating the relevant \ttbar
    background.

   \item \textbf{Experimental uncertainties:} The dominant experimental systematic
    uncertainties are associated with the jet energy scale, jet energy resolution,
    and flavor tagging. Nuisance parameters related to the jet energy scale are correlated between the analyses with the
    exception of the uncertainty in the fractions of jets initiated by quarks and
    by gluons, which differs between the channels.  The jet energy
    resolution is correlated between all channels except for the control
    regions of the \Htobb analysis, to avoid constraining this systematic
    uncertainty in the signal regions; this gives a conservative estimate of the
    impact.  The \Htoyy and $H \to 4\ell$ analyses use a
    different calibration for the flavor-tagging efficiencies and mistag
    rates compared to the \Htobb and multilepton analyses. Due to this, the flavor-tagging
    uncertainties are correlated between \Htoyy and $H\to 4\ell$ and between \Htobb
    and multilepton analyses, but are uncorrelated between the two pairs.
    The flavor-tagging uncertainties are constrained significantly by the \Htobb
    analysis, due to its large samples of $b$- and $c$-jets, which carries
    over to the multilepton analysis.

    Other experimental
    systematic uncertainties such as luminosity, pileup effects, lepton identification, isolation, and trigger efficiencies are treated as
    correlated, except for statistical uncertainties associated with
    efficiency measurements for different working points.
  \end{itemize}
None of the NPs in the fit are strongly constrained by more than one analysis,
    and the value of $\mu$ obtained from the combined fit does not depend on the choice of the correlation scheme.

The best-fit value of the \ttH signal strength, as determined from the combined
likelihood function, is
\[ \mu = 1.17 \pm 0.19\ \textrm{(stat.)}\ ^{+0.27}_{-0.23}\ \textrm{(syst.)}. \]
The background-only hypothesis ($\mu = 0$) is excluded at $4.2\sigma$, with an
expectation of $3.8\sigma$ in the case of a SM signal.  This constitutes
evidence for \ttH production.

The values of $\mu$ obtained in each analysis, and the result of the
combination, are shown in Figure~\ref{fig:combo} and Table~\ref{tbl:combo}.
The probability that the signal strengths from the individual analyses are compatible with the combined value of $\mu$ is 38\%.
The impact of various uncertainties on the combination is shown in
Table~\ref{tbl:combosyst}.  The leading systematic uncertainties are those
associated with the \ttH signal
modeling and cross section and the \ttbar background modeling in the \Htobb
analysis. The cross section for \ttH production corresponding to the best-fit
value of $\mu$ is
$590\ ^{+160}_{-150}$~fb, as compared to the SM prediction of
$\sigma(\ttH) = 507\ ^{+35}_{-50}$~fb.

\begin{figure}
  \centering
  \includegraphics[width=.6\linewidth]{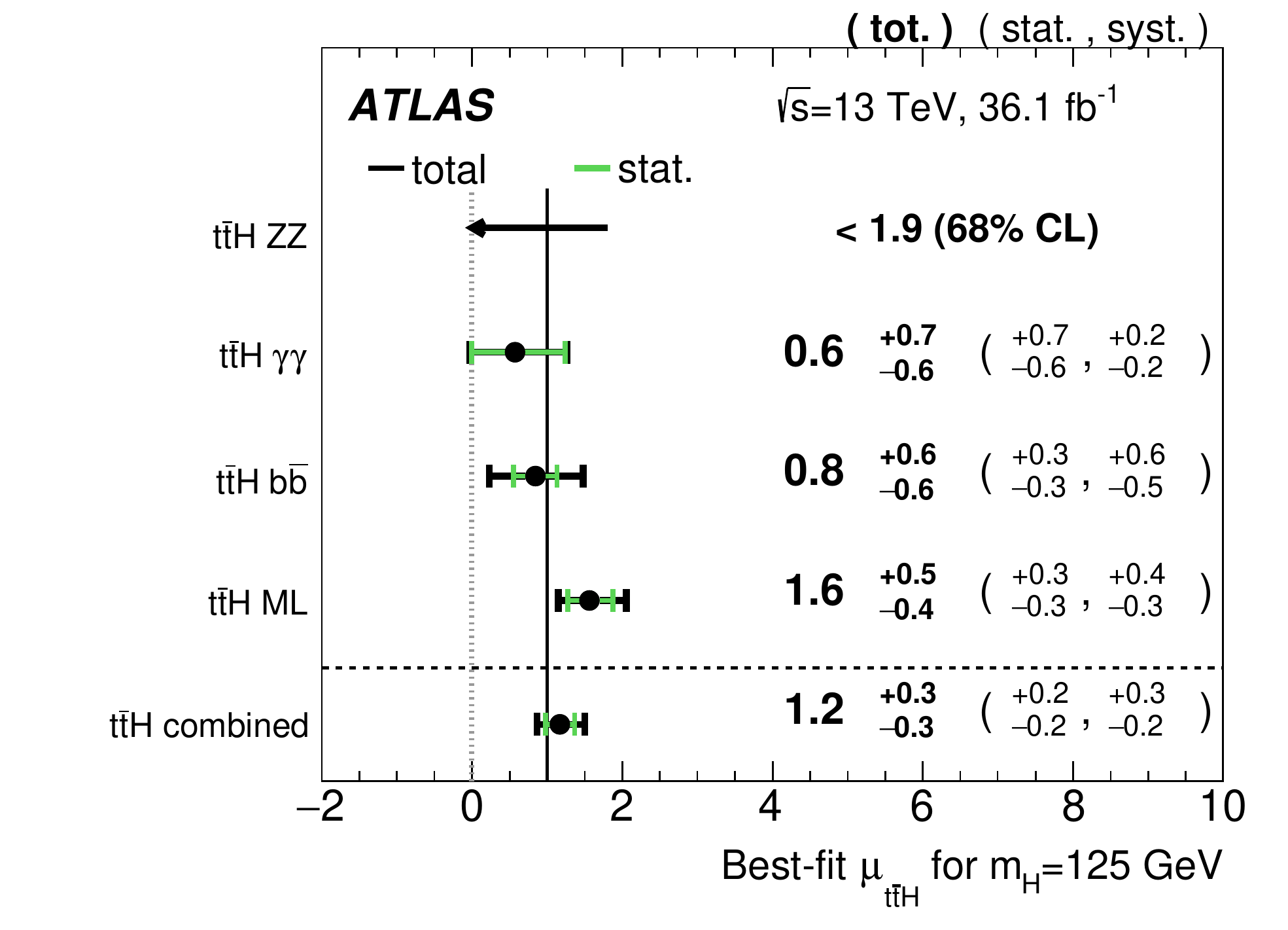}
  \caption{\label{fig:combo} Summary of the measurements of $\mu$ from
    individual analyses and the combined result.  ``ML'' refers to the
  multileptonic decay channels discussed in Section~\ref{sec:result}. The best-fit values of $\mu$ for the
    individual analyses are extracted independently, and systematic uncertainty nuisance parameters are only
    correlated for the combination.  As no events are
    observed in the $H\to 4\ell$ analysis, a 68\% confidence level (CL)\ upper
  limit on $\mu$, computed using the CL$_\mathrm{s}$ method \cite{cls},
    is reported.}
 \end{figure}
\begin{table}
  \centering
  \caption{\label{tbl:combo} Summary of the observed and expected $\mu$
    measurements and \ttH production significance from individual analyses and
  the combination.  As no events are observed in the $H\to4\ell$ analysis, a
  68\% confidence level (CL)\ upper limit on $\mu$, computed using the CL$_\mathrm{s}$ method \cite{cls}, is reported.}
  \begin{tabular}{lcccc}
    \hline\hline
    Channel & \multicolumn{2}{c}{Best-fit $\mu$}
  & \multicolumn{2}{c}{Significance} \\
    & Observed & Expected & Observed & Expected \\
    \hline\noalign{\vskip 1mm}
    Multilepton & $1.6\ ^{+0.5}_{-0.4}$ & $1.0\ ^{+0.4}_{-0.4}$ & 4.1$\sigma$
  & 2.8$\sigma$\\\noalign{\vskip 1mm}
    \Htobb & $0.8\ ^{+0.6}_{-0.6}$ & $1.0\ ^{+0.6}_{-0.6}$ & 1.4$\sigma$ & 1.6$\sigma$\\\noalign{\vskip 1mm}
    \Htoyy & $0.6\ ^{+0.7}_{-0.6}$ & $1.0\ ^{+0.8}_{-0.6}$ & 0.9$\sigma$ & 1.7$\sigma$\\\noalign{\vskip 1mm}
    $H \to 4\ell$ & $<1.9$ & $1.0\ ^{+3.2}_{-1.0}$ & --- & 0.6$\sigma$ \\\noalign{\vskip 1mm}
    \hline\noalign{\vskip 1mm}
    Combined & $1.2\ ^{+0.3}_{-0.3}$ & $1.0\ ^{+0.3}_{-0.3}$ & 4.2$\sigma$
  & 3.8$\sigma$\\\noalign{\vskip 1mm}
    \hline\hline
    \end{tabular}
\end{table}

\begin{table}
  \centering
  \caption{\label{tbl:combosyst} Summary of the uncertainties
    affecting the combined value of $\mu$. 
      }
  \begin{tabular}{lcc}
    \hline\hline
    Uncertainty Source & \multicolumn{2}{c}{$\Delta\mu$} \\
    \hline
    \ttbar modeling in \Htobb analysis & $+0.15$ & $-0.14$\\
    \ttH\ modeling (cross section) & $+0.13$ & $-0.06$\\
    Non-prompt light-lepton and fake \tauh\ estimates & $+0.09$ & $-0.09$\\
    Simulation statistics & $+0.08$ & $-0.08$ \\
    Jet energy scale and resolution & $+0.08$ & $-0.07$ \\
    \ttV\ modeling & $+0.07$ & $-0.07$\\
    \ttH\ modeling (acceptance) & $+0.07$ & $-0.04$\\
    Other non-Higgs boson backgrounds & $+0.06$ & $-0.05$ \\
    Other experimental uncertainties & $+0.05$ & $-0.05$\\
    Luminosity & $+0.05$ & $-0.04$ \\
    Jet flavor tagging & $+0.03$ & $-0.02$ \\
    Modeling of other Higgs boson production modes & $+0.01$ & $-0.01$\\
    \hline
    Total systematic uncertainty & $+0.27$ & $-0.23$ \\
    \hline\hline
    Statistical uncertainty & $+0.19$ & $-0.19$ \\
    \hline\hline
    Total uncertainty & $+0.34$ & $-0.30$ \\
    \hline\hline
  \end{tabular}
\end{table}

Due to the different acceptances for the different analysis categories for different Higgs
boson decay modes, it is possible to independently determine $\mu$ in
different Higgs boson decay modes.  In particular the multilepton analysis has
categories with zero and $\ge 1$ \tauh\ candidates, which are enriched in
\HtoWW and \Htott, respectively (see Figure~\ref{fig:soverb}).  The result of a fit for four signal
strengths is shown in Figure~\ref{fig:comboBR}.  Due to very weak sensitivity
for \HtoZZ, the ratio of branching fractions of \HtoZZ and \HtoWW are assumed
to be as in the SM and a single combined signal strength for $H \to VV$ is computed. For \Htobb and \Htoyy the result is
essentially the same as for the individual analyses, due to the high purity of
those signal regions for the respective Higgs boson decays.  The \HtoWW and \Htott
decays are distinguished only by their different contributions to the various
multilepton signal regions, resulting in a significant
anticorrelation.  Two-dimensional scans of the
signal strengths are shown in Figure~\ref{fig:comboBRbymode} for \Htobb versus
$H \to VV$ and for \Htott versus $H \to VV$; in these plots the two signal
strengths not shown are profiled in the scan.

\begin{figure}
  \centering
  \includegraphics[width=.6\linewidth]{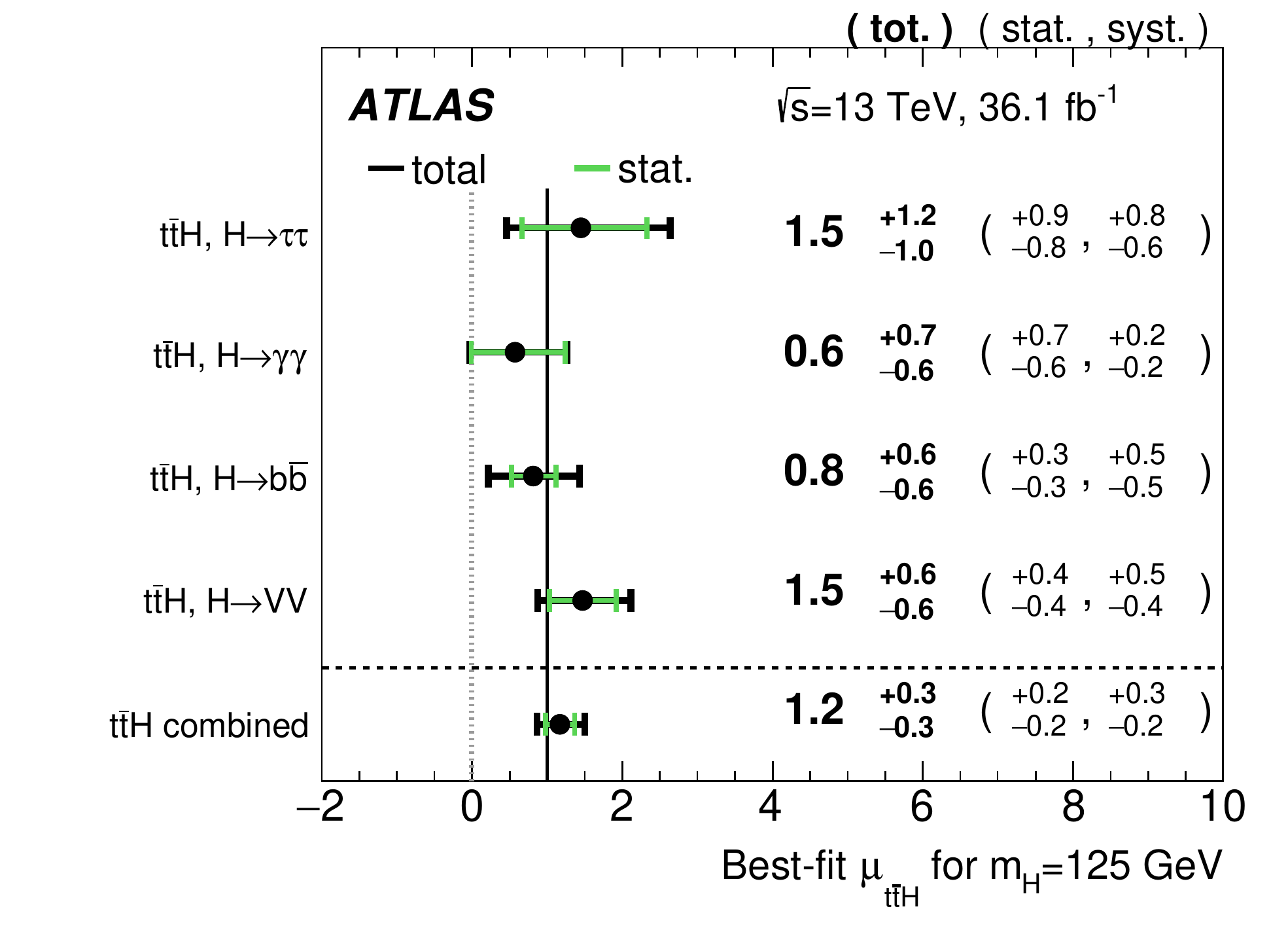} 
  \caption{\label{fig:comboBR} Summary of the best-fit values of $\mu$ broken
  down by Higgs boson decay mode.  The decays $H \to WW^*$ and $H \to ZZ^*$
  are assumed to have the same signal-strength modification factor and are
  shown together as $VV$. All systematic uncertainties are correlated as in
  the nominal result.}
 \end{figure}

\begin{figure}
  \centering
\includegraphics[width=.5\linewidth]{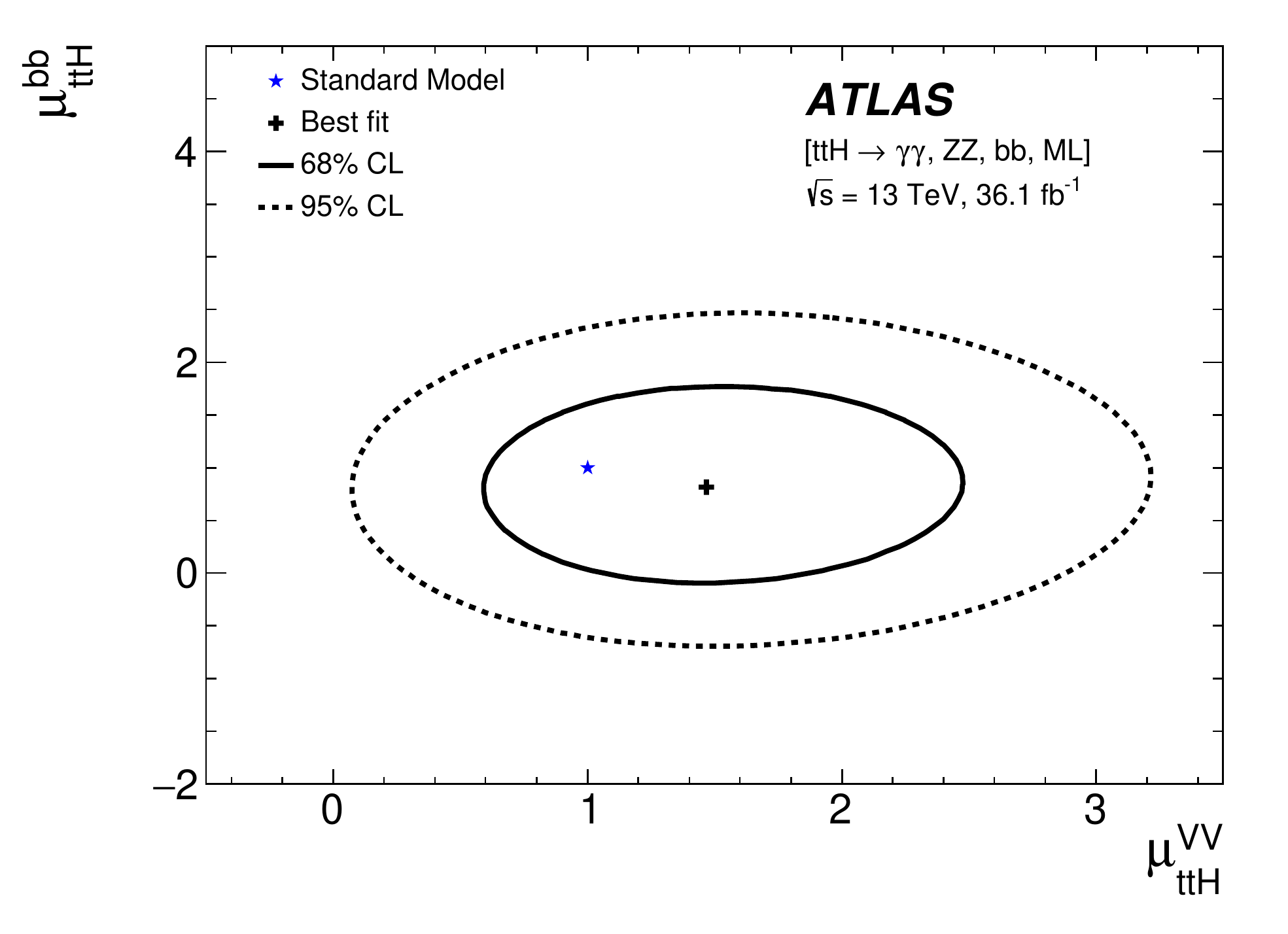}
\includegraphics[width=.5\linewidth]{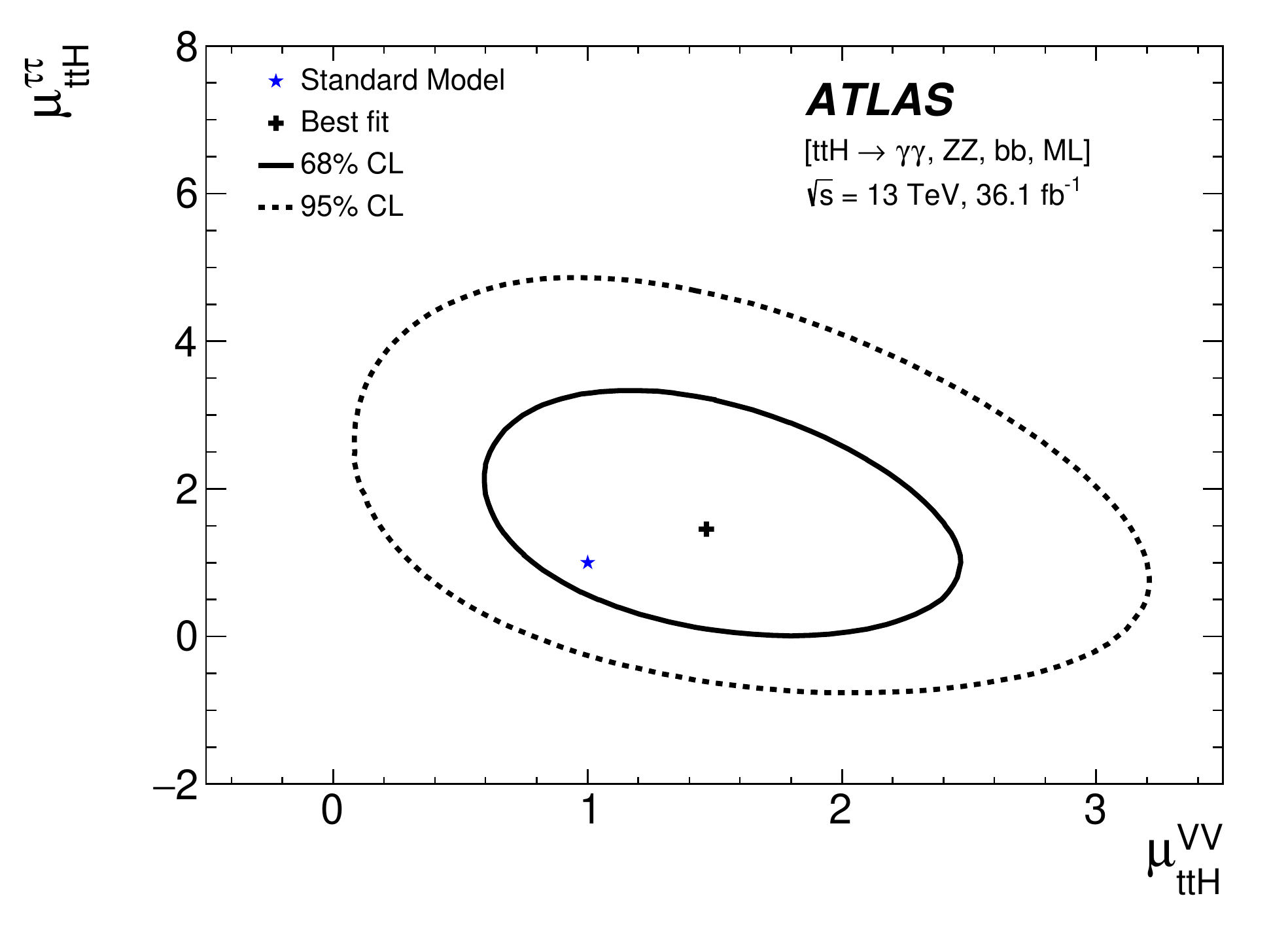}
\caption{\label{fig:comboBRbymode} Two-dimensional scans of the
signal-strength modifiers for the processes (left) $\ttH, \Htobb$ versus $\ttH, H \to
  WW^*/ZZ^*$ and (right) $\ttH, \Htott$ versus $\ttH, H \to
  WW^*/ZZ^*$.  The two signal strengths not appearing in each plot are
  profiled.  The decays $H \to WW^*$ and $H \to ZZ^*$
  are assumed to have the same signal-strength modification factor $\mu^{VV}$.}
 \end{figure}

The \ttH analyses are sensitive to the $Htt$, $Hbb$, and $H\tau\tau$ fermion
couplings, the $HWW$ and $HZZ$ gauge boson couplings, and the effective
$H\gamma\gamma$ coupling.  Accordingly, constraints can be
placed on deviations of these couplings from the SM.  An interpretation is
made using the $\kappa$-parameterization, in which
Higgs boson couplings to particle species $i$ are linearly scaled by factors
$\kappa_i$.  Here, all fermion couplings are assumed to scale by a common
factor $\kappa_F$ and the $WW/ZZ$ couplings by a common factor $\kappa_V$. As only the relative sign of the $\kappa$ factors is meaningful, the convention that $\kappa_V \ge 0$ is chosen.
Modifications to loop-induced processes are determined by multiplying the contributing SM amplitudes by
the relevant $\kappa$-factors; no contributions from non-SM particles are
considered and no non-SM Higgs boson decay modes are allowed. The relevant parameterizations are given in Ref.~\cite{\ifIsML HIGG-2015-07\else Khachatryan:2016vau\fi}.  In particular the factor $\kappa_\gamma$
modifying the effective $H\gamma\gamma$ coupling
is expressed in terms of $\kappa_V$ and $\kappa_F$, and $\kappa_g$ is set equal
to $\kappa_F$.  The total width of the Higgs boson is modified
appropriately.

The \ttH analyses, especially the \Htoyy, multilepton, and $H \to 4\ell$ channels,
have acceptance for $tHqb$ and $WtH$ production.  The amplitudes for the \Htoyy decay and the production of $tHqb$
and $WtH$ involve interference between the $Htt$ and $HWW$ couplings.  In the
SM, the interference is destructive, almost
completely in the case of $tHqb$ and $WtH$.  As a
result, a global analysis of the \ttH channels, in this parameterization, is
able to resolve the relative sign of the two couplings.

A likelihood scan is performed in the
$\kappa_V$--$\kappa_F$ plane.  The analysis acceptances for all Higgs boson
production mechanisms
and decays are assumed to be constant as the $\kappa$ parameters are varied
over the scanned region, with only rates
being modified.  The results are shown in
Figure~\ref{fig:kfkv}, and are in good agreement with the Standard Model values
$\kappa_F, \kappa_V = 1$. The possibility that $\kappa_F < 0$ is excluded at 95\% CL in this parameterization.

\begin{figure}
\centering
  \includegraphics[width=.6\linewidth]{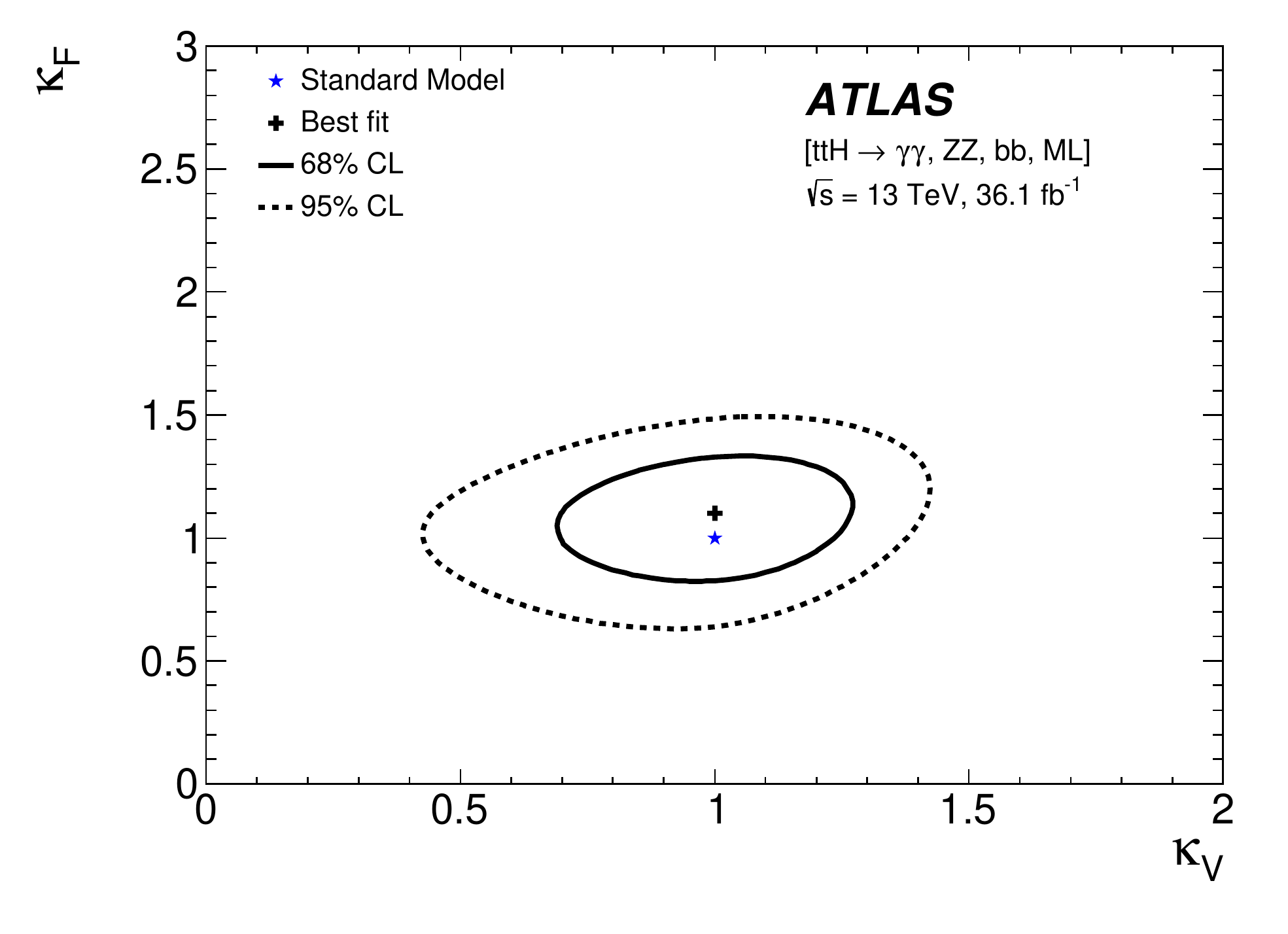}
  \caption{\label{fig:kfkv}Allowed regions at 68\% and 95\% CL\ in the
    $\kappa_V$--$\kappa_F$ plane from the combination of all \ttH
    channels. The Higgs boson is assumed to not couple to any particles beyond
  the Standard Model, and the $H \to \gamma\gamma$ and $H \to gg$ couplings are
  expressed in terms of $\kappa_F$ and $\kappa_V$.}
\end{figure}

\FloatBarrier

\section{Conclusions}
\label{sec:conclusion}

A search for \ttH production in multilepton final states using a dataset corresponding to an integrated luminosity of 36.1~\ifb\ of proton--proton collision at \ThirTeV\ recorded by the ATLAS experiment at the LHC is presented. Seven final states, targeting Higgs boson decays to ${WW^*}$, \ensuremath{\tau\tau}, and $\ensuremath{ZZ^*}$,  categorized by the number and flavor of charged-lepton candidates, are analyzed.
An excess of events over the expected background from SM processes is found, which is interpreted as an observed significance of 4.1 standard deviations for a SM Higgs boson of mass 125~GeV.
The expected significance for a SM Higgs boson is 2.8 standard deviations. The best-fit result of the observed production cross section is $\sigma(\ttH) = 790^{+230}_{-210}$~fb, in agreement with the SM prediction of $507^{+35}_{-50}$~fb.

The combination of this result with other \ttH studies from the ATLAS experiment using the Higgs boson decay modes to $b\bbar$, $\gamma\gamma$ and $ZZ^* \to 4\ell$ is presented. The combination has an observed significance of 4.2 standard deviations, compared to an expectation of 3.8 standard deviations. The cross section for \ttH production is measured to be $\sigma(\ttH) = 590^{+160}_{-150}$~fb, in agreement with the SM prediction.
This provides evidence for the \ttH production mode.

\section*{Acknowledgments}

We thank CERN for the very successful operation of the LHC, as well as the
support staff from our institutions without whom ATLAS could not be
operated efficiently.

We acknowledge the support of ANPCyT, Argentina; YerPhI, Armenia; ARC, Australia; BMWFW and FWF, Austria; ANAS, Azerbaijan; SSTC, Belarus; CNPq and FAPESP, Brazil; NSERC, NRC and CFI, Canada; CERN; CONICYT, Chile; CAS, MOST and NSFC, China; COLCIENCIAS, Colombia; MSMT CR, MPO CR and VSC CR, Czech Republic; DNRF and DNSRC, Denmark; IN2P3-CNRS, CEA-DRF/IRFU, France; SRNSF, Georgia; BMBF, HGF, and MPG, Germany; GSRT, Greece; RGC, Hong Kong SAR, China; ISF, I-CORE and Benoziyo Center, Israel; INFN, Italy; MEXT and JSPS, Japan; CNRST, Morocco; NWO, Netherlands; RCN, Norway; MNiSW and NCN, Poland; FCT, Portugal; MNE/IFA, Romania; MES of Russia and NRC KI, Russian Federation; JINR; MESTD, Serbia; MSSR, Slovakia; ARRS and MIZ\v{S}, Slovenia; DST/NRF, South Africa; MINECO, Spain; SRC and Wallenberg Foundation, Sweden; SERI, SNSF and Cantons of Bern and Geneva, Switzerland; MOST, Taiwan; TAEK, Turkey; STFC, United Kingdom; DOE and NSF, United States of America. In addition, individual groups and members have received support from BCKDF, the Canada Council, CANARIE, CRC, Compute Canada, FQRNT, and the Ontario Innovation Trust, Canada; EPLANET, ERC, ERDF, FP7, Horizon 2020 and Marie Sk{\l}odowska-Curie Actions, European Union; Investissements d'Avenir Labex and Idex, ANR, R{\'e}gion Auvergne and Fondation Partager le Savoir, France; DFG and AvH Foundation, Germany; Herakleitos, Thales and Aristeia programmes co-financed by EU-ESF and the Greek NSRF; BSF, GIF and Minerva, Israel; BRF, Norway; CERCA Programme Generalitat de Catalunya, Generalitat Valenciana, Spain; the Royal Society and Leverhulme Trust, United Kingdom.

The crucial computing support from all WLCG partners is acknowledged gratefully, in particular from CERN, the ATLAS Tier-1 facilities at TRIUMF (Canada), NDGF (Denmark, Norway, Sweden), CC-IN2P3 (France), KIT/GridKA (Germany), INFN-CNAF (Italy), NL-T1 (Netherlands), PIC (Spain), ASGC (Taiwan), RAL (UK) and BNL (USA), the Tier-2 facilities worldwide and large non-WLCG resource providers. Major contributors of computing resources are listed in Ref.~\cite{ATL-GEN-PUB-2016-002}.

\printbibliography

\clearpage
\begin{flushleft}
{\Large The ATLAS Collaboration}

\bigskip

M.~Aaboud$^\textrm{\scriptsize 137d}$,
G.~Aad$^\textrm{\scriptsize 88}$,
B.~Abbott$^\textrm{\scriptsize 115}$,
O.~Abdinov$^\textrm{\scriptsize 12}$$^{,*}$,
B.~Abeloos$^\textrm{\scriptsize 119}$,
S.H.~Abidi$^\textrm{\scriptsize 161}$,
O.S.~AbouZeid$^\textrm{\scriptsize 139}$,
N.L.~Abraham$^\textrm{\scriptsize 151}$,
H.~Abramowicz$^\textrm{\scriptsize 155}$,
H.~Abreu$^\textrm{\scriptsize 154}$
% ...
\bigskip
\\
% ...
$^{12}$ Institute of Physics, Azerbaijan Academy of Sciences, Baku, Azerbaijan\\
% ...
$^{88}$ CPPM, Aix-Marseille Universit{\'e} and CNRS/IN2P3, Marseille, France\\
% ...
$^{115}$ Homer L. Dodge Department of Physics and Astronomy, University of Oklahoma, Norman OK, United States of America\\
% ...
$^{119}$ LAL, Univ. Paris-Sud, CNRS/IN2P3, Universit{\'e} Paris-Saclay, Orsay, France\\
% ...
$^{137}$ $^{(a)}$ Facult{\'e} des Sciences Ain Chock, R{\'e}seau Universitaire de Physique des Hautes Energies - Universit{\'e} Hassan II, Casablanca; $^{(b)}$ Centre National de l'Energie des Sciences Techniques Nucleaires, Rabat; $^{(c)}$ Facult{\'e} des Sciences Semlalia, Universit{\'e} Cadi Ayyad, LPHEA-Marrakech; $^{(d)}$ Facult{\'e} des Sciences, Universit{\'e} Mohamed Premier and LPTPM, Oujda; $^{(e)}$ Facult{\'e} des sciences, Universit{\'e} Mohammed V, Rabat, Morocco\\
% ...
$^{139}$ Santa Cruz Institute for Particle Physics, University of California Santa Cruz, Santa Cruz CA, United States of America\\
% ...
$^{151}$ Department of Physics and Astronomy, University of Sussex, Brighton, United Kingdom\\
% ...
$^{154}$ Department of Physics, Technion: Israel Institute of Technology, Haifa, Israel\\
$^{155}$ Raymond and Beverly Sackler School of Physics and Astronomy, Tel Aviv University, Tel Aviv, Israel\\
% ...
$^{161}$ Department of Physics, University of Toronto, Toronto ON, Canada\\
% ...
$^{a}$ Also at Department of Physics, King's College London, London, United Kingdom\\
$^{b}$ Also at Institute of Physics, Azerbaijan Academy of Sciences, Baku, Azerbaijan\\
$^{c}$ Also at Novosibirsk State University, Novosibirsk, Russia\\
$^{d}$ Also at TRIUMF, Vancouver BC, Canada\\
$^{e}$ Also at Department of Physics {\&} Astronomy, University of Louisville, Louisville, KY, United States of America\\
% ...
$^{*}$ Deceased
\end{flushleft}


\clearpage
\appendix

\end{document}